\pdfoutput=1
\documentclass[cernpreprint, atlasdraft=false, texlive=2023, UKenglish, texmf, orcidlogo]{atlasdoc}
 
\usepackage{atlaspackage}
\usepackage{atlasbiblatex}
\usepackage{multirow}
\usepackage{array}
\usepackage{xcolor}
\usepackage{bbding}
\usepackage{pifont}
\usepackage{wasysym}
\newcolumntype{~}{>{\global\let\currentrowstyle\relax}}
\newcolumntype{^}{>{\currentrowstyle}}
\newcommand{\rowstyle}[1]{\gdef\currentrowstyle{#1}
#1\ignorespaces
}
\usepackage{atlasphysics}

\graphicspath{{logos/}{figures/}}
 
\usepackage{ANA-HDBS-2020-03-PAPER-defs}
 
\LEcontact{Bob Kowalewski bob.kowalewski@cern.ch}

\AtlasTitle{Search for heavy Higgs bosons with flavour-violating couplings in multi-lepton plus $b$-jets final states in $pp$ collisions at $13\,\mathrm{TeV}$ with the ATLAS detector}

\AtlasAbstract{
A search for new heavy scalars with flavour-violating decays in final states with multiple leptons and $b$-tagged jets is presented.
The results are interpreted in terms of a general two-Higgs-doublet model involving an additional scalar with couplings to the top-quark and the three up-type quarks ($\rho_{tt}$, $\rho_{tc}$, and $\rho_{tu}$).
The targeted signals lead to final states with either a same-sign top-quark pair, three top-quarks, or four top-quarks.
The search is based on a data sample of proton--proton collisions at $\sqrt{s}=13~\tev$ recorded with the ATLAS detector during Run 2 of the Large Hadron Collider, corresponding to an integrated luminosity of 139~fb$^{-1}$.
Events are categorised depending on the multiplicity of light charged leptons (electrons or muons), total lepton charge, and a deep-neural-network output to enhance the purity of each of the signals.
Masses of an additional scalar boson $m_{H}$ between $200-630$ \gev\ with couplings $\rho_{tt}=0.4$, $\rho_{tc}=0.2$, and $\rho_{tu}=0.2$ are excluded at 95\% confidence level.
Additional interpretations are provided in models of $R$-parity violating supersymmetry, motivated by the recent flavour and $(g-2)_\mu$ anomalies.
}
 
\author{The ATLAS Collaboration}
 
\AtlasRefCode{HDBS-2020-03}

\PreprintIdNumber{CERN-EP-2023-040}

\AtlasJournal{JHEP}
\AtlasJournalRef{JHEP 12 (2023) 081}
\AtlasDOI{10.1007/JHEP12(2023)081}

\hypersetup{pdftitle={ATLAS document},pdfauthor={The ATLAS Collaboration}}
 
\begin{document}
 
\maketitle

\section{Introduction}
\label{sec:intro}

% The next lines are included from the .//Tex/intro.tex input file
Several extensions of the Standard Model (SM) propose the augmentation of the Higgs sector by the addition of a second
complex Higgs doublet~\cite{Branco:2011iw,Gunion:2002zf} (2HDM), giving rise to five Higgs bosons: two CP-even scalar fields $h$ and $H$, one CP-odd pseudo-scalar
$A$, and two charged fields $H^\pm$.
The two CP-even scalars are expected to mix; however, the measurement of Higgs boson properties has revealed no deviations from the expectations of the Standard Model~\cite{HIGG-2021-23, CMS-HIG-22-001}. This implies that extra scalars from 2HDMs have to be either very heavy (decoupling limit) or have a vanishingly small mixing with the SM Higgs (alignment limit). To avoid flavour changing neutral Higgs (FCNH) couplings mediated by the SM Higgs, a discrete $Z_2$ symmetry is usually imposed~\cite{Branco:2011iw,Gunion:2002zf}.
A large set of searches for heavy scalars or pseudo-scalars with flavour-conserving decays has been performed
in ATLAS~\cite{HDBS-2018-34, EXOT-2018-46, HIGG-2018-27, HDBS-2018-51, HDBS-2018-13, HIGG-2018-09, HDBS-2018-10, HDBS-2018-46} and CMS~\cite{CMS-B2G-19-002, CMS-HIG-20-014, CMS-HIG-20-017, CMS-HIG-18-013, CMS-HIG-18-021, CMS-HIG-18-015, CMS-HIG-17-033, CMS-HIG-18-012, CMS-HIG-18-023, CMS-HIG-18-004, CMS-HIG-17-027, CMS-HIG-18-014, CMS-HIG-18-005}.
However, if the $Z_2$ symmetry is dropped, alignment automatically emerges when all heavy Higgs quartic couplings are $\mathcal{O}(1)$~\cite{Hou:2017hiw}. Therefore, models without $Z_2$ symmetry can lead naturally to the alignment limit and predict FCNH couplings in the heavy Higgs sector, while respecting the SM-like nature of the $h(125)$ discovered at the LHC.
 
The search presented here targets a general two Higgs doublet model (g2HDM) without $Z_2$ symmetry~\cite{Hou:2019gpn}, where the heavy Higgs bosons feature FCNH couplings. Only couplings involving top-quarks are considered: $\rhott$, $\rhotc$, and $\rhotu$. The $\rho_{AB}$ parameters indicate the coupling of the heavy Higgs boson to particles $A$ and $B$. The notation \rhotq\ is used to refer to both the \rhotc\ and \rhotu\ couplings.
The g2HDM model with its additional top Yukawa couplings is phenomenologically interesting since it can explain the generation of the baryon asymmetry through the couplings $\rhott$ or $\rhotc$~\cite{Fuyuto:2017ewj}. No distinction is performed between the different chiralities in the coupling, and an effective coupling $\rhotq = \sqrt{\hat{\rho}_{t_Lq_R}^2+\hat{\rho}_{q_Lt_R}^2}\,/\sqrt{2}$ is used, where the hat symbol is used to denote the original couplings in the g2HDM Lagrangian.
 
\begin{figure}[t]
\centering
\subfloat[]{\includegraphics[width=0.30\textwidth]{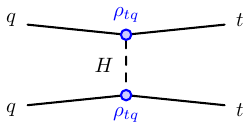}\label{fig:qq_tt}}
\subfloat[]{\includegraphics[width=0.30\textwidth]{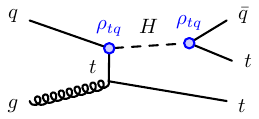}\label{fig:gq_tA_tq}}
\subfloat[]{\includegraphics[width=0.30\textwidth]{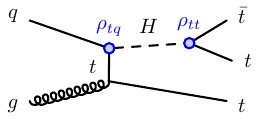}\label{fig:gq_tA_tt}}\\
\subfloat[]{\includegraphics[width=0.30\textwidth]{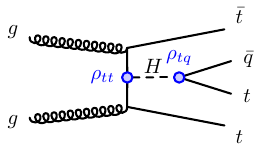}\label{fig:gg_ttA_tq}}
\subfloat[]{\includegraphics[width=0.30\textwidth]{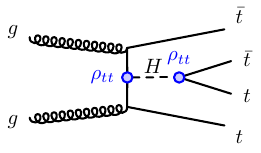}\label{fig:gg_ttA_tt}}
\caption{
Signal diagrams for the dominant production and decay modes of the heavy scalar considered in the analysis. The subsequent decay can lead to a final state with high multiplicity of leptons and $b$-jets that is targeted by the search. Single $H$ production through gluon--gluon fusion, $gg \rightarrow H \rightarrow t\bar{t} / tq$, is not considered since the decay does not lead to the relevant multi-lepton final state.}
\label{fig:diagrams_with_decay}
\end{figure}

The production and decay modes at tree level considered in the analysis are shown in Figure~\ref{fig:diagrams_with_decay}.
The presence of the $\rhotq$ coupling opens the possibility of same-sign top production, as shown in Figures~\ref{fig:qq_tt} (sstt) and~\ref{fig:gq_tA_tq} (ttq), and also three-top production, as shown in Figures~\ref{fig:gq_tA_tt} (ttt) and~\ref{fig:gg_ttA_tq} (tttq). The three-top signature is
a sensitive probe of beyond-the-SM (BSM) physics~\cite{Altmannshofer:2016zrn,Kohda:2017fkn,Hou:2019gpn,Hou:2019vgf}. Additionally, four-top quarks can be produced, as shown in Figure~\ref{fig:gg_ttA_tt} (tttt)~\footnote{The g2HDM signal processes sstt, ttq, ttt, tttq, and tttt include $tt$ and $\bar{t}\bar{t}$, $ttq$ and $\bar{t}\bar{t}q$, $tt\bar{t}$ and $\bar{t}\bar{t}t$, $t\bar{t}tq$ and $t\bar{t}\bar{t}q$, and $t\bar{t}t\bar{t}$, respectively, where the $q$ can be an up/charm or anti-up/anti-charm quark.}.
The targeted final state is characterised by multiple leptons (electrons and muons) and multiple jets containing $b$-flavoured hadrons ($b$-jets).
Many of the production modes are expected to be charge-asymmetric (with preference to positively charged), and this feature is exploited in the search.
The relevance of each production mode depends on the chosen coupling. A benchmark of $\rhott=0.4$ and $\rhotq=0.2$ is chosen to guide the analysis design and optimisation. The values are chosen so that the signal could account for the higher \ttW and \tttt yields observed in ATLAS analyses~\cite{ATLAS-CONF-2019-045,TOPQ-2016-11, TOPQ-2018-05,TOPQ-2020-10,SUSY-2019-04}.
Significant kinematic differences and a much stronger charge asymmetry are expected from the targeted signals, which allow them to be differentiated from simple rescalings of both processes.
For the chosen couplings, the $tH$ production (Figures~\ref{fig:gq_tA_tq} and~\ref{fig:gq_tA_tt}) cross section is two orders of magnitude larger than $ttH$ production (Figures~\ref{fig:gg_ttA_tq} and~\ref{fig:gg_ttA_tt}), and three orders of magnitude larger than same-sign tops production via $t$-channel $H$ (Figures~\ref{fig:qq_tt}).
 
This analysis is the first to target BSM production leading to three-top final states and the first to probe the g2HDM. The production of four-tops in the SM or via heavy scalars was
explored previously by ATLAS~\cite{TOPQ-2020-10,EXOT-2017-11,SUSY-2019-04,ATLAS-CONF-2021-048,ATLAS-CONF-2022-008} and CMS~\cite{CMS-TOP-17-009, CMS-TOP-17-019, CMS-TOP-18-003, CMS-SUS-16-035}. Limits on the g2HDM model couplings can be derived from LHC Higgs measurements, $B$ physics, and assuming the couplings stay perturbative~\cite{Altunkaynak:2015twa}, leading to $\rhott < 2$, $\hat{\rho}_{t_Lc_R} < 1.5$, and $\hat{\rho}_{c_Lt_R} < 0.1$, implying $\rhotc < 1.06$.
In addition, $K - \bar{K}$ mixing provides the constraint $\hat{\rho}_{c_Lt_R} < 0.14$~\cite{Crivellin:2013wna}, and $D - \bar{D}$ mixing provides the constraint $|\hat{\rho}_{t_Lc_R}\hat{\rho}_{t_Lu_R}^*|< 0.02$~\cite{Crivellin:2013wna}, which translates to $|\rhotc\rhotu^*|< 0.01$ assuming a negligible $\hat{\rho}_{u_Lt_R}$. These constraints are derived assuming $m_{H} \approx m_{H^+} = 500$ \gev, and become weaker for higher masses.
 
The event selections optimised for the heavy scalar signal models are also sensitive to models based on $R$-parity-violating (RPV) supersymmetry (SUSY).  These models are motivated by the recent flavour anomalies~\cite{Belle:2019rba,,LHCb:2020gog,LHCb:2020lmf,LHCb:2017smo,LHCb:2021trn} and $(g-2)_\mu$ anomaly~\cite{Muong-2:2021ojo}, and can provide a successful explanation with different choices of particles, masses, and couplings~\cite{Chakraborty:2015bsk,Zheng:2021wnu,Altmannshofer:2020axr,Das:2017kfo,Earl:2018snx,Deshpande:2016yrv,Hu:2020yvs,Hu:2019ahp,Altmannshofer:2017poe,Trifinopoulos:2018rna}.  The RPV SUSY models discussed below are also used to interpret the results of this search.
 
\begin{figure}[t]
\centering
\subfloat[]{\includegraphics[width=0.30\textwidth]{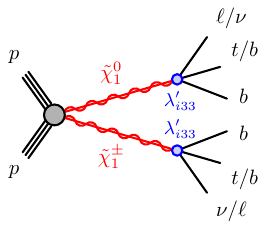}\label{fig:C1N1_LQD}}
\subfloat[]{\includegraphics[width=0.30\textwidth]{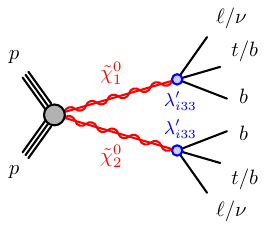}\label{fig:N1N2_LQD}}
\subfloat[]{\includegraphics[width=0.30\textwidth]{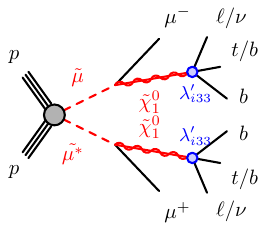}\label{fig:smuon_LQD}}
\caption{
Signal diagrams for the RPV SUSY signals used as additional interpretation in the analysis. The subsequent decay can lead to a final state with high multiplicity of leptons and $b$-jets that is targeted by the search.}
\label{fig:RPV_diagrams_with_decay}
\end{figure}

The first model features production of electroweakinos (wino or Higgsino) that decay via a lepton-number-violating RPV coupling of the $LQ\bar{D}$ type to a lepton and third-generation quarks. The corresponding term in the superpotential has the form $\lambda^\prime_{i33}L_iQ_3\bar{D}_3$, where $i \in {2,3}$ is a generation index, and $L$, $Q$, $\bar{D}$ are the lepton doublet, quark doublet, and down-type quark singlet superfields, respectively. Relevant diagrams for the production and decay are shown in Figures~\ref{fig:C1N1_LQD} and~\ref{fig:N1N2_LQD}. The second model features direct smuon production and decay to a bino-like neutralino, which in turn decays via the same RPV coupling ($\lambda^\prime_{i33}$), as shown in Figure~\ref{fig:smuon_LQD}.
 
% End of text imported from the .//Tex/intro.tex input file

\section{ATLAS detector}
\label{sec:detector}

% The next lines are included from the .//Tex/detector.tex input file
\newcommand{\AtlasCoordFootnote}{
ATLAS uses a right-handed coordinate system with its origin at the nominal interaction point (IP)
in the centre of the detector and the \(z\)-axis along the beam pipe.
The \(x\)-axis points from the IP to the centre of the LHC ring,
and the \(y\)-axis points upwards.
Cylindrical coordinates \((r,\phi)\) are used in the transverse plane,
\(\phi\) being the azimuthal angle around the \(z\)-axis.
The pseudorapidity is defined in terms of the polar angle \(\theta\) as \(\eta = -\ln \tan(\theta/2)\).
Angular distance is measured in units of \(\Delta R \equiv \sqrt{(\Delta\eta)^{2} + (\Delta\phi)^{2}}\).}
 
The ATLAS detector~\cite{PERF-2007-01} at the LHC covers nearly the entire solid angle around the collision point.\footnote{\AtlasCoordFootnote}
It consists of an inner tracking detector surrounded by a thin superconducting solenoid, electromagnetic and hadron calorimeters,
and a muon spectrometer incorporating three large superconducting air-core toroidal magnets.
 
The inner-detector system (ID) is immersed in a \SI{2}{\tesla} axial magnetic field
and provides charged-particle tracking in the region \(|\eta| < 2.5\).
The high-granularity silicon pixel detector covers the vertex region and typically provides four measurements per track,
the first hit normally being in the insertable B-layer (IBL) installed before Run~2~\cite{ATLAS-TDR-19,PIX-2018-001}.
It is followed by the silicon microstrip tracker (SCT), which usually provides eight measurements per track.
These silicon detectors are complemented by the transition radiation tracker (TRT),
which enables radially extended track reconstruction up to \(|\eta| = 2.0\).
The TRT also provides electron identification information
based on the fraction of hits (typically 30 in total) above a higher energy-deposit threshold corresponding to transition radiation.
 
The calorimeter system covers the pseudorapidity range \(|\eta| < 4.9\).
Within the region \(|\eta|< 3.2\), electromagnetic calorimetry is provided by barrel and
endcap high-granularity lead/liquid-argon (LAr) calorimeters,
with an additional thin LAr presampler covering \(|\eta| < 1.8\)
to correct for energy loss in material upstream of the calorimeters.
Hadron calorimetry is provided by the steel/scintillator-tile calorimeter,
segmented into three barrel structures within \(|\eta| < 1.7\), and two copper/LAr hadron endcap calorimeters.
The solid angle coverage is completed with forward copper/LAr and tungsten/LAr calorimeter modules
optimised for electromagnetic and hadronic energy measurements, respectively.
 
The muon spectrometer (MS) comprises separate trigger and
high-precision tracking chambers measuring the deflection of muons in a magnetic field generated by the superconducting air-core toroidal magnets.
The field integral of the toroids ranges between \num{2.0} and \SI{6.0}{\tesla\metre}
across most of the detector.
Three layers of precision chambers, each consisting of layers of monitored drift tubes, cover the region \(|\eta| < 2.7\),
complemented by cathode-strip chambers in the forward region, where the background is highest.
The muon trigger system covers the range \(|\eta| < 2.4\) with resistive-plate chambers in the barrel, and thin-gap chambers in the endcap regions.
 
Interesting events are selected by the first-level trigger system implemented in custom hardware,
followed by selections made by algorithms implemented in software in the high-level trigger~\cite{TRIG-2016-01}.
The first-level trigger accepts events from the \SI{40}{\MHz} bunch crossings at a rate below \SI{100}{\kHz},
which the high-level trigger reduces further to record events to disk at about \SI{1}{\kHz}.
 
An extensive software suite~\cite{ATL-SOFT-PUB-2021-001} is used in the reconstruction and analysis of real
and simulated data, in detector operations, and in the trigger and data acquisition systems of the experiment.

% End of text imported from the .//Tex/detector.tex input file

\section{Data and simulated event samples}
\label{sec:datamc}

% The next lines are included from the .//Tex/datamc.tex input file
This analysis uses data from $pp$ collisions at $\sqrt{s}=13~\tev$ collected by the ATLAS experiment during 2015--2018. After the application of data-quality requirements~\cite{DAPR-2018-01}, the data sample corresponds to an integrated luminosity of $139~\ifb$~\cite{ATLAS-CONF-2019-021}.
The number of additional $pp$ interactions per bunch crossing (pile-up) in this sample ranges from about 8 to 70, with an average of 34. Only events recorded under stable beam conditions and for which all detector subsystems were known to be in a good operating condition are used. The trigger requirements are discussed in Section~\ref{sec:events}.
 
Monte Carlo (MC) simulation samples were produced for the different signal and background processes. Table~\ref{tab:mcconfig} shows the configurations used in this analysis, with the samples in parentheses and in grey indicating those used to estimate the systematic uncertainties.
All simulated samples, except those produced with the \textsc{Sherpa}~\cite{Gleisberg:2008ta} event generator, utilised \textsc{EvtGen} 1.2.0~\cite{Lange:2001uf} to model the decays of heavy-flavour hadrons. All samples showered with \textsc{Pythia} use the A14 set of tuned parameters~\cite{ATL-PHYS-PUB-2014-021} (referred to as `tune'), whereas those showered with \textsc{Herwig} use the H7-UE tune~\cite{Bellm:2015jjp}.
Pile-up was modelled using events from minimum-bias interactions generated with \textsc{Pythia}~8.186~\cite{Sjostrand:2007gs} with the A3 tune~\cite{ATL-PHYS-PUB-2011-014}, and overlaid onto the simulated hard-scatter events according to the luminosity profile of the recorded data.
The generated events were processed through either a full simulation of the ATLAS detector geometry and response using \textsc{Geant4}~\cite{Agostinelli:2002hh}, or a faster simulation where the full \textsc{Geant4} simulation of the calorimeter response is replaced by a detailed parameterisation of the shower shapes~\cite{SOFT-2010-01}. Both types of simulated events were processed through the same reconstruction software used for the $pp$ collision data.
Corrections were applied to the simulated events so that the particle candidates' selection efficiencies, energy scales and energy resolutions match those determined from data control samples.
The simulated samples are normalised to their cross sections, and generated to the highest order available in perturbation theory.
 
Samples used to model the g2HDM signal were generated at leading-order (LO) in QCD with \textsc{Madgraph}~v2.9.3~\cite{Alwall:2014hca} with the \NNPDF[3.1nlo]~\cite{NNPDF:2017mvq} parton distribution function (PDF) set. Samples were generated for masses in the range of $200~\gev$ to $1.5~\tev$ with a $100~\gev$ step, and processed with the ATLAS Fast Simulation~\cite{SOFT-2010-01}.
All signals were produced with the set of couplings $\rhott=\rhotc=\rhotu=0.1$. Each signal process described in Section~\ref{sec:intro} was generated as a separate MC sample.
The LO cross section obtained from Madgraph is used for the normalisation of the signals.
Simulated events for different coupling values are obtained by rescaling the samples to match the target cross section and branching ratio of each subprocess. For a given choice of couplings all the processes are taken into account and rescaled.
The RPV SUSY signal samples were generated with \textsc{Madgraph}~v2.9.3, with up to two extra jets at LO in QCD. The matching scale is set at 1/4 of the mass of the SUSY particle being produced.
Supersymmetric particle decays via the RPV coupling are simulated with 25\% branching ratio to $\mu/\tau/\nu_{\mu}/\nu_{\tau}$ each, a $b$-quark, and a $b$- or $t$-quark depending on the lepton charge. The identical branching ratio to second- and third-generation leptons follows from the choice of $\lambda^\prime_{233}=\lambda^\prime_{333}$, while the balance in charged and neutral leptons is an assumption. This assumption originates naturally from the presence of a left-handed lepton superfield in the $LQ\bar{D}$ coupling, but is distorted by the large mass difference between top and bottom quarks and also affected by the choice of $\tan(\beta)$.
Signal cross sections are calculated to next-to-leading order in the strong coupling constant, adding the resummation of soft gluon emission at next-to-leading-logarithmic accuracy (NLO+NLL)~\cite{Beenakker:1999xh,Debove:2010kf,Fuks:2012qx,Fuks:2013vua,Fiaschi:2018hgm}. The nominal cross section and the uncertainty are taken from an envelope of cross-section predictions using different PDF sets and factorisation and renormalisation scales, as described in Ref.~\cite{Borschensky:2014cia}.
All signal events were showered with \textsc{Pythia}~8.245~\cite{Sjostrand:2007gs} using the \NNPDF[2.3lo]~\cite{Ball:2012cx} PDF set.
 
The samples used to model the \ttW and the \ttll backgrounds were generated using \Sherpa-2.2.10~\cite{Bothmann:2019yzt} and \Sherpa-2.2.11, where the matrix element (ME) were calculated for up to one and zero additional partons at next-to-leading-order (NLO) in QCD, respectively, and up to two partons at LO in QCD using \textsc{Comix}~\cite{Gleisberg:2008fv} and \textsc{OpenLoops}~\cite{Cascioli:2011va}. The ME were merged with the \textsc{Sherpa} parton shower (PS)~\cite{Schumann:2007mg} using the \textsc{MePs@Nlo} prescription~\cite{Hoeche:2012yf}, with a CKKW merging scale of $30~\gev$ for the \ttW sample.
These samples are generated using the \NNPDF[3.0nnlo]~\cite{Ball:2014uwa} PDF set, along with the dedicated set of tuned parton-shower parameters developed by the \SHERPA authors.
The invariant mass of the lepton pair ($\mll$) in the \ttll sample is set to be greater than $1~\gev$.
Both the factorisation and renormalisation scales are set to $\mu_r = \mu_f = m_T / 2$ in the \ttW sample, where $m_T$ is defined as the scalar sum of the transverse masses $\sqrt{m^2 + p_{T}^2}$ of the particles generated from the ME calculation.
In addition to this \ttW prediction at NLO in QCD, higher-order corrections relating to electroweak (EW) contributions are also included. First, event-by-event correction factors are applied that provide virtual NLO EW corrections of the order $\alpha^{2}\alphas^{2}$ derived using the formalism described in Ref.~\cite{Kallweit:2015dum} along with LO
corrections of order $\alpha^3$. Second, real emission contributions from the sub-leading EW corrections at order $\alpha^{3}\alphas$\,\cite{Frederix:2017wme} are simulated with an independent \Sherpa-2.2.10 sample produced at LO in QCD.
The complete \ttW simulation is normalised to the total cross section of $\sigma(\ttW) = 614.7$~fb that comes from the \Sherpa configuration outlined above considering NLO QCD and NLO EWK effects, based on a similar strategy as used in Ref.~\cite{Frederix:2021agh}. The \ttdy sample is normalised to the cross section $\sigma(\ttdy) = 839$~fb, calculated at NLO QCD and NLO EW accuracy using \MGNLO~\cite{deFlorian:2016spz} and scaled by an off-shell correction estimated at one-loop level in \alphas.
 
The production of SM \tttt events was modelled using the \textsc{Madgraph5\_aMC@NLO}~v2.6.2 generator that provides matrix elements at NLO in QCD with the \NNPDF[3.1nlo] PDF set. The functional form of the renormalisation and factorisation scales are set to $\mu_r = \mu_f = m_T / 4$. Top quarks are decayed at LO using \textsc{MadSpin} to preserve all spin correlations. The events are interfaced with \textsc{Pythia}~8.230 for the parton shower and hadronisation, using the \NNPDF[2.3lo] PDF set. The production of \tttt\  events is normalised to a cross section of 12 fb computed at NLO in QCD including EW corrections~\cite{Frederix:2017wme}.
 
Diboson ($\VV$) background processes were simulated with \textsc{Sherpa}~2.2.2~\cite{Bothmann:2019yzt}.  The matrix element was calculated using $\textsc{Comix}$~\cite{Gleisberg:2008fv} and \textsc{OpenLoops}~\cite{Cascioli:2011va} with NLO accuracy in QCD for up to one additional parton and at LO accuracy for up to three additional partons, and merged with the \textsc{Sherpa} PS using \textsc{MePs@Nlo} prescription~\cite{Hoeche:2012yf}. The \NNPDF[3.0nnlo] set of PDFs was used, along with the dedicated parton-shower tune for \SHERPA. The cross section of $\sigma(\VV) = 104$~pb used to normalise the sample was computed by \textsc{Sherpa}~2.2.2.
 
Samples for $\tth$, \ttbar, and single top production were generated using the NLO generator \textsc{Powheg-Box-v2}~\cite{Frixione:2007nw,Nason:2004rx,Frixione:2007vw,Alioli:2010xd,Alioli:2009je,Re:2010bp} and interfaced with \textsc{Pythia}~8 for the parton showering and fragmentation. These samples used the \NNPDF[3.0nlo] PDF set. The $h_\text{damp}$ parameter, which controls the transverse momentum of the first additional emission beyond the LO Feynman diagram in the PS and therefore regulates the high-$\pt$ radiation, is set to $3/4 \times (m_t + m_{\bar{t}} + m_{h})$ in the \ttH sample and to $1.5 \times m_t$ in the \ttbar and single top samples, where $m_t$ ($m_h$) denotes the mass of the top quark (SM Higgs boson).
 
A dedicated $\ttbar$ sample including rare $t \to Wb \gamma^*(\to \ell^{+}\ell^{-})$ radiative decays, $\ttbar \to W^+bW^-\bar{b}\ell^{+}\ell^{-}$, was generated using a ME calculated at LO in QCD and requiring $\mll > 1~\gev$. In this sample the photon can be radiated from the top quark, the $W$ boson, or the $b$-quark. Both the \ttll and $\ttbar \to W^+bW^-\bar{b}\ell^{+}\ell^{-}$ samples are combined and together form the ``$\ttdy$'' sample.
The contribution from internal photon conversions ($\gamma^* \to \ell^{+}\ell^{-}$) with $\mll < 1~\gev$ were modelled by QED multi-photon radiation via the PS in an inclusive $\ttbar$ sample and is referred to as ``$\ttbar\gamma^*$ (LM)''.
Dedicated $Z$+jets samples containing electrons from material photon conversion ($\gamma \to e^+e^-$) or internal photon conversion were generated with \textsc{Powheg-Box} and interfaced with \textsc{Pythia}~8 for the parton showering and fragmentation. These samples are used to model the data in control regions enriched in material and internal conversion electrons, as explained in Section~\ref{sec:events}.
 
The remaining rare background contributions listed in Table~\ref{tab:mcconfig} are normalised using their NLO theoretical cross sections, except for the $t\bar t t$, $t\bar t W^+ W^-$, $t \bar{t} ZZ$, $t \bar{t} hh$, and $t \bar{t} Wh$ processes, for which a LO cross section is used.
 
\begin{table}[htb]
\begin{center}
\caption{
The configurations used for event generation of signal and background processes.
The samples used to estimate the systematic uncertainties are indicated in parentheses and grey.
$V$ refers to production of an electroweak boson ($W$ or $Z/\gamma^*$).
The matrix element order refers to the order in the strong coupling constant of the perturbative calculation.
The ``$\ttbar W$ (EW)'' sample also includes next-to-leading-order electroweak corrections.
Tune refers to the underlying-event tune of the parton shower generator.
MG5\_aMC refers to \textsc{MadGraph5\_aMC@NLO} 2.2, 2.3, or 2.6;
\textsc{Pythia} 8 refers to version 8.2~\cite{Sjostrand:2000wi};
\textsc{MePs@Nlo} refers to the method used in \textsc{Sherpa} to match the matrix element to the parton shower.
All samples include leading-logarithm photon emission, either modelled by the parton shower generator or by \textsc{Photos}~\cite{Golonka:2005pn}.
The mass of the top quark ($m_t$) and SM Higgs boson were set to $172.5~\gev$ and $125~\gev$, respectively.}
\label{tab:mcconfig}
\vspace{0.25cm}
\small
\setlength\tabcolsep{1.5pt}
\begin{tabular}{~l^l^l^l^l^l}
\hline\hline
Process & Generator & ME order & Parton shower & PDF & Tune  \\
\hline
g2HDM signal           & MG5\_aMC & LO & \textsc{Pythia} 8\ & \NNPDF[3.1nlo] & A14 \\
SUSY signal            & MG5\_aMC & LO & \textsc{Pythia} 8\ & \NNPDF[3.1nlo] & A14 \\
$\ttbar W$             & \textsc{Sherpa 2.2.10} & \textsc{MePs@Nlo} & \textsc{Sherpa} &  \NNPDF[3.0nnlo] & \textsc{Sherpa} default \\
\rowstyle{\color{gray}}
& (MG5\_aMC)  & (NLO) & (\textsc{Pythia} 8) & (\NNPDF[3.0nlo]) & (A14)   \\
$\ttbar W$ (EW)        & \textsc{Sherpa 2.2.10} & LO & \textsc{Sherpa} &  \NNPDF[3.0nnlo] & \textsc{Sherpa} default \\ 
\rowstyle{\color{gray}}
& (MG5\_aMC)  & (LO) & (\textsc{Pythia} 8) & (\NNPDF[3.0nlo]) & (A14)   \\
$t\bar t t\bar t$      & MG5\_aMC & NLO & \textsc{Pythia} 8 & \NNPDF[3.1nlo] & A14 \\
\rowstyle{\color{gray}}
& (\textsc{Sherpa 2.2.10}) & (\textsc{MePs@Nlo}) & (\textsc{Sherpa}) &  (\NNPDF[3.0nnlo]) & (\textsc{Sherpa} default) \\
\ttH                   & \textsc{Powheg-BOX}  & NLO & \textsc{Pythia} 8\ & \NNPDF[3.0nlo] & A14 \\
\rowstyle{\color{gray}}
& (\textsc{Powheg-BOX}) & (NLO) & (\textsc{Herwig7.0.4}) &  (\NNPDF[3.0nlo]) &  (H7-UE-MMHT) \\
\rowstyle{\color{gray}}
& (MG5\_aMC)  & (NLO) & (\textsc{Pythia} 8) & (\NNPDF[3.0nlo]) & (A14)   \\
\ttll                  & \textsc{Sherpa 2.2.11} & \textsc{MePs@Nlo} & \textsc{Sherpa} &  \NNPDF[3.0nnlo] & \textsc{Sherpa} default \\ 
\rowstyle{\color{gray}}
& (MG5\_aMC) & (NLO) & (\textsc{Pythia} 8) & (\NNPDF[3.0nlo]) & (A14)  \\
$\ttbar \to W^+bW^-\bar{b}\ell^{+}\ell^{-}$ & MG5\_aMC & LO & \textsc{Pythia} 8 & \NNPDF[3.0lo] & A14   \\
$t (Z/\gamma^*)$       & MG5\_aMC & NLO & \textsc{Pythia} 8 & \NNPDF[2.3lo]  & A14  \\
$t W (Z/\gamma^*)$     & MG5\_aMC & NLO & \textsc{Pythia} 8 & \NNPDF[2.3lo]  & A14  \\
$\ttbar$               & \textsc{Powheg-BOX} & NLO & \textsc{Pythia} 8 & \NNPDF[3.0nlo] & A14  \\
\rowstyle{\color{gray}}
& (\textsc{Powheg-BOX}) & NLO & (\textsc{Herwig7.1.3}) & (\NNPDF[3.0nlo]) &  (H7-UE-MMHT) \\
$t\bar t t$            & MG5\_aMC & LO & \textsc{Pythia} 8 & \NNPDF[2.3lo] & A14 \\
Single top             & \textsc{Powheg-Box} & NLO & \textsc{Pythia} 8 & \NNPDF[3.0nlo] & A14   \\
($t$-, $Wt$-, $s$-channel) &  &  &                                     &  &  \\
$VV$, $qqVV$, $VVV$    & \textsc{Sherpa} 2.2.2  & \textsc{MePs@Nlo} & \textsc{Sherpa} & \NNPDF[3.0nnlo] & \textsc{Sherpa} default  \\
$Z \to \ell^{+}\ell^{-}$         & \textsc{Sherpa} 2.2.1 & \textsc{MePs@Nlo} & \textsc{Sherpa} & \NNPDF[3.0nlo] & \textsc{Sherpa} default \\ 
$Z \to \ell^{+}\ell^{-} (\gamma \to e^{+}e^{-})$ & \textsc{Powheg-BOX} & NLO & \textsc{Pythia} 8 & \CTEQ[6L1nlo] & A14 \\
$Z \to \ell^{+}\ell^{-} (\gamma* \to e^{+}e^{-})$ & \textsc{Powheg-BOX} & NLO & \textsc{Pythia} 8 & \CTEQ[6L1nlo] & A14 \\
$W+$jets               & \textsc{Sherpa} 2.2.1 & \textsc{MePs@Nlo} & \textsc{Sherpa} & \NNPDF[3.0nlo] & \textsc{Sherpa} default \\ 
$Vh$                   & \textsc{Powheg-BOX} & NLO  & \textsc{Pythia} 8 & \NNPDF[3.0nlo] & A14 \\
$t\bar t W^+ W^-$      & MG5\_aMC & LO  & \textsc{Pythia} 8 & \NNPDF[2.3lo]  & A14  \\
$t \bar{t} ZZ$         & MG5\_aMC & LO  & \textsc{Pythia} 8 & \NNPDF[2.3lo] & A14 \\
$t \bar{t} hh$         & MG5\_aMC & LO  & \textsc{Pythia} 8 & \NNPDF[2.3lo] & A14 \\
$t \bar{t} Wh$         & MG5\_aMC & LO  & \textsc{Pythia} 8 & \NNPDF[2.3lo] & A14 \\
\hline\hline
\end{tabular}
\end{center}
\end{table}

% End of text imported from the .//Tex/datamc.tex input file

\clearpage
\section{Event reconstruction and object identification}
\label{sec:objects}

% The next lines are included from the .//Tex/objects.tex input file
Interaction vertices from the $pp$ collisions are reconstructed from at least two tracks with transverse momentum ($\pt$) larger than $500~\mev$ that are consistent with originating from the beam collision region in the $x$--$y$ plane. If more than one primary vertex candidate is found in the event, the candidate for which the associated tracks form the largest sum of squared $\pt$ is selected as the hard-scatter primary vertex~\cite{ATL-PHYS-PUB-2019-015}.
 
Electron candidates are reconstructed from energy clusters in the electromagnetic calorimeter matched to a track in the ID~\cite{EGAM-2018-01}. They are required to satisfy $\pt>10 \GeV$ and $|\eta_{\mathrm{cluster}}|<2.47$, with the
transition region between the endcap and barrel calorimeters ($1.37<|\eta_{\mathrm{cluster}}|<1.52$) excluded.
Loose and tight electron identification working points are used~\cite{EGAM-2018-01}, based on a likelihood discriminant
that employs 
calorimeter, tracking and combined variables
to 
distinguish between electrons and jets. The associated track of an electron candidate is required to have at least two hits in the pixel detector and seven hits total in the pixel and silicon-strip detectors combined. For the tight identification working point, one of these pixel hits must be in the innermost layer, or the next-to-innermost layer if the module traversed in the innermost layer is non-operational, and there must be no association with a vertex from a reconstructed photon conversion~\cite{PERF-2017-02} in the detector material (denoted as `material conversion').
 
Muon candidates are reconstructed by combining tracks in the ID with tracks in the MS~\cite{MUON-2018-03}. The resulting muon candidates are re-fit using the complete track information from both detector systems~\cite{PERF-2015-10}. They are required to satisfy $\pt > 10~\gev$ and $|\eta| < 2.5$. Loose and medium muon identification working points are used~\cite{PERF-2015-10}.
 
Electron (muon) candidates are matched to the primary vertex by requiring that their transverse impact parameter, $d_0$, satisfies $|d_0/\sigma(d_0)|<5\,(3)$, where $\sigma(d_0)$ is the measured uncertainty in $d_0$, and requiring that the longitudinal impact parameter, $z_0$, satisfies $|z_0 \sin\theta|<0.5$~mm, where $\theta$ is the polar angle of the track.
 
To further suppress leptons from heavy-flavour hadron decays, misidentified jets, or photon conversions (collectively referred to as `non-prompt leptons'), lepton candidates are also required to be isolated in the tracker and in the calorimeter~\cite{EleGammaPerf_2019}. A track-based lepton isolation criterion is defined by calculating the quantity $I_R = \sum \pt^{\textrm{trk}}$, where the scalar sum includes all tracks (excluding the lepton candidate itself) within the cone defined by $\Delta R<R_{\textrm{cut}}$ around the direction of the lepton.  The value of $R_{\textrm{cut}}$ is the smaller of $r_{\textrm{min}}$ and $10~\gev/\pt^\ell$, where $r_{\textrm{min}}$ is set to 0.2 (0.3) for electron (muon) candidates and where $\pt^\ell$ is the lepton $\pt$. All lepton candidates must satisfy $I_R/\pt^\ell < 0.15$.
Additionally, electrons (muons) are required to satisfy a calorimeter-based isolation criterion: the sum of the transverse energy within a cone of size $\Delta R=0.2$ around the lepton, after subtracting the contributions from pile-up and the energy deposit of the lepton itself, is required to be less than 20\% (30\%) of $\pt^\ell$.
Muons are required to be separated by $\Delta R > 0.2$ from any selected jets (defined below). If two electrons are closer than $\Delta R=0.1$, only the one with the higher \pt\ is considered. Electrons within $\Delta R=0.1$ of a selected muon are removed.
 
The selection criteria described above greatly suppress the contribution from non-prompt leptons. However, several channels considered in this search have additional suppression requirements targeting the main non-prompt lepton types. Non-prompt leptons from hadron decays that contain bottom- and charm-quarks, denoted by `heavy-flavour (HF) non-prompt leptons', are further rejected using a boosted decision tree (BDT) discriminant, referred to as the non-prompt lepton BDT~\cite{HIGG-2017-02,MUON-2018-03}, which uses isolation and displacement information associated with a track jet that matches the selected light lepton. Three working points (WPs) are used: \textit{Tight}, \textit{VeryTight}, and \textit{Tight-not-VeryTight}.
The \textit{Tight} WP allows to select prompt-like leptons with an efficiency for muons (barrel/endcap electrons) that satisfy the calorimeter- and track-based isolation criteria of about 60\% (60/70\%) for $\pt \sim 20~\gev$ and reaches a plateau of 95\% (95/90\%) at $\pt \sim 40\ (40/65) ~\gev$.
The prompt lepton efficiency of the \textit{VeryTight} WP for muons (barrel/endcap electrons) that satisfy the calorimeter- and track-based isolation criteria is about 55\% (55/60\%) for $\pt \sim 20~\gev$ and reaches a plateau of 90\% (85/83\%) at $\pt \sim 40\ (40/65) ~\gev$. The corresponding rejection factor\footnote{The rejection factor is defined as the reciprocal of the efficiency.} against muons (electrons) from the decay of $b$-hadrons ranges from 33 to 50 (20 to 50) for the \textit{Tight} WP, and from 50 to 100 (33 to 66) for the \textit{VeryTight} WP, depending on \pt and $\eta$, after resolving ambiguities between overlapping reconstructed objects.
The \textit{Tight-not-VeryTight} WP allows to select non-prompt 
leptons and is part of the event selection of the control regions enriched in HF non-prompt lepton background, as described in Section~\ref{sec:background}.
 
To further suppress electrons with incorrect charge assignment, a BDT discriminant based on calorimeter and tracking quantities~\cite{PERF-2017-01} is used. An efficiency of approximately 96\% in the barrel region and 81\% in the endcaps is obtained, with rejection factors of 19 in the barrel region and 40 in the endcaps.
Material and internal conversion candidates are identified based on a combination of requirements on the invariant mass of tracks and the radius from the reconstructed displaced vertex to the primary vertex.
Material conversion candidates have a reconstructed displaced vertex with radius $r > 20$~mm that includes the track associated with the electron.\footnote{The beampipe and insertable B-layer inner radii are 23.5~mm and 33~mm, respectively.} The invariant mass of the associated track and the closest (in $\Delta\eta$) opposite-charge track, calculated at the conversion vertex, is required to be less than $100~\mev$. Internal conversion candidates, which correspond to the internal photon conversions (see Section~\ref{sec:datamc}), must fail the requirements for material conversions, and the
invariant mass of the track pair,
calculated 
at the primary vertex, is also required to be less than $100~\mev$.
 
The lepton working points used in this analysis are summarised in Table~\ref{tbl:tightleps}. After the initial categorisation based on loose leptons (corresponding to ``$L$''), the most optimal lepton working point to further optimise the event selection is chosen depending on the main background processes and the expected number of events in each category. The defined working points are medium inclusive (``$M$''), medium exclusive (``$M_{\mathrm{ex}}$''), and tight (``$T$'').
The various choices can be seen for the signal and control regions in Section~\ref{sec:events}.
Electron candidates from internal or material conversion are rejected from the $M$, $M_{\mathrm{ex}}$, and $T$ electron selections.  They are used to define control regions enriched in internal or material conversions, and are collectively denoted $e^{*}$ (see Section~\ref{sec:events}).
 
\begin{table}
\begin{center}
\caption{Description of the loose inclusive (``$L$''), medium inclusive (``$M$''), medium exclusive (``$M_{\mathrm{ex}}$''), and tight (``$T$'')  lepton definitions. The electron $e^{*}$ is required to fulfil, in addition to the corresponding lepton definition requirements, those corresponding to an internal or material conversion candidate.}
\label{tbl:tightleps}
\resizebox{\textwidth}{!}{
\begin{tabular}{l||c|c|c|c||c|c|c|c}
\hline\hline
& \multicolumn{4}{c||}{$e$} & \multicolumn{4}{c}{$\mu$} \\
\hline
Lepton categorization  & $L$ & $M$ & $M_{\mathrm{ex}}$ & $T$ &  $L$ & $M$ & $M_{\mathrm{ex}}$ & $T$ \\
\hline
Isolation                          &  \multicolumn{4}{c||}{Yes} & \multicolumn{4}{c}{Yes} \\
\hline
Non-prompt lepton BDT WP               &  No & \textit{Tight} & \textit{Tight-not-}  & \textit{VeryTight} &  No & \textit{Tight} & \textit{Tight-not-}  & \textit{VeryTight}  \\
&  &   & \textit{VeryTight} & &  &  & \textit{VeryTight} &  \\
\hline
Identification                     & Loose & \multicolumn{3}{c||}{Tight} & Loose & \multicolumn{3}{c}{Medium} \\
\hline
Electron charge-misassignment veto &  No & \multicolumn{3}{c||}{Yes} & \multicolumn{4}{c}{Not applicable} \\
Electron conversion candidate veto &  No & \multicolumn{3}{c||}{Yes (except $e^{*}$)} & \multicolumn{4}{c}{Not applicable} \\
\hline
Transverse impact parameter        &  \multicolumn{4}{c||}{$<5$} & \multicolumn{4}{c}{$<3$ } \\
significance $|d_0|/\sigma_{d_0}$  & \multicolumn{4}{c||}{} &  \multicolumn{4}{c}{}  \\
\hline
Longitudinal impact parameter      &  \multicolumn{8}{c}{$< 0.5$ mm} \\
$|z_0 \sin \theta|$                &  \multicolumn{8}{c}{} \\
\hline\hline
\end{tabular}
}
\end{center}
\end{table}
 
The constituents for jet reconstruction are identified by combining measurements from both the ID and the calorimeter using a particle flow (PFlow) algorithm~\cite{PERF-2015-09,JETM-2018-05}. Jet candidates are reconstructed from these PFlow objects using the anti-$k_t$ algorithm~\cite{Cacciari:2008gp,Cacciari:2011ma} with a radius parameter of $R=0.4$. They are calibrated using simulation with corrections obtained from in situ techniques in data~\cite{JETM-2018-05}. Only jet candidates with $\pT> 25$~\GeV~that satisfy $|\eta|<2.5$ are selected.  To reduce the effects of pile-up, each jet with $\pT<60$~\GeV~and $|\eta|<2.4$ is required to satisfy the ``Tight''  working point of the Jet Vertex Tagger (JVT)~\cite{ATLAS-CONF-2014-018} criteria used to identify the jets as originating from the selected primary vertex. A set of quality criteria is also applied to reject events containing at least one jet arising from non-collision sources or detector noise~\cite{ATLAS-CONF-2015-029}.
 
Jets containing $b$-hadrons are identified ($b$-tagged) via the DL1r algorithm~\cite{FTAG-2019-07,ATL-PHYS-PUB-2017-013} that uses a deep-learning neural network based on the distinctive features of the $b$-hadrons in terms of the impact parameters of tracks and the displaced vertices reconstructed in the ID. Additional input to this network is provided by discriminant variables constructed by a recurrent neural network~\cite{ATL-PHYS-PUB-2017-003}, which exploits the spatial and kinematic correlations between tracks originating from the same $b$-hadron. For each jet, a value for the multivariate $b$-tagging discriminant is calculated. A jet is $b$-tagged if the $b$-tagging score is above a certain threshold, referred to as a working point (WP). Four WPs are defined with average expected efficiencies for $b$-jets of 60\%, 70\%, 77\% and 85\%, as determined in simulated \ttbar\ events. The $b$-tagging distribution obtained by ordering the resulting five exclusive bins from the four WPs from higher to lower $b$-jet efficiency is referred to as ``pseudo-continuous'' $b$-tagging score, and it is used as input to the multivariate analysis discriminant described in Section~\ref{sec:events}. In this search, a jet is considered $b$-tagged if it passes the WP corresponding to 77\% or 60\% efficiency to tag a $b$-jet, with a light-jet\footnote{`Light jet' refers to a jet originating from the hadronisation of a light quark ($u$, $d$, $s$) or a gluon.} rejection factor of about 200 or 2500, and a charm-jet ($c$-jet) rejection factor of about 6 or 40, as determined for jets with $\pt >20~\gev$ and $|\eta|<2.5$ in simulated $\ttbar$ events~\cite{FTAG-2019-07}.
Correction factors derived from dedicated calibration samples enriched in $b$-jets, $c$-jets, or light jets, are applied to the simulated samples~\cite{FTAG-2018-01,FTAG-2020-08,FTAG-2019-02}.
The notation \bloose and \btight is used to denote the number of $b$-tagged jets with the corresponding WP.
 
Ambiguities between independently reconstructed electrons, muons and jets can arise. A sequential procedure, referred to as `overlap removal', is performed to resolve these ambiguities and, thus, avoids double counting of particle candidates. This procedure is applied to leptons satisfying the $L$ criteria.
If two electrons are closer than $\Delta R=0.1$, only the one with the higher \pt\ is considered. If an electron and a muon overlap within $\Delta R=0.1$, the muon is removed if it is reconstructed from a track and calorimeter deposits consistent with a minimum ionising particle, else the electron is removed. If an electron and a selected jet are within $\Delta R < 0.2$ of each other, the jet is removed if it is not a $b$-tagged jet~\footnote{For the overlap removal, a jet is considered $b$-tagged if it passes the 70\% working point. However, the choice of the $b$-tagging working point does not have a sizeable impact on the signal acceptance.} or if it has $\pt > 200~\gev$. Muons are required to be separated by $\Delta R > 0.4$ from any jet that is ghost-associated~\cite{Cacciari:2008gn} to it. If the jet satisfying the $\Delta R < 0.4$ requirement is not a $b$-tagged jet and contains less than three tracks with $\pt > 500~\mev$, the overlapping jet is rejected from the event, otherwise, the muon is rejected. A lepton lying within a variable-size cone depending on the lepton \pt and with a maximum radius of $R=0.4$ around a selected jet that survived all previous overlap criteria is rejected.
 
The missing transverse momentum $\vec{p}_{\mathrm{T}}^{ \mathrm{\, miss}}$ (with magnitude $\met$) is defined as the negative vector sum of the $\pt$ of all selected and calibrated objects in the event, including a term to account for the momentum from soft particles that are not associated with any of the selected objects~\cite{PERF-2016-07}.
This soft term is calculated from inner-detector tracks matched to the selected primary vertex, which makes it more resilient to contamination from pile-up interactions. The $\met$ distribution is used as an input variable to the machine learning training discussed in Section~\ref{sec:events}.

% End of text imported from the .//Tex/objects.tex input file

\section{Search strategy}
\label{sec:events}

% The next lines are included from the .//Tex/strategy.tex input file
Events are required to satisfy a minimal preselection and are categorised into orthogonal signal regions (SRs) based on different criteria such as number of leptons, total lepton charge (indicated by $Q$), and a multi-output deep neural network classifier (\dnncat).
This categorisation provides a set of regions that are sensitive to all the possible signal production and decay modes
considered in this search.
A deep neural network is trained in each of the signal regions to discriminate the signal from the backgrounds (\dnnsb).
Additional orthogonal control regions (CRs) are defined in order to fit the normalisation of the main backgrounds.
Dedicated kinematic selections are applied to the control regions to improve the purity of the targeted backgrounds.
A maximum-likelihood fit is performed across categories to test for a possible signal and constrain in-situ the leading backgrounds simultaneously.
 
At trigger level, events were selected for read-out using a combination of single-lepton and dilepton triggers, requiring the electrons or muons to
satisfy identification criteria similar to those used in the offline reconstruction and isolation requirements~\cite{TRIG-2018-01,TRIG-2018-05}.
Single-electron triggers require a minimum $\pt$ threshold of 24 (26)~$\gev$ in the 2015 (2016, 2017 and 2018) data-taking period(s), while single-muon triggers have a lowest $\pt$ threshold of 20 (26)~$\gev$ in 2015 (2016--2018).
The dielectron triggers require two electrons with minimum $\pt$ thresholds ranging from 12~$\gev$ in 2015 to 24~$\gev$ in 2017--2018, whereas the dimuon triggers use asymmetric $\pt$ thresholds for leading (subleading) muons: 18 (8)~$\gev$ in 2015 and 22 (8)~$\gev$ in 2016--2018. Finally, an electron+muon trigger requires events to have an electron candidate with a 17~$\gev$ threshold and a muon candidate with a 14~$\gev$ threshold for all periods.
 
For the analysis selection, at least two jets and at least two leptons are required in the event, and leptons are required to match, with $\Delta R < 0.15$, the corresponding leptons reconstructed by the trigger and to have a $\pt$ exceeding the trigger $\pt$ threshold by 1~$\gev$.
Events are required to contain at least one $b$-tagged jet with the 60\% efficiency working point, or at least two $b$-tagged jets with the 77\% efficiency working point.
If events contain pairs of opposite-sign charge and same-flavour leptons (OS-SF), all pairs are required to satisfy a mass requirement on the dilepton system mass of $\mllOSSF > 12$ \GeV\ and $|\mllOSSF - m_Z| > 10$ \GeV.
Three disjoint event categories are defined according to the number of loose leptons in the event: same-charge dilepton (\llSS), three-lepton (\lll), and four-lepton (\llll) categories. The four-lepton category is inclusive and contains events with higher lepton multiplicity, while the other two are exclusive.
Leptons are ordered by \pt in the \llSS and \llll regions. In the \lll regions the lepton with opposite-sign charge is taken first, followed by the two same-sign leptons in \pt order.
The \pt and identification requirements of each lepton in each category are optimised based on a compromise between non-prompt lepton background suppression and signal acceptance enhancement, and are summarized in Table~\ref{tab:selectionSR}.
 
\begin{table}[htb]
\begin{center}
\caption{Event selection summary in the signal regions.
Leptons are ordered by \pt in the \llSS and \llll regions. In the \lll regions the lepton with opposite-sign charge is taken first, followed by the two same-sign leptons in \pt order.
In the lepton selection, \textit{T, M, L} stand for Tight, Medium and Loose lepton definitions.
In the region naming, the ``CAT ttX'' denotes the category based on the \dnncat output enriched in the signal process ``ttX''. Each of these regions is split according to the lepton charge of the same-sign lepton pair (``++'' or ``- -'').
}
\resizebox{\textwidth}{!}{
\def\arraystretch{1.4}
\begin{tabular}{l||c|c|c}
\hline\hline
Lepton category                    & \llSS                                 & \lll                                         & \llll \\
\hline\hline
\multirow{2}{*}{Lepton definition} & $(T,T)$ with $\geq$ 1 \btight ||         & $(L,T,M)$ with $\geq$ 1 \btight ||         & \multirow{2}{*}{$(L,L,L,L)$} \\
& $(T,M)$ with $\geq$ 2 \bloose & $(L,M,M)$ with $\geq$ 2 \bloose &  \\
Lepton \pt[\GeV]                  & (20, 20)                              & (10, 20, 20)                                 & (10, 10, 10, 10) \\
\mllOSSF[\GeV]                    & --                                    & \multicolumn{2}{c}{$>12$} \\
$|\mllOSSF - m_Z|$ [\GeV]          & --                                    & \multicolumn{2}{c}{$>10$} \\
\njets                             & \multicolumn{3}{c}{$\ge$ 2} \\
\nbjets                            & \multicolumn{3}{c}{$\geq$ 1 \btight || $\geq$ 2 \bloose} \\
\hline\hline
Region split                       & {\small(sstt, ttq, ttt, tttq, tttt) $\times$ ($Q^{++}, Q^{--}$)} & {\small(ttt, tttq, tttt) $\times$ ($Q^+, Q^-$)} & -- \\
\hline
Region naming                      & \twopptt      &   \threeppttt     & \fourlep \\
& \twoppttq     &   \threepptttq    & \\
& \twoppttt     &   \threepptttt    & \\
& \twopptttq    &   \threennttt     & \\
& \twopptttt    &   \threenntttq    & \\
& \twonntt      &   \threenntttt    & \\
& \twonnttq     &    & \\
& \twonnttt     &    & \\
& \twonntttq    &    & \\
& \twonntttt    &    & \\
\hline\hline
\end{tabular}
}
\label{tab:selectionSR}
\end{center}
\end{table}
 
Multiple control regions (CRs) are defined in order to fit the normalisation of the leading backgrounds. These regions are orthogonal to the signal regions and with one another based on different requirements on the lepton working points, dilepton invariant mass, and jet and $b$-jet multiplicities. Two regions enriched in diboson and \ttZ are defined by requiring one OS-SF pair compatible with a $Z$ boson, $|\mllOSSF - m_Z| < 10~\GeV$, differing in the jet multiplicity requirement. Two control regions enriched in photon conversions from $Z \rightarrow \mu \mu \gamma^*(\rightarrow ee)$ are defined, according to the identification of the electron as a material conversion or internal conversion candidate. Finally, six control regions are defined enriched in HF non-prompt leptons, making use of the exclusive lepton identification $M_{\mathrm{ex}}$ to be orthogonal to the signal regions. Events with two same-sign leptons are split according to the criteria $(T,M_{\mathrm{ex}})$, $(M_{\mathrm{ex}},T)$, $(M_{\mathrm{ex}},M_{\mathrm{ex}})$ for the leading and subleading leptons in $\pt$, and further split according to the fake-lepton-candidate flavour. The fake-lepton candidate is assumed to be the subleading lepton in the $(T,M_{\mathrm{ex}})$, $(M_{\mathrm{ex}},M_{\mathrm{ex}})$ regions, and the leading lepton in the $(M_{\mathrm{ex}},T)$ region. This splitting creates six control regions sensitive to different relative composition of electron and muon non-prompt lepton backgrounds. Additionally, the transverse mass of the leading lepton and the missing transverse energy, $m_T$($\ell_0$, \met), defined as
$\sqrt{2 \met  p_T^{\ell_0} \left(1- \cos(\phi^{\text{miss}}-\phi^{\ell_0})\right)}$,
is required to be lower than $250~\GeV$ in the $(T,M_{\mathrm{ex}})$ and $(M_{\mathrm{ex}},T)$ regions to reduce the \ttW contribution in these CRs. The full definition of the kinematic selection applied to each control region is given in Table~\ref{tab:selectionCR}.
Figure~\ref{fig:sketch_strategy} illustrates the categorisation and definition of the signal and control regions that are fit simultaneously. The signal contamination is found to be at most 3\% of the total prediction in the control regions, assuming $m_{H}=400~\GeV$ and $\rho_{tt}=0.4$, $\rho_{tc}=0.2$, and $\rho_{tu}=0.2$.
 
\begin{table}[htb]
\begin{center}
\caption{Event selection summary in the control regions. The notation $e^*$ is used to denote material conversion or internal conversion candidates, as described in Section~\ref{sec:objects}.
In the HF non-prompt lepton region naming, ``$2\ell$SStt(e)'' (``$2\ell$SStt($\mu$)'') refers to the control region enriched in non-prompt electrons (muons) from semileptonic $b$-decays originating mostly from \ttbar and with the lepton flavours for the leading and subleading leptons corresponding to ``$ee, \mu e$'' (``$\mu\mu, e \mu$''). The additional $(T,M_{\mathrm{ex}})$, $(M_{\mathrm{ex}},T)$, and $(M_{\mathrm{ex}},M_{\mathrm{ex}})$ subscripts refer to the lepton definitions required for the leading and subleading leptons in each region.
}
\resizebox{\textwidth}{!}{
\def\arraystretch{1.4}
\begin{tabular}{l|cccc}
\hline\hline
Control regions                    & WZ     & \ttZ     & Conversions & HF non-prompt \\
\hline\hline
\njets                             & 2 or 3 & $\geq 4$ &  $\geq 0$           & $\geq 2$ \\
\nbjets                            & \multicolumn{2}{c}{$\geq$ 1 \btight || $\geq$ 2 \bloose} & 0 \bloose & 1 \bloose\\
Lepton requirement                 & \multicolumn{2}{c}{\lll}         & $\mu\mu e^*$ & \llSS \\
Lepton definition                  & \multicolumn{3}{c}{$(L,M,M)$} & {\small$(T,M_{\mathrm{ex}})$ || $(M_{\mathrm{ex}},T)$ || $(M_{\mathrm{ex}},M_{\mathrm{ex}})$} \\
Lepton \pt[\GeV]                  & \multicolumn{3}{c}{(10, 20, 20)}                                 & (20, 20) \\
\mllOSSF[\GeV]                    & \multicolumn{2}{c}{$>12$} & $>12$ & --\\
$|\mllOSSF - m_Z|$ [\GeV]          & \multicolumn{2}{c}{$<10$} & $>10$ & --\\
$|\mlll - m_Z|$ [\GeV]             & \multicolumn{2}{c}{--}  & $<10$ & --\\
$m_T$($\ell_0$, \met) [\GeV]       & \multicolumn{3}{c}{--}  & $< 250$ \\
\hline\hline
Region split                       & -- & -- & internal / material & subleading $e/\mu$ $\times$ {\small[$(T,M_{\mathrm{ex}})$, $(M_{\mathrm{ex}},T)$, $(M_{\mathrm{ex}},M_{\mathrm{ex}})$]} \\
\hline
Region naming                      & \lllVV  & \lllttZ & \lllintc & \llttetm, \llttemt, \llttemm \\
&         &         & \lllmatc & \llttmutm, \llttmumt, \llttmumm \\
\hline\hline
\end{tabular}
}
\label{tab:selectionCR}
\end{center}
\end{table}

\begin{figure}[h!]
\centering
\includegraphics[width=1.1\textwidth]{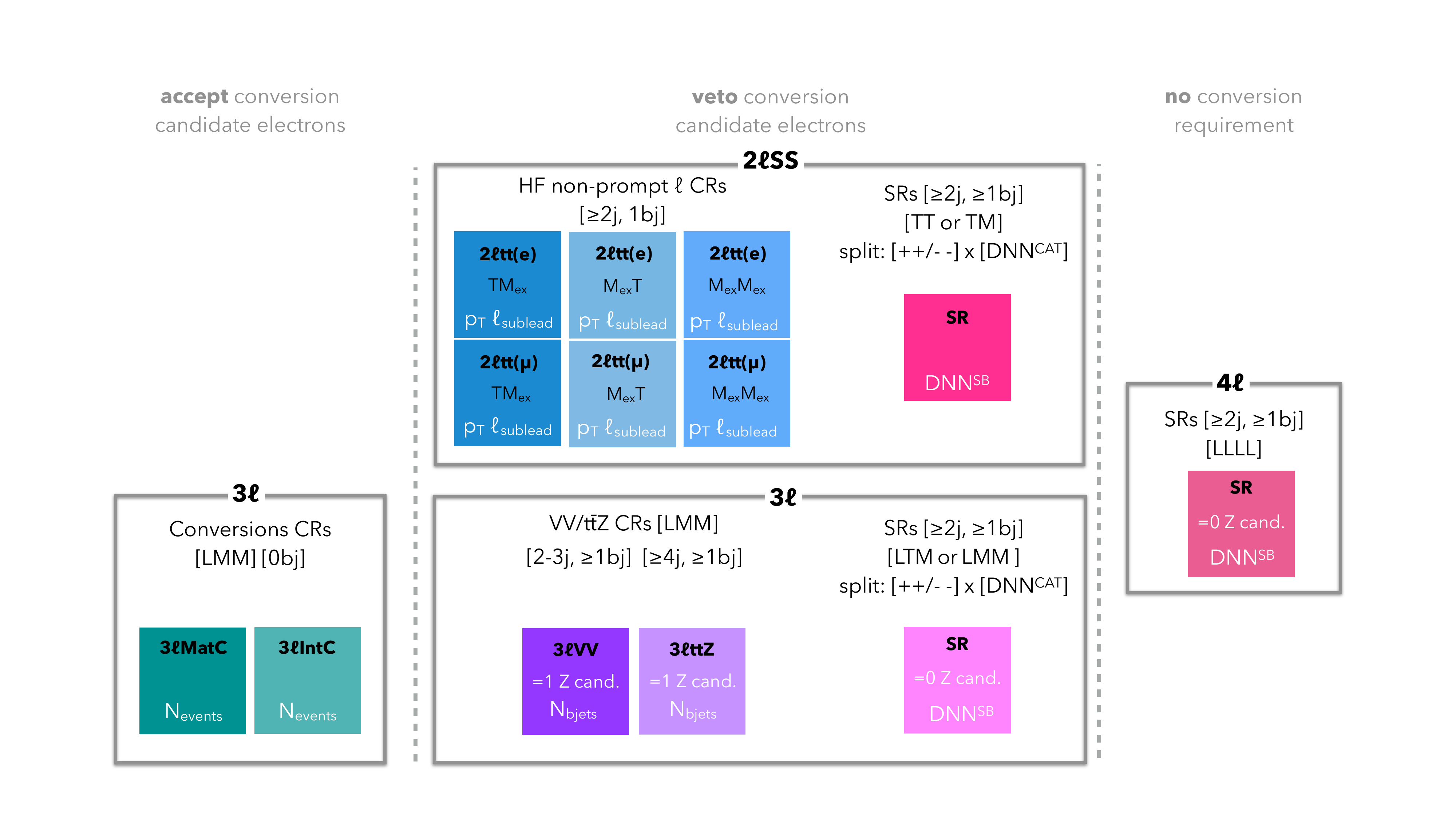}\\
\caption{Illustrative sketch of the definition of the signal and control regions.  The corresponding observable used in the simultaneous fit, as described in Section~\ref{sec:results}, is given at the bottom of each region box. The notation Nj (Nbj) is used to denote selections on the jet ($b$-jet) multiplicity.}
\label{fig:sketch_strategy}
\end{figure}
 
In order to better target each of the possible signals, a \dnncat is trained to identify each of the five possible production and decay modes of the g2HDM signal. Two \dnncat are trained individually for the \llSS and \lll channels using the \textsc{Keras} library~\cite{chollet2015keras} with \textsc{Tensorflow} as a backend~\cite{tensorflow2015-whitepaper} and Adam optimiser~\cite{kingma2017adam}. Hyperparameters are optimised with the Talos library~\cite{talos}. The networks consist of nine input features, two dense fully connected layers of 33 nodes with rectified linear units as activation functions, interleaved with a drop-out layer with 20\% rate, and five (three) output nodes with a soft-max activation function for the categorisation of \llSS (\lll) events. The output categories correspond to the five production modes considered, ignoring in the \lll category signals that cannot produce three leptons. Each event is categorised according to the highest class probability.
The nine input features are the number of jets, $b$-tagging score of the three leading jets, sum of $b$-tagging score of all jets, sum of all pair-wise angular distances between leptons, scalar sum of jet \pt, scalar sum of lepton \pt, and the event \met.
The network is trained with batch size of 2000 and up to 100 epochs, using all the available signal mass points. To avoid discarding signal events in the evaluation, cross-training is used with the events divided by even/odd event number.

\begin{table}[t!]
\small
\begin{center}
\caption{Input variables to the training of the \dnncat and \dnnsb discriminants.}
\begin{tabular}{l|c|c} \hline
Variable                                             & \dnncat &      \dnnsb      \\
\hline
Number of jets ($N_\text{jets}$)                           &   \ding{51}  &      \ding{51}        \\
Sum of pseudo-continuous $b$-tagging scores of jets     &   \ding{51}  &      \ding{51}        \\
Pseudo-continuous $b$-tagging score of 1st, 2nd, 3rd leading jet in \pt &   \ding{51}  &      \ding{51}        \\
Sum of \pt of the jets and leptons ($H_\text{T,jets}$, $H_\text{T,lep}$)        &   \ding{51}  &      \ding{51}        \\
Angular distance of leptons (sum in the case of \lll and \llll)       &   \ding{51}  &      \ding{51}        \\
Missing transverse energy                             &   \ding{51}  &      \ding{51}        \\
Leading transverse momentum of jet                    &   -       &      \ding{51}        \\
Invariant mass of leading lepton and missing transverse energy &   -  &      \ding{51}        \\
Di/tri/quad-lepton type variable (associated with the number of electrons/muons in event) &   -  &      \ding{51}        \\
\hline
\end{tabular}
\label{table:varDNNs}
\end{center}
\end{table}
 
Since several of the probed signal processes are expected to be charge-asymmetric, all the \llSS and \lll regions are further split into two categories each, corresponding to the positive and negative total lepton charge selections. Figures~\ref{fig:summary400} and~\ref{fig:summary1000} show the normalized distributions of the targeted signals with a scalar mass of $400~\GeV$ or $1000~\GeV$, compared to the expected background distribution across the various categories described in Table~\ref{tab:selectionSR}. At high signal mass, a strong migration is observed from the ttt to the tttq category, due to the high probability of additional radiation.
Figures~\ref{fig:composition400} and~\ref{fig:composition1000} show the expected fractional signal contribution in each category for the benchmark coupling. The signals originating from top-Higgs associated production (ttq and ttt) are expected to dominate across all regions, including the categories designed to target other processes, due to the much larger production cross section. This contribution is however strongly dependent on the coupling choice. For the benchmark coupling of $\rhott=0.4$, $\rhotc=0.2$, $\rhotu=0.2$, decays to top-quark pairs dominate when not suppressed by the available phase-space.
 
\begin{figure*}[htpb]
\centering
\subfloat[]{\includegraphics[width=0.49\textwidth,trim={0.2cm 0.2cm 1.3cm 0.5cm},clip]{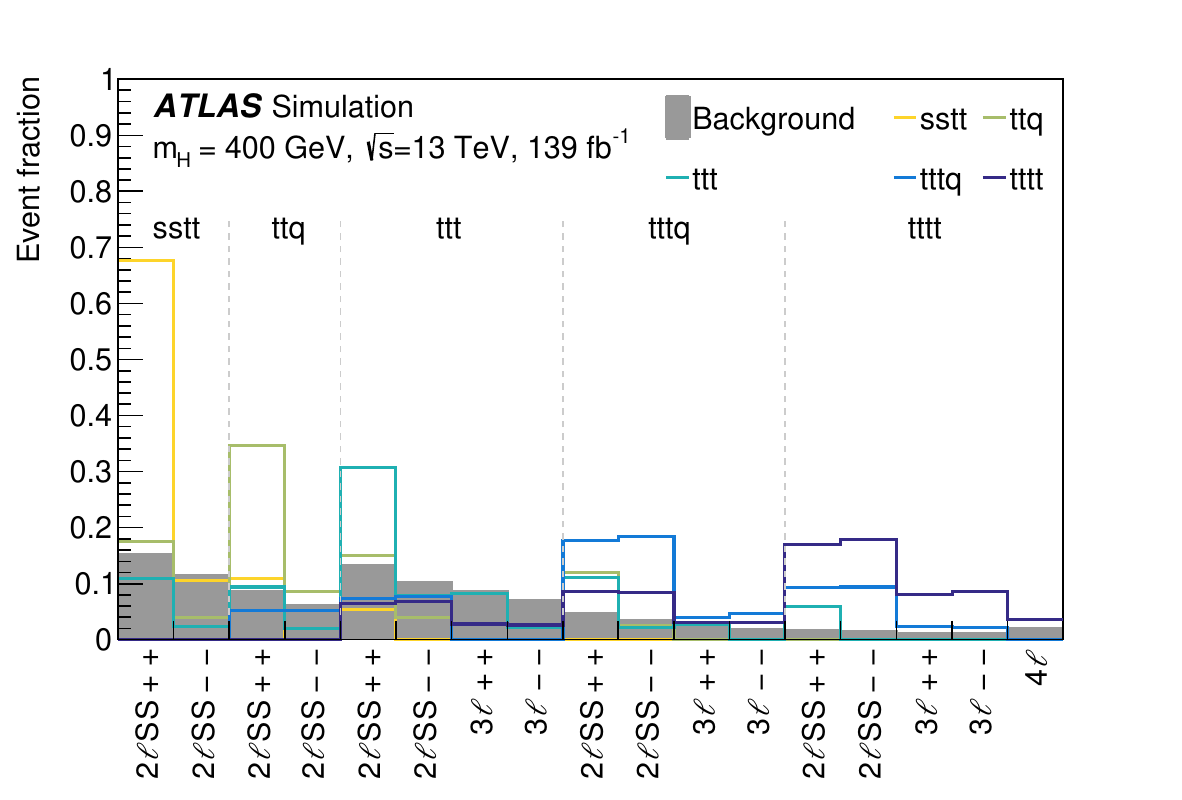}\label{fig:summary400}}
\subfloat[]{\includegraphics[width=0.49\textwidth,trim={0cm 0.2cm 1.5cm 0.5cm},clip]{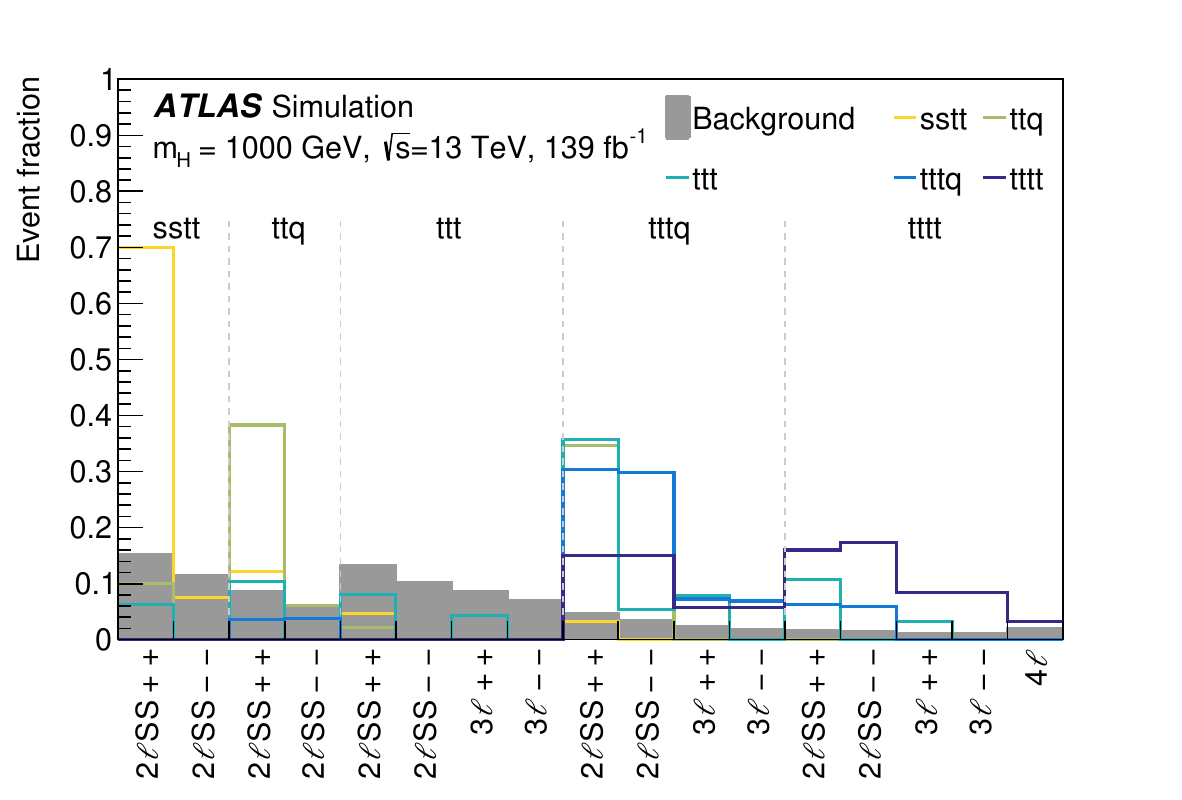}\label{fig:summary1000}} \\
\subfloat[]{\includegraphics[width=0.49\textwidth,trim={0.2cm 0.2cm 1.3cm 0.5cm},clip]{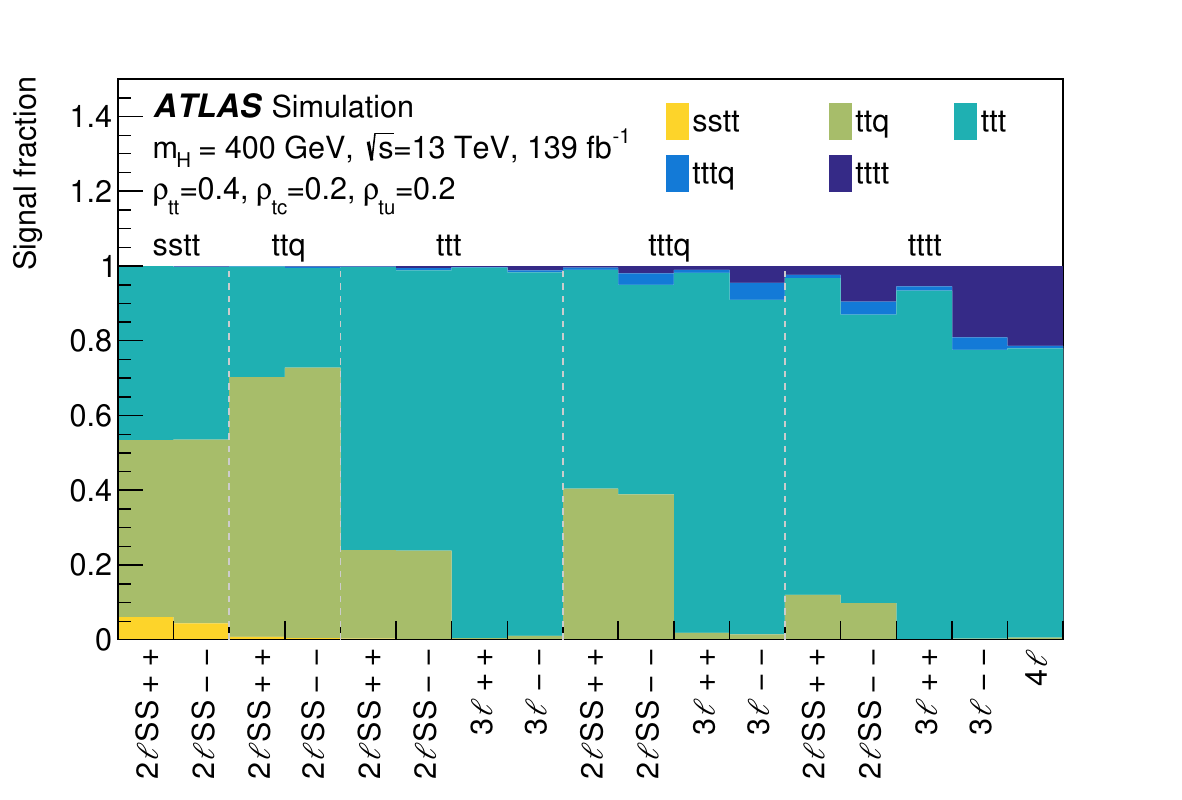}\label{fig:composition400}}
\subfloat[]{\includegraphics[width=0.49\textwidth,trim={0cm 0.2cm 1.5cm 0.5cm},clip]{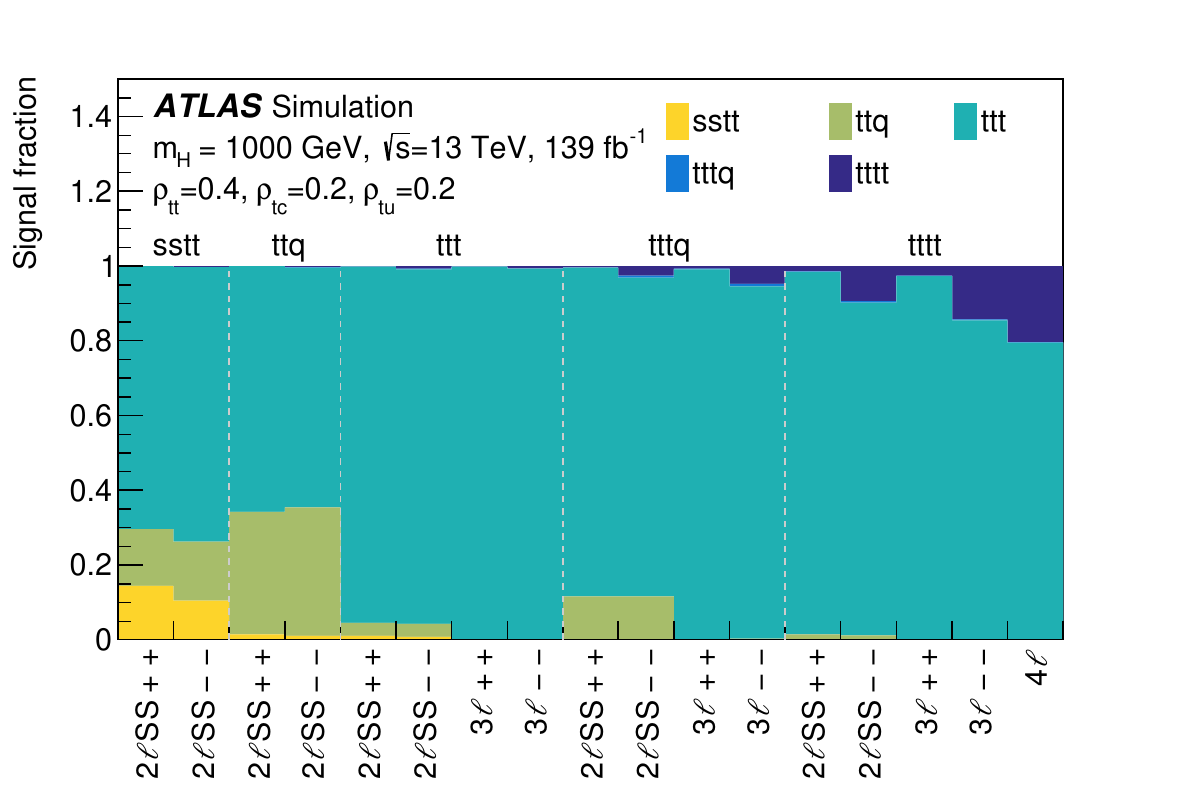}\label{fig:composition1000}}
\caption{Distribution of signal processes and total background between the different DNN categories (top row), for (a) $400~\GeV$ and (b) $1000~\GeV$ scalar masses. All distributions are normalised to unity. The vertical dashed grey lines separate categories targeting each of the main signal processes: sstt, ttq, ttt, tttq, and tttt. Signal contributions below 2\% in a single bin are omitted for clarity. The expected fractional signal contribution in each category (bottom row) is shown for (c) $400~\GeV$ and (d) $1000~\GeV$ scalar masses for the coupling set $\rhott=0.4$, $\rhotc=0.2$, $\rhotu=0.2$.}
\label{fig:cat_fraction}
\end{figure*}
 
A total of 27 analysis regions are defined, with 17 signal regions (10 with \llSS, 6 with \lll, and one \llll) and 10 control regions. In each region, a given kinematic variable is fit to improve the sensitivity to the targeted signal process (signal regions) or to improve the modelling of a particular background process (control regions).
 
A \dnnsb classifier is trained in each signal region to separate the targeted signal from the sum of backgrounds.
The networks consist of 12 input features, two dense fully connected layers of 36 and 48 nodes respectively with sigmoid activation functions, interleaved with a drop-out layer with 40\% rate, and one output node with a sigmoid activation function.
The 12 input features are the leading jet \pt, number of muons, transverse mass of leading lepton and \met system, and the nine variables that are used in the \dnncat.
Table~\ref{table:varDNNs} summarises the input variables used for each multivariate discriminant.
To achieve good sensitivity over the large range of masses that are tested, the output of the classifier is decorrelated from the signal mass introducing an additional term to the loss function via distance correlation~\cite{szekely2007,Kasieczka:2020yyl}.
A hyperparameter $\lambda$ controls the weight of the additional penalty term, with a value of $\lambda = 0.5$.
The value was optimised to achieve a minimal signal mass dependence without compromising the discrimination power.
A separate training is performed in each lepton category and signal category. The same \dnnsb is used in both positive- and negative-charge regions.
The same \dnncat and \dnnsb classifiers trained under the g2HDM signal hypothesis are also used in the interpretation based on the SUSY models.
Figure~\ref{fig:SvsBDNNcomp} shows the \dnnsb distribution of the targeted signal in each signal-enriched category, the total signal, and the background in the
\twopptt, \twoppttq, \threeppttt, \threepptttq, \threepptttt, and \fourlep\ categories.

\begin{figure*}[hbtp]
\centering
\subfloat[]{\includegraphics[width=0.32\textwidth]{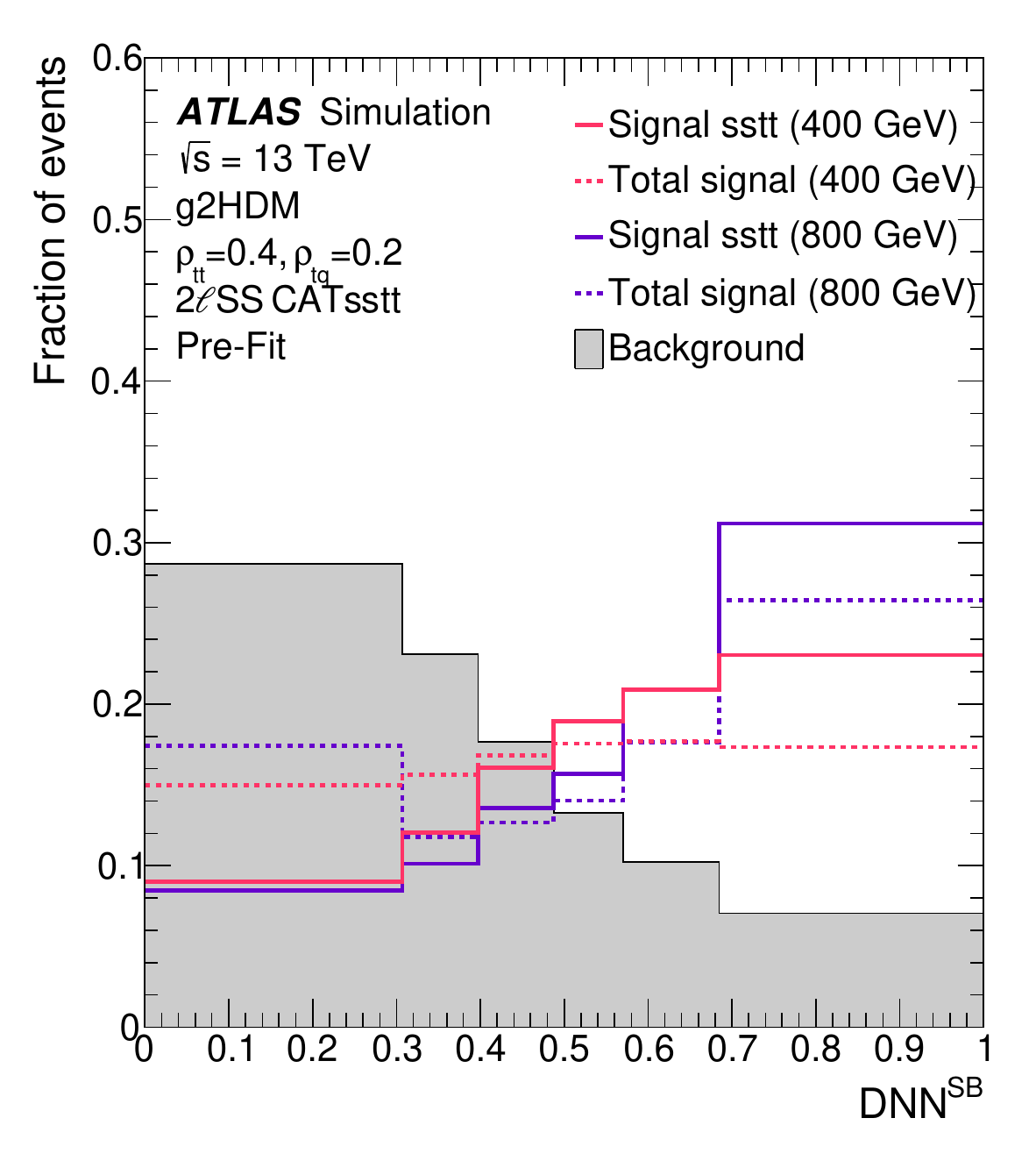}\label{fig:SvsBDNN-2lsstt}}
\subfloat[]{\includegraphics[width=0.32\textwidth]{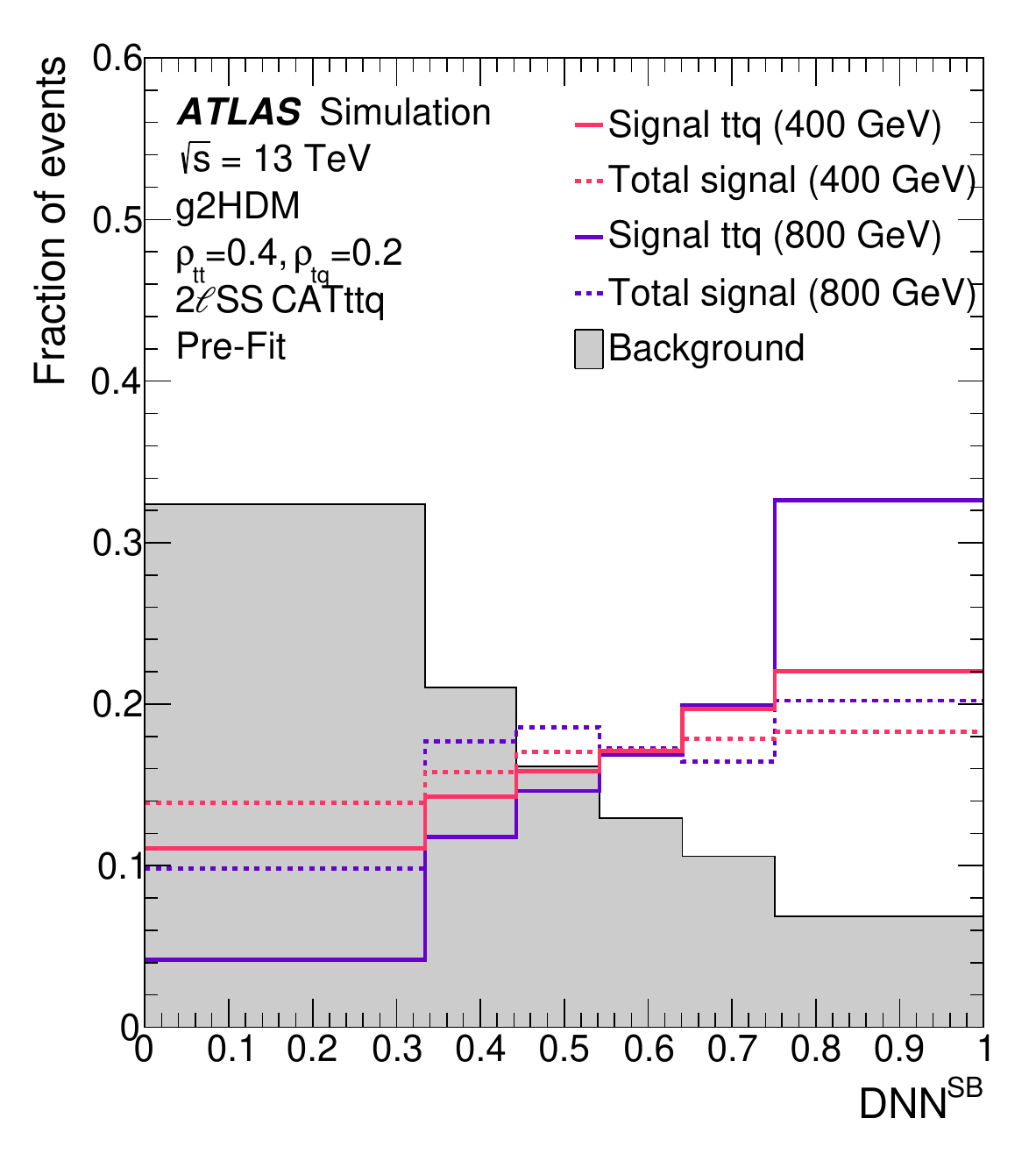}\label{fig:SvsBDNN-2lttq}}
\subfloat[]{\includegraphics[width=0.32\textwidth]{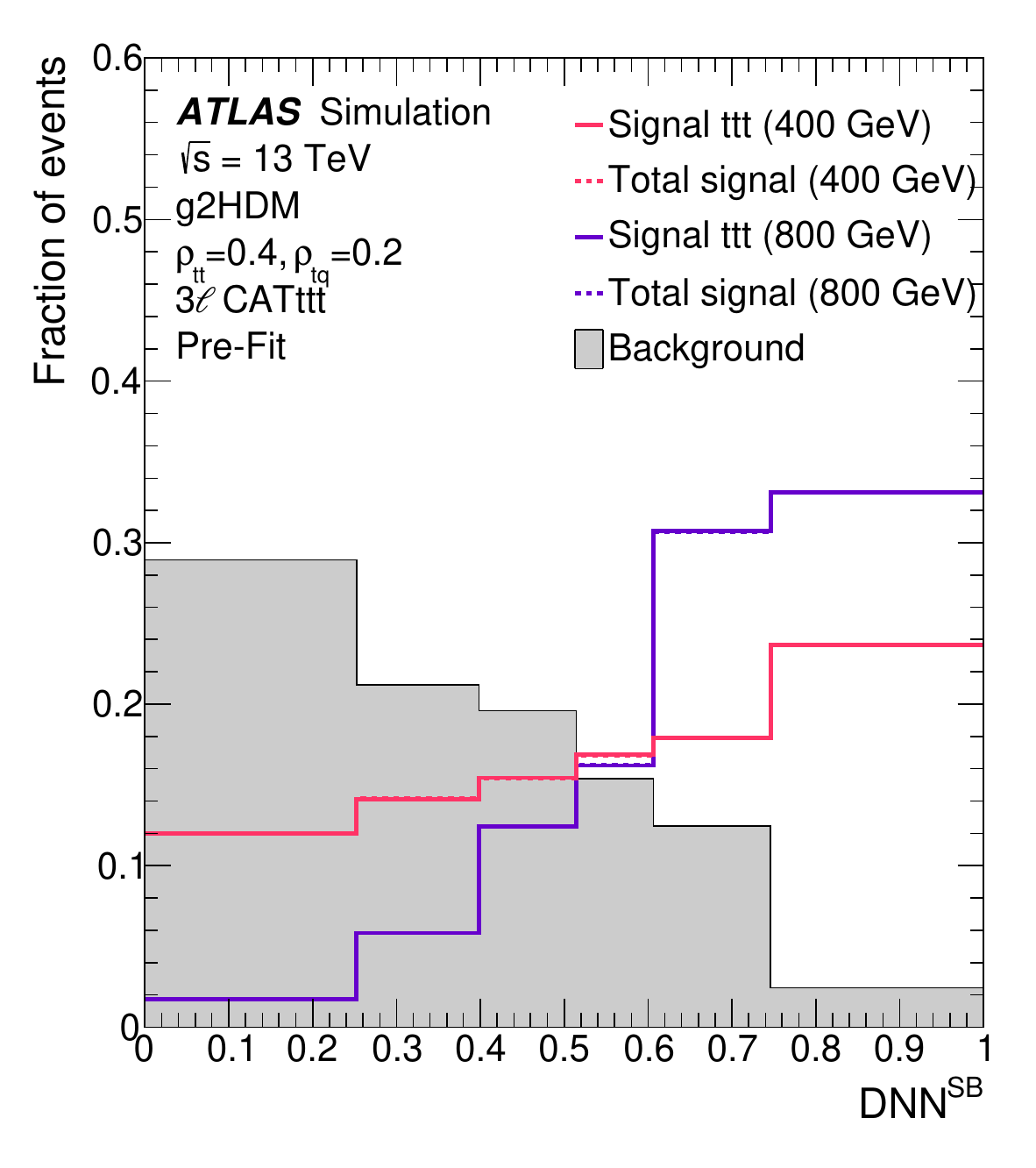}\label{fig:SvsBDNN-3lttt}}\\
\subfloat[]{\includegraphics[width=0.32\textwidth]{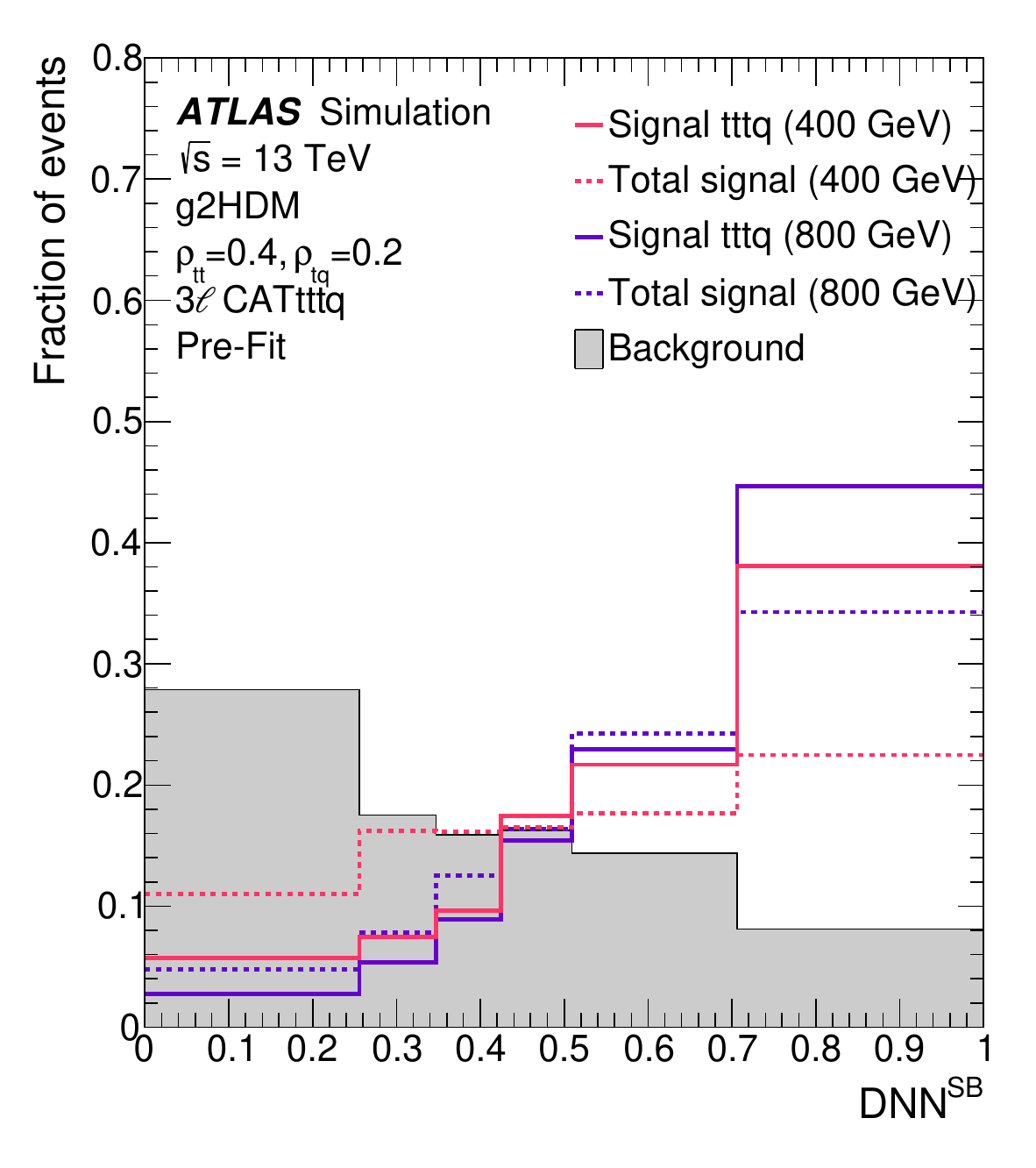}\label{fig:SvsBDNN-3ltttq}}
\subfloat[]{\includegraphics[width=0.32\textwidth]{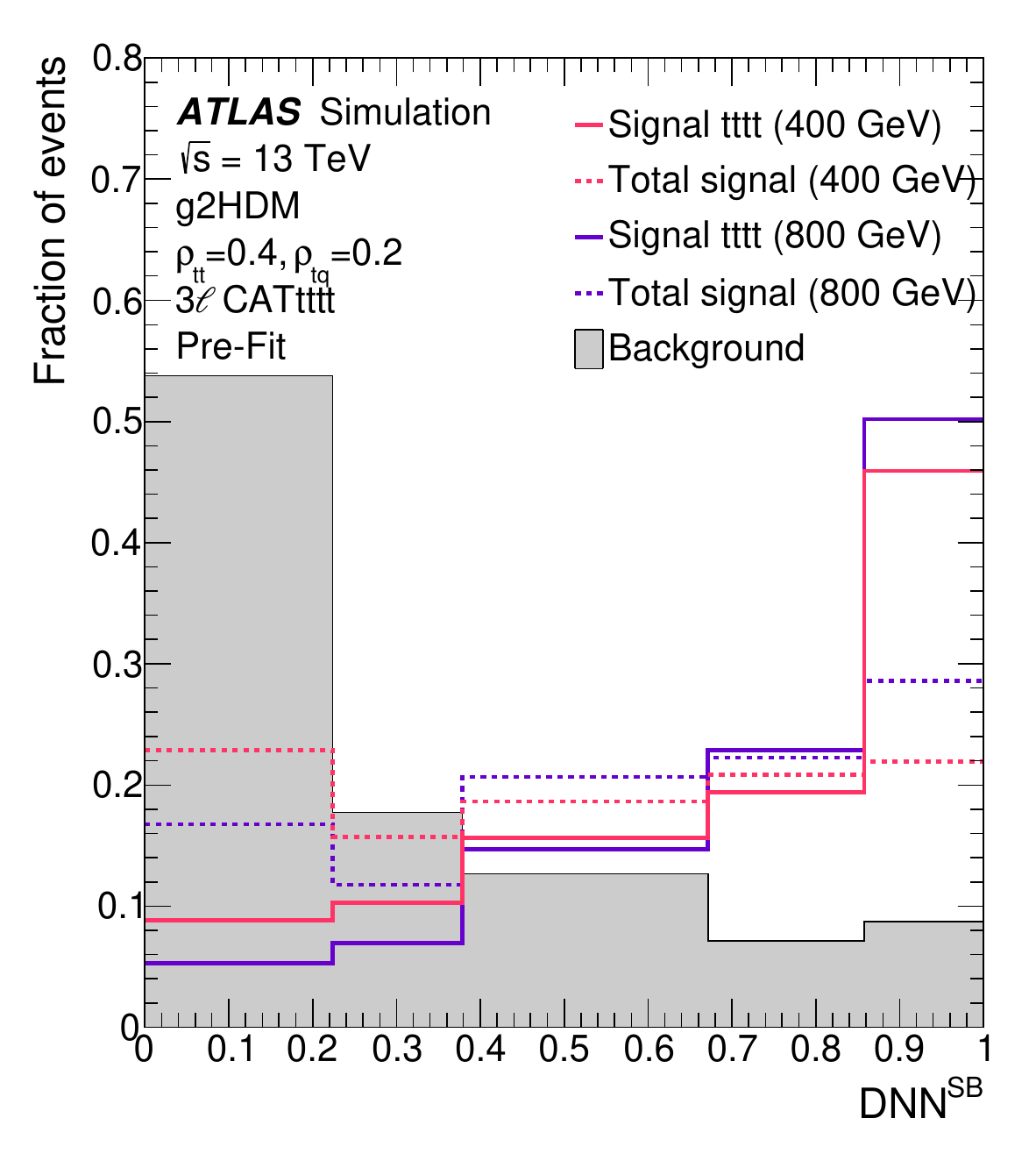}\label{fig:SvsBDNN-3ltttt}}
\subfloat[]{\includegraphics[width=0.32\textwidth]{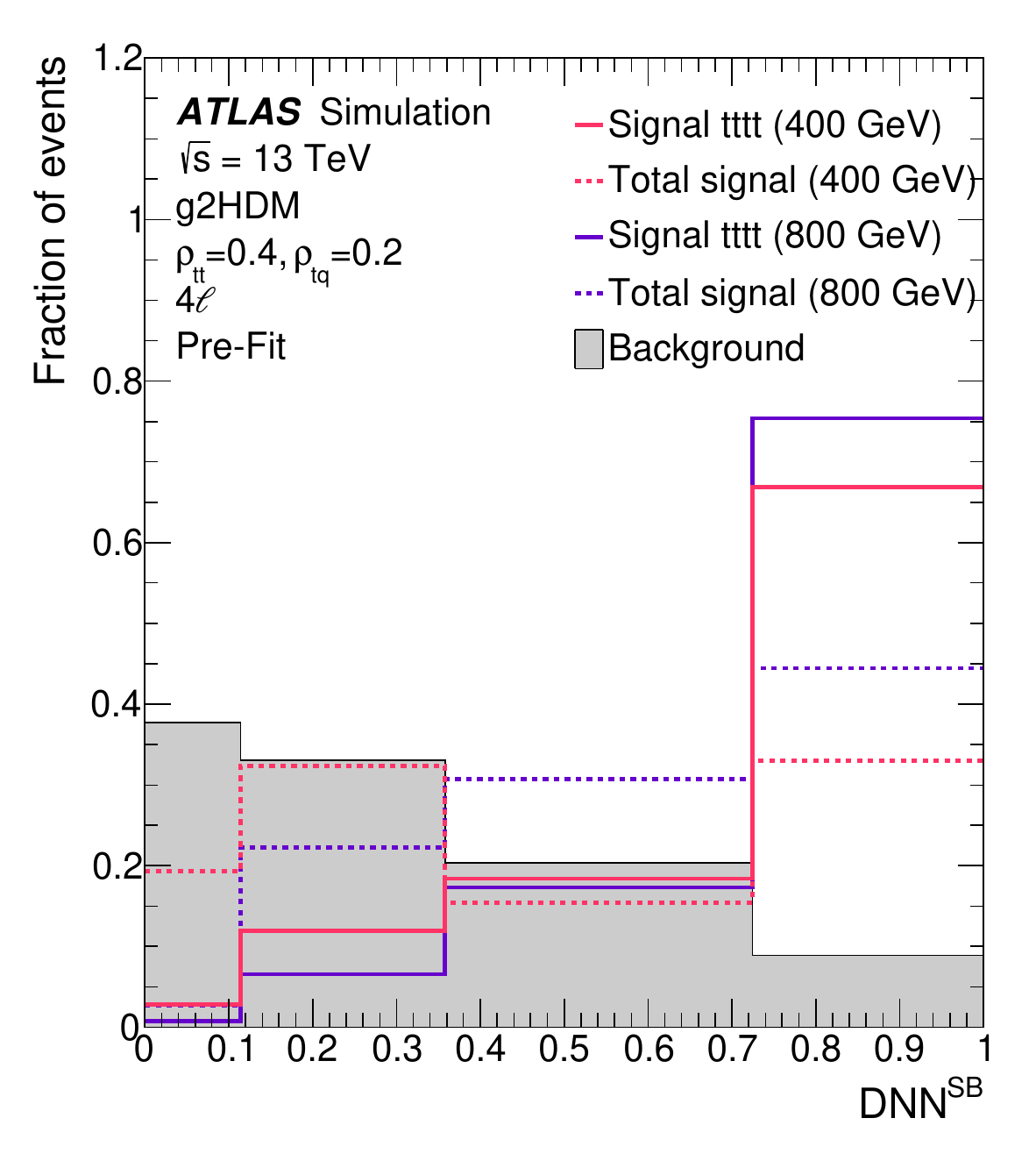}\label{fig:SvsBDNN-4l}}
\caption{Comparison of the \dnnsb distribution of the targeted signal (solid line), the total signal (dashed line) and the background (filled grey area) in the (a) \twott, (b) \twottq, (c) \threettt, (d) \threetttq, (e) \threetttt, and (f) \fourlep\ categories, for an assumed $m_{H} = 400~\GeV$ (pink) and $m_{H} = 800~\GeV$ (violet), with couplings $\rhott=0.4$ and $\rhotq=0.2$. All distributions are normalised to unity.}
\label{fig:SvsBDNNcomp}
\end{figure*}
 
In the diboson and \ttZ control regions the fitted variable is the $b$-jet multiplicity, \nbjets, where the distribution is binned with an upper limit of $\geq 2$ $b$-jets and $\geq 3$ $b$-jets respectively. The subleading lepton \pt spectrum is used in the HF non-prompt control regions. Finally, the total event yield is fit in the control regions enriched in electrons from photon conversion.

% End of text imported from the .//Tex/strategy.tex input file

\section{Background estimation}
\label{sec:background}

% The next lines are included from the .//Tex/background.tex input file
The background processes passing the signal region selections are categorised into irreducible and reducible backgrounds.
Irreducible backgrounds (Section~\ref{sec:bkg_irreducible}) produce prompt leptons in their decay, i.e.\ leptons originating from $W/Z$ boson decays, leptonic $\tau$-lepton decays, or internal conversions. Reducible backgrounds (Section~\ref{sec:bkg_reducible}) have prompt leptons with misassigned charge or at least one non-prompt lepton.
 
Except for the background from electrons with misassigned charge (denoted as QMisID), all other backgrounds are estimated using the simulated samples described in Section~\ref{sec:datamc}. In some cases, the simulation is improved using additional corrections derived from data control samples before the simultaneous fit to data. In particular, the event kinematics of the simulated $\ttbar$ and $\VV$ backgrounds require dedicated corrections to better describe the data. In addition, the yields of some simulated backgrounds, in particular $\ttW$, $\ttZ$, $\VV$ and non-prompt-lepton backgrounds, are adjusted via normalisation factors that are determined by performing a likelihood fit to data across all event categories (signal and control regions as defined in Tables~\ref{tab:selectionSR} and~\ref{tab:selectionCR}) as discussed in Section~\ref{sec:results}.
 
\subsection{Irreducible backgrounds}
\label{sec:bkg_irreducible}
 
Background contributions with prompt leptons originate from a wide range of physics processes with the relative importance of individual processes varying by channel. The main irreducible backgrounds originate from $\ttW$, \tttt, and $\ttdy$ production, followed by $\VV$ (in particular $WZ$) and $\ttH$ production, and have final states and kinematic properties similar to the g2HDM signal. Smaller contributions originate from the following rare processes: $\tZ$, $\tW$, $\tWZ$, $\ttWW$, $\VVV$, and $\ttt$ production.
 
\subsubsection{\ttW background}
 
The $\ttW$ background represents the leading background in several event categories. Despite the use of state-of-the-art simulations, accurate modelling of additional QCD and QED radiation in $\ttW$ production remains challenging. Given the excellent discriminating power of the $\dnnsb$ in the signal regions, the events at lower values of the $\dnnsb$ score are enriched in and sensitive to the $\ttW$ background. Additionally, the signal regions in the \ll and \lll categories are split by the sign of the total lepton charge ($Q$) to better discriminate some g2HDM signal processes and the \ttW process, which have a large charge asymmetry, from other SM backgrounds that are charge symmetric. This discrimination increases the sensitivity to this background in the simultaneous fit. Finally, the \dnncat categories with negative total lepton charge, which are depleted in signals with large charged lepton asymmetry, provide additional constraints on the \ttW background, in particular at high values of the \dnnsb distribution tail.
 
Disagreement between the data and the prefit prediction from the simulation is observed, which is accommodated by an overall normalisation factor that is assigned to the \ttW background, and that is determined during the likelihood fit. The measured normalisation factor for the background-only hypothesis is $\hat{\lambda}_{\ttW} = 1.50 \pm 0.14$, which is compatible with that determined in the SM $t\bar{t}t\bar{t}$ analysis~\cite{Aad:2020klt}, and with a previous measurement of the \ttW production cross section~\cite{Aaboud:2019njj}.

\subsubsection{\VV and \ttdy backgrounds}
 
A data-driven correction to the jet multiplicity spectrum is derived from an inclusive trilepton diboson-enriched region with zero $b$-jets defined with the 85\% WP for $b$-jet efficiency and at least one jet (denoted as \lllVVnob region). The events are required to have three leptons that satisfy the same selection as in the $\lllVV$ CR.
 
Figure~\ref{fig:VV00_NJets_before} shows the jet multiplicity distribution in the \lllVVnob region before the correction. After the correction is applied to \VV, a good modelling of the $\njets$ distribution is found in a \lll region with at least one jet and exactly one $b$-jet defined with the 60\% WP, as shown in Figure~\ref{fig:CR_NJets_after}.

\begin{figure}[t]
\centering
\subfloat[]{\includegraphics[width=0.40\textwidth]{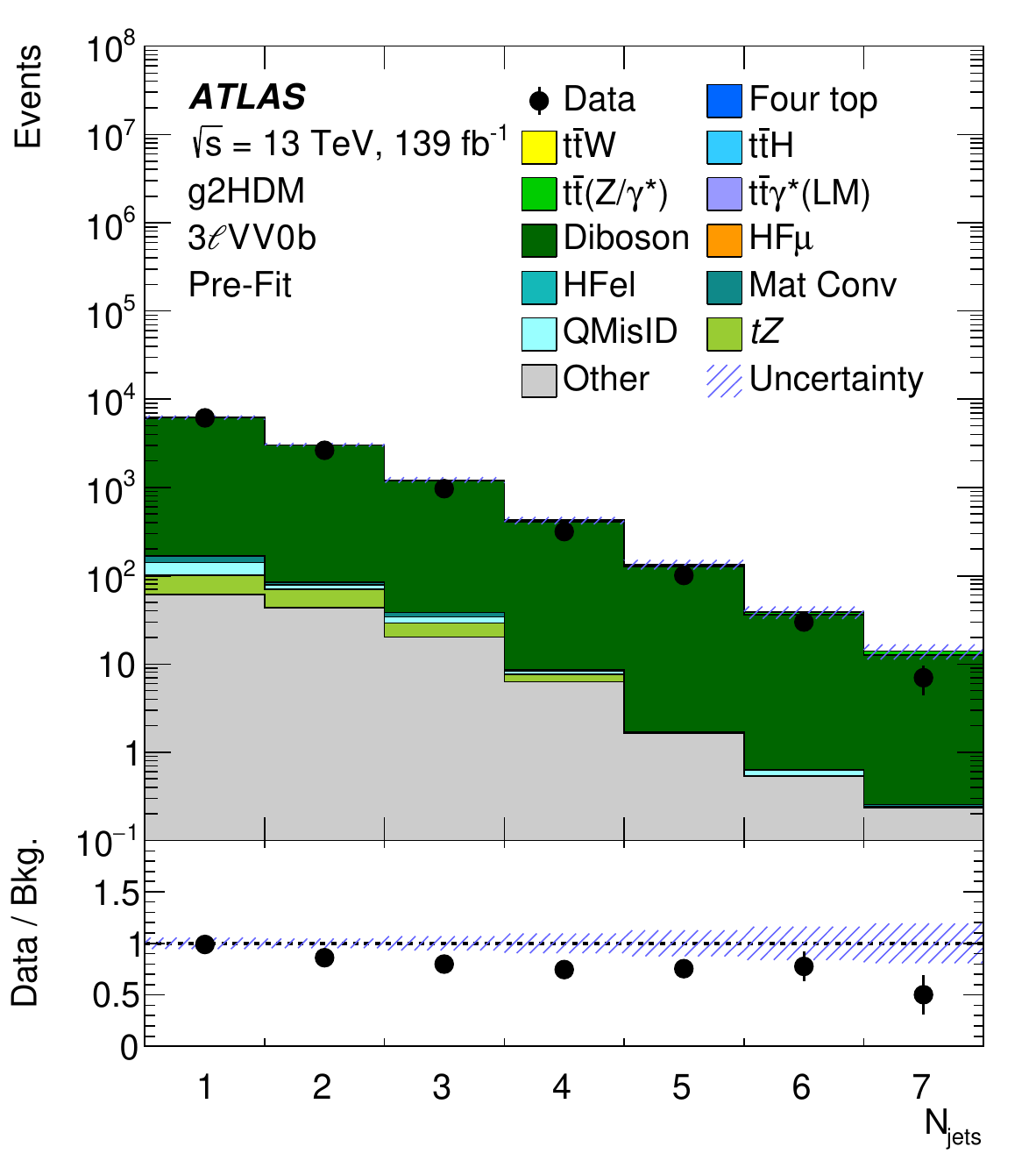}\label{fig:VV00_NJets_before}}
\subfloat[]{\includegraphics[width=0.40\textwidth]{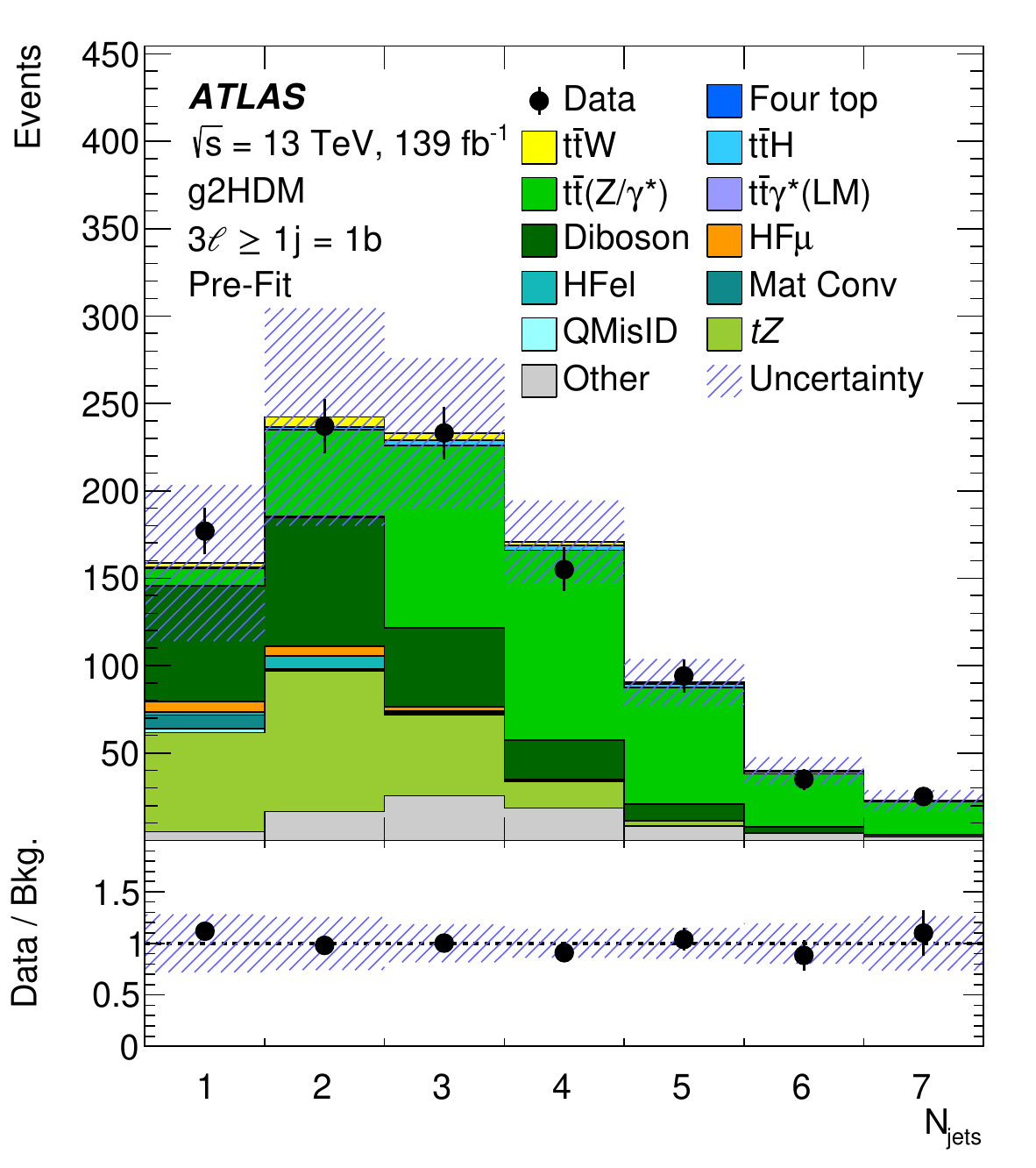}\label{fig:CR_NJets_after}}
\caption{
Comparison between data and the background prediction for the distribution of (a) the number of jets in the \lllVVnob region before the \VV jet multiplicity correction and (b) the number of jets in a \lll region with at least one jet and exactly one $b$-jet defined with the 60\% WP after the \VV jet multiplicity correction.
The ratio of the data to the background prediction (``Pred.'') is shown in the lower panel.
The size of the combined statistical and systematic uncertainty in the background prediction is indicated by the blue hatched band.
The last bin in each figure contains the overflow.}
\end{figure}
 
The \lllVV and \lllttZ CRs are used in the likelihood fit to improve the prediction of the background contribution from the $\VV$ and $\ttdy$ processes; these processes have purities of 15\% and 75\% in the corresponding CRs. The numbers of jets and $b$-jets provide good discrimination between these two processes and are used to build the control regions (number of jets) and as variables 
in the fit (number of $b$-jets). The measured normalisation factors for the background-only hypothesis are: $\hat{\lambda}_\textrm{\VV} = 0.85 \pm 0.30$ and $\hat{\lambda}_\textrm{\ttZ} = 0.97 \pm 0.19$, where the uncertainty includes both statistical and systematic contributions.
 
Figures~\ref{fig:VVCR_Nbjpost} and~\ref{fig:ttZCR_Nbjpost} show the $b$-jet multiplicity distribution in the \lllVV and \lllttZ CRs after the likelihood fit to data.
 
\begin{figure}[t]
\centering
\subfloat[]{\includegraphics[width=0.40\textwidth]{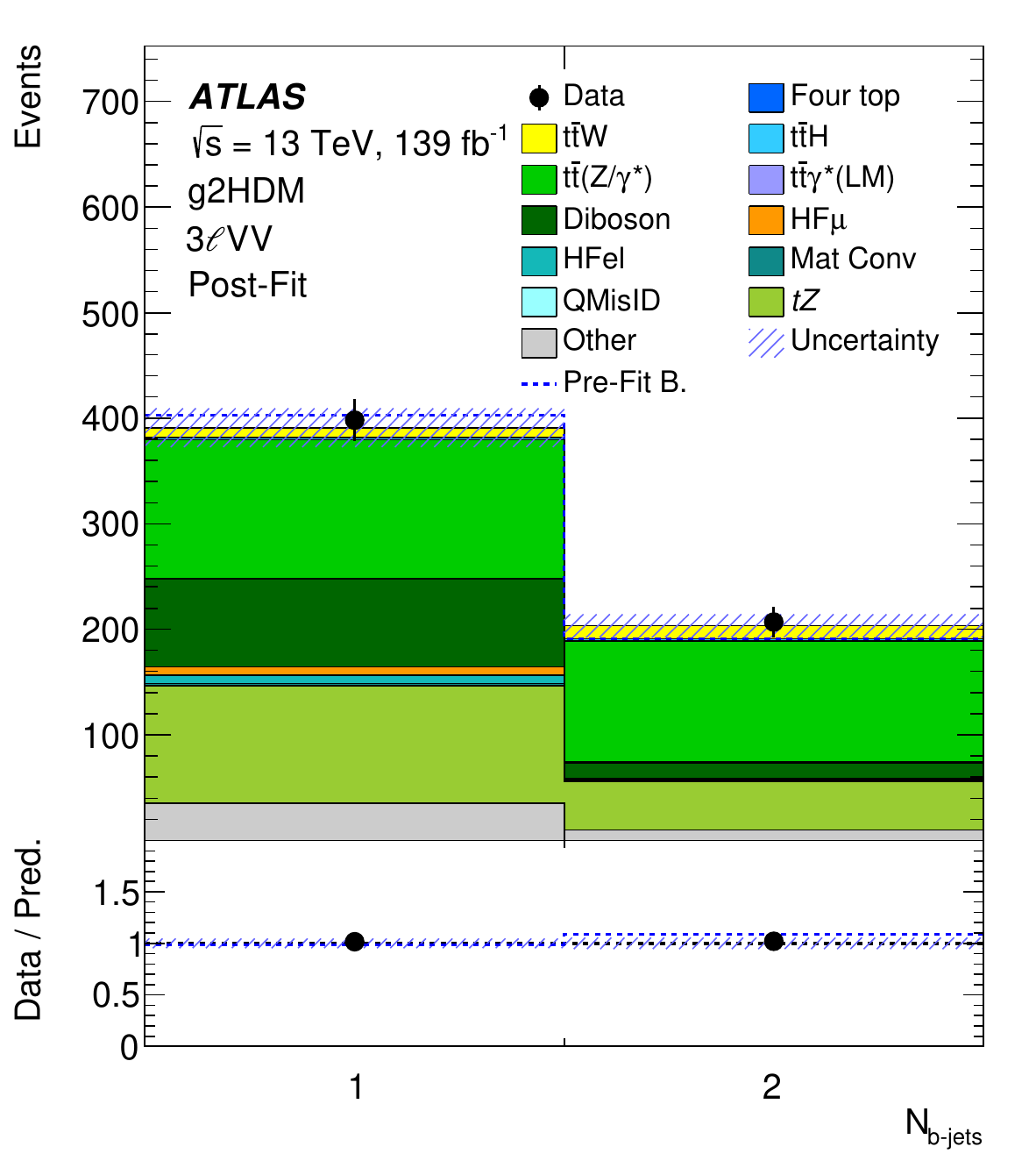}\label{fig:VVCR_Nbjpost}}
\subfloat[]{\includegraphics[width=0.40\textwidth]{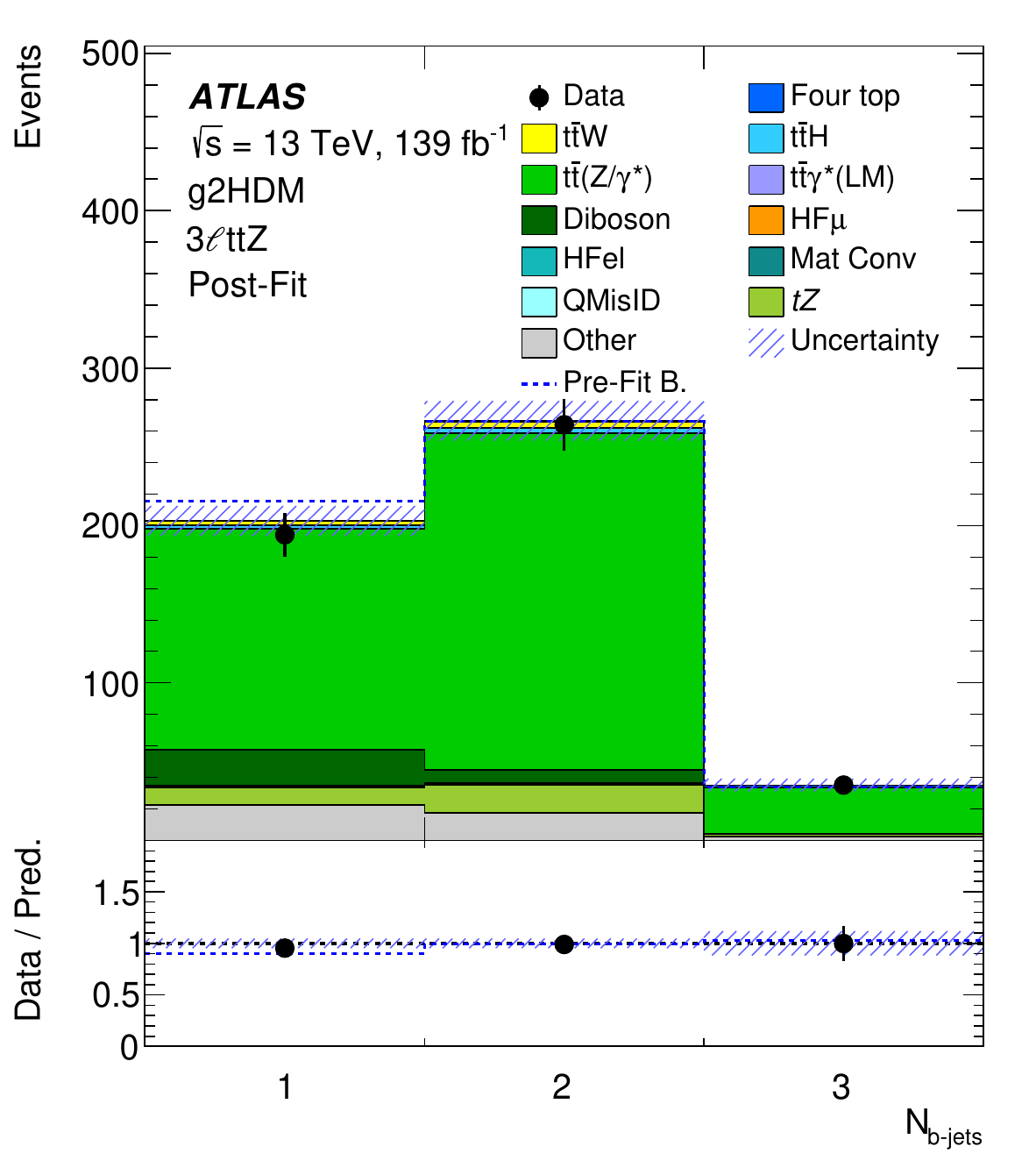}\label{fig:ttZCR_Nbjpost}}
\caption{
Comparison between data and the background prediction for the distribution of the $b$-jet multiplicity in the (a) \lllVV and (b) \lllttZ CRs after the \VV jet multiplicity correction.
The background contributions after the likelihood fit to data (``Post-Fit'') for the background-only hypothesis are shown as filled histograms.
The ratio of the data to the post-fit background prediction (``Pred.'') is shown in the lower panel, separately for post-fit background (black points) and pre-fit background (dashed blue line).
The size of the combined statistical and systematic uncertainty in the background prediction is indicated by the blue hatched band.
The last bin in each figure contains the overflow.
}
\end{figure}
 
\subsubsection{Other irreducible backgrounds}
\label{sec:other_irreducible_backgrounds}
 
The rate of the background from internal conversions with $m(e^+e^-)<1~\gev$ is estimated using the two dedicated CRs, \lllintc and \lllmatc, with a purity of 98\% and 30\%, respectively. The total yield in each category is used in the likelihood fit to determine the normalisation factor, which is measured for the background-only hypothesis to be $\hat{\lambda}^\textrm{IntC}_e = 1.06 \pm 0.23$, where the uncertainty is dominated by the statistical uncertainty.

\subsection{Reducible backgrounds}
\label{sec:bkg_reducible}
 
\subsubsection{Non-prompt leptons}
\label{sec:bkg_reducible_nonprompt}
 
Non-prompt leptons originate from material conversions, heavy-flavour hadron decays, or the improper reconstruction of other particles, with an admixture that depends strongly on the lepton quality requirements and varies across event categories. These backgrounds are small in all \ll and \lll SRs and thus are estimated from simulation, with the normalisation determined by the likelihood fit. The non-prompt lepton background contribution in the \llll SR is very small and is therefore taken from simulation without dedicated data-driven corrections. The main contribution to the non-prompt-lepton background is from $\ttbar$ production, with much smaller contributions from $V$+jets and single-top-quark processes. The non-prompt leptons in the simulated samples are labelled according to whether they originate from heavy-flavour (HF) or light-flavour (LF) hadron decays, or from a material conversion candidate (Mat. Conv.). The HF category includes leptons from both bottom and charm decays.
 
Two corrections are applied to the \ttbar and the overall non-prompt lepton background simulation before the fit. First, the \ttbar+$\geq 1$ $b$-jet contribution from simulation is known to be mismodelled and is therefore corrected by a factor of 1.3 as measured by a previous ATLAS analysis sensitive to the in-situ measurement of this contribution in the single- and opposite-sign dilepton final states~\cite{HIGG-2020-23}. This correction is well motivated since the mismodelling of additional $b$-jets in \ttbar is not expected to depend on the presence of additional non-prompt leptons in the event. Second, the shape of the $b$-jet multiplicity in the non-prompt lepton background simulation is corrected for electrons and muons separately to match data in an orthogonal \llSS validation region enriched with non-prompt leptons, where one of the leptons must satisfy a looser requirement on the non-prompt lepton BDT score but fail the $M$ lepton WP criteria.
 
Several of the event categories introduced in Section~\ref{sec:events} were designed to be enriched in specific processes and are used to derive normalisation factors to improve their modelling by the simulation.
The \lllmatc CR is enriched in material conversions with a purity of 70\% and only the total event yield is used. There are six \ll CRs enriched in contributions from HF non-prompt leptons in \ttbar events, i.e.\ \llttetm, \llttemt, \llttemm, \llttmutm, \llttmumt, and \llttmumm. In these CRs, the transverse momentum of the fake-lepton-candidate distribution is used to be able to correct for a possible mismodelling in the \pt of the non-prompt lepton.  The fake-lepton candidate is assumed to be the subleading lepton in the $(T,M_{\mathrm{ex}})$, $(M_{\mathrm{ex}},M_{\mathrm{ex}})$ regions, and the leading lepton in the $(M_{\mathrm{ex}},T)$ region. The event requirement to have at least one $M_{\mathrm{ex}}$ lepton provides separation from the irreducible backgrounds, in particular $\ttW$, and thus increases the sensitivity to the HF non-prompt electron and muon contributions. Normalisation factors for three non-prompt-lepton background contributions are estimated from the likelihood fit. The normalisation factor for HF non-prompt leptons is estimated separately for electrons and muons, $\lambda^\textrm{had}_e$ and $\lambda^\textrm{had}_\mu$, respectively. An additional normalisation factor is determined for the material conversions background, $\lambda^\textrm{Mat Conv}_e$. The measured normalisation factors for the background-only hypothesis are: $\hat{\lambda}^\textrm{had}_e = 1.05 \pm 0.31$,  $\hat{\lambda}^\textrm{had}_\mu = 0.92 \pm 0.18$, and $\hat{\lambda}^\textrm{Mat Conv}_e = 1.16 \pm 0.29$, where the uncertainties include systematic effects but are dominated by the statistical uncertainty.
 
Figures~\ref{fig:HFe_TMCR} and ~\ref{fig:HFmu_TMCR} display the fake-lepton-candidate \pt distribution in the \llttetm and \llttmutm CRs after the likelihood fit to data. As shown in the figures, the purity of HF non-prompt lepton background is 45\% and 55\%, respectively, which was possible to achieve with the usage of the exclusive $M_{\mathrm{ex}}$ lepton working point.
 
\begin{figure}[t]
\centering
\subfloat[]{\includegraphics[width=0.40\textwidth]{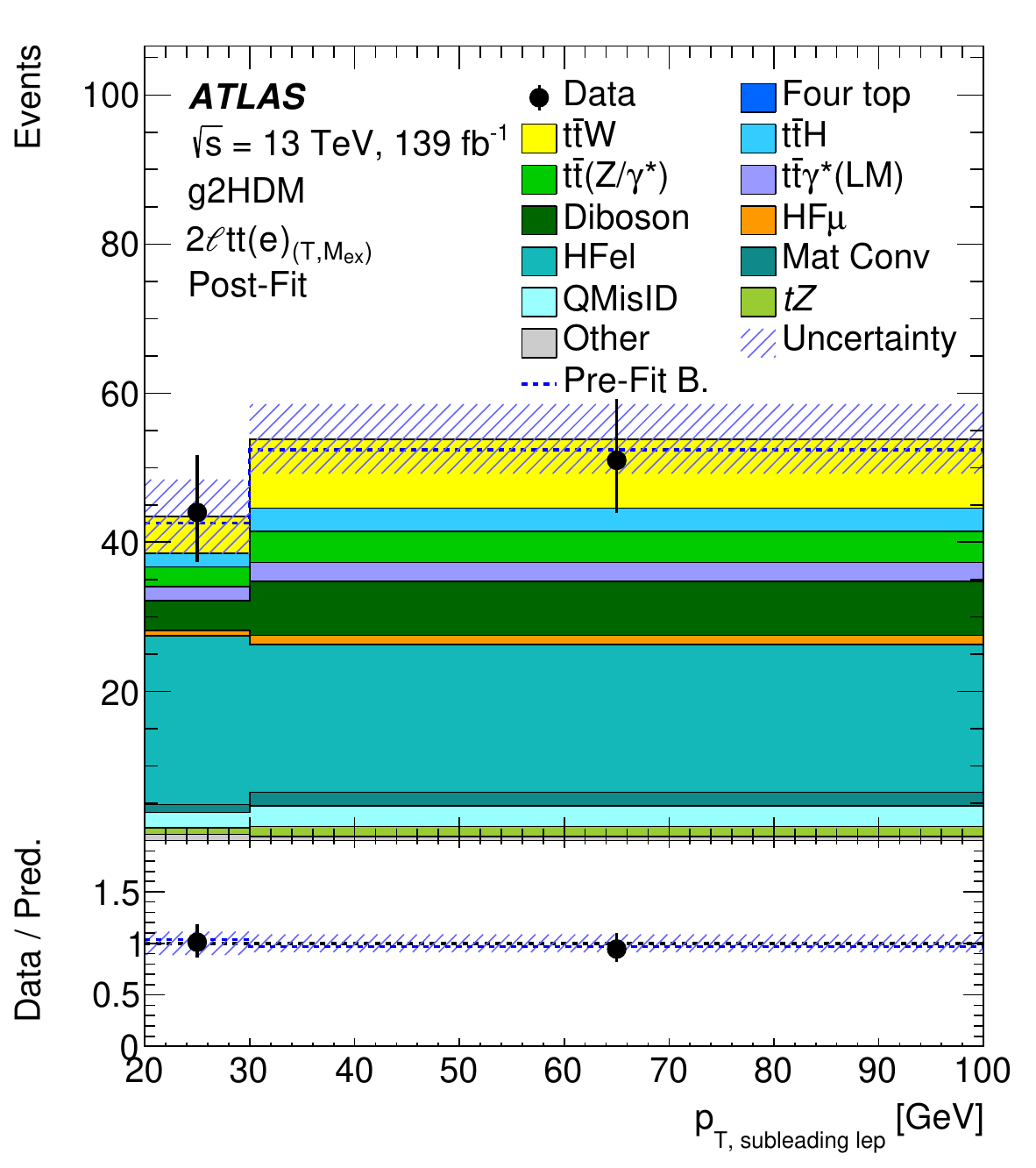}\label{fig:HFe_TMCR}}
\subfloat[]{\includegraphics[width=0.40\textwidth]{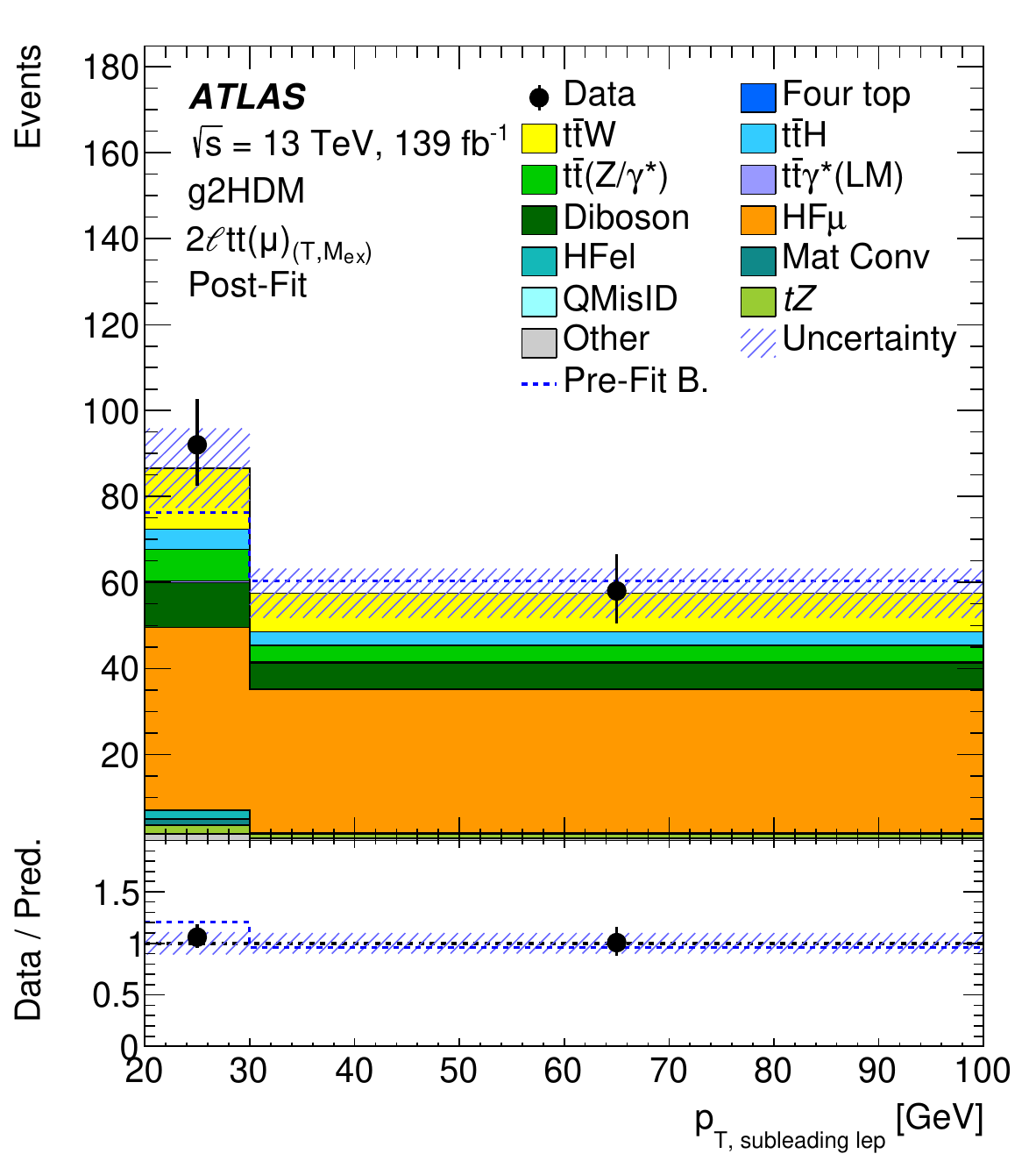}\label{fig:HFmu_TMCR}}
\caption{
Comparison between data and the background prediction for the distribution of the transverse momentum of the sub-leading lepton ($p_\textrm{T, subleading lep}$) in (a) the \llttetm CR and (b) the \llttmutm CR.
The background contributions after the likelihood fit to data (``Post-Fit'') for the background-only hypothesis are shown as filled histograms.
The ratio of the data to the post-fit background prediction (``Pred.'') is shown in the lower panel, separately for post-fit background (black points) and pre-fit background (dashed blue line).
The size of the combined statistical and systematic uncertainty in the background prediction is indicated by the blue hatched band.
The last bin in each figure contains the overflow.}
\end{figure}
 
\subsubsection{Charge misassignment}
\label{sec:bkg_reducible_chg_miss}
 
Backgrounds with leptons with the charge incorrectly assigned affect primarily the $\ll$ channel and predominantly arise from $t\bar{t}$ production, where one electron udergoes a hard bremsstrahlung and an asymmetric conversion ($e^\pm \to e^\pm \gamma^* \to e^\pm e^+ e^-$) or a mismeasured track curvature.
The muon charge misassignment rate is negligible in the $\pt$ range relevant to this analysis. The electron charge misassignment rate is measured in data using samples of
$Z \to e^+e^-$ events reconstructed as same-charge pairs and as opposite-charge pairs, with the background subtracted via a sideband method~\cite{EGAM-2018-01}.
 
The charge misassignment rate is parameterised as a function of electron $\pt$ and $|\eta|$. It varies from about $10^{-5}$ for low-$\pt$ electrons ($17 \leq \pt \leq 50~\gev$) that satisfy $|\eta| \leq 1.37$, to about $4 \times 10^{-3}$ for high-$\pt$ electrons ($\pt \geq 100~\gev$) in the region $2 \leq  |\eta| \leq 2.47$.
To estimate the QMisID background in each of the corresponding event categories, the measured charge misassignment rate is then applied to data events satisfying the requirements of the $\ll$ channels, except that the two leptons are required to be of opposite charge.

\subsection{Background modelling}
 
The modelling of some representative variables at preselection level is showed in Figure~\ref{fig:bkgvalidation}. The background prediction includes all the corrections previously described as well as the normalisation factors  determined through a likelihood fit to data as discussed in Section~\ref{sec:results}.
Good modelling is observed across all variables and lepton categories.
 
\begin{figure}[pht]
\centering
\subfloat[]{\includegraphics[width=0.33\textwidth]{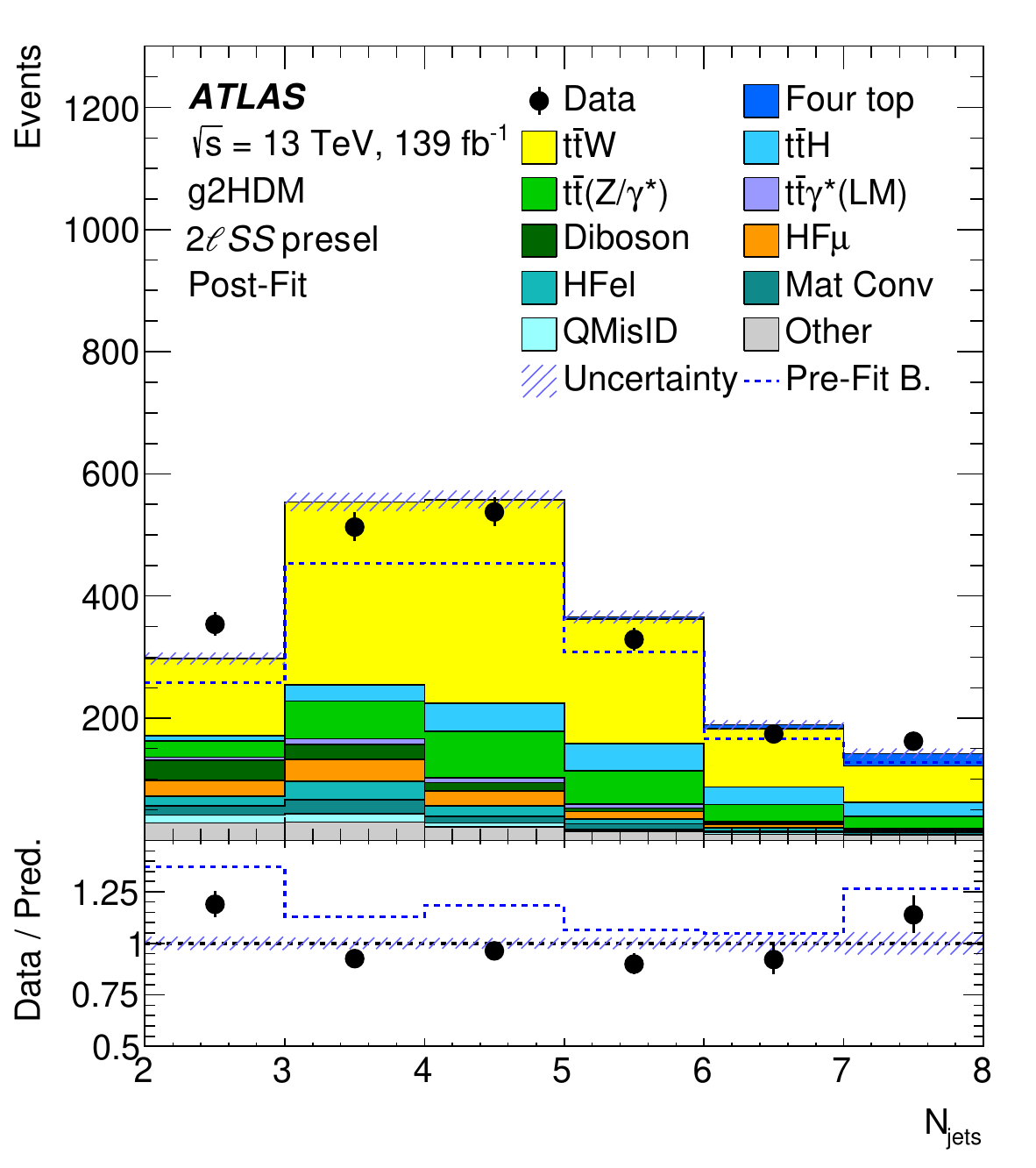}}
\subfloat[]{\includegraphics[width=0.33\textwidth]{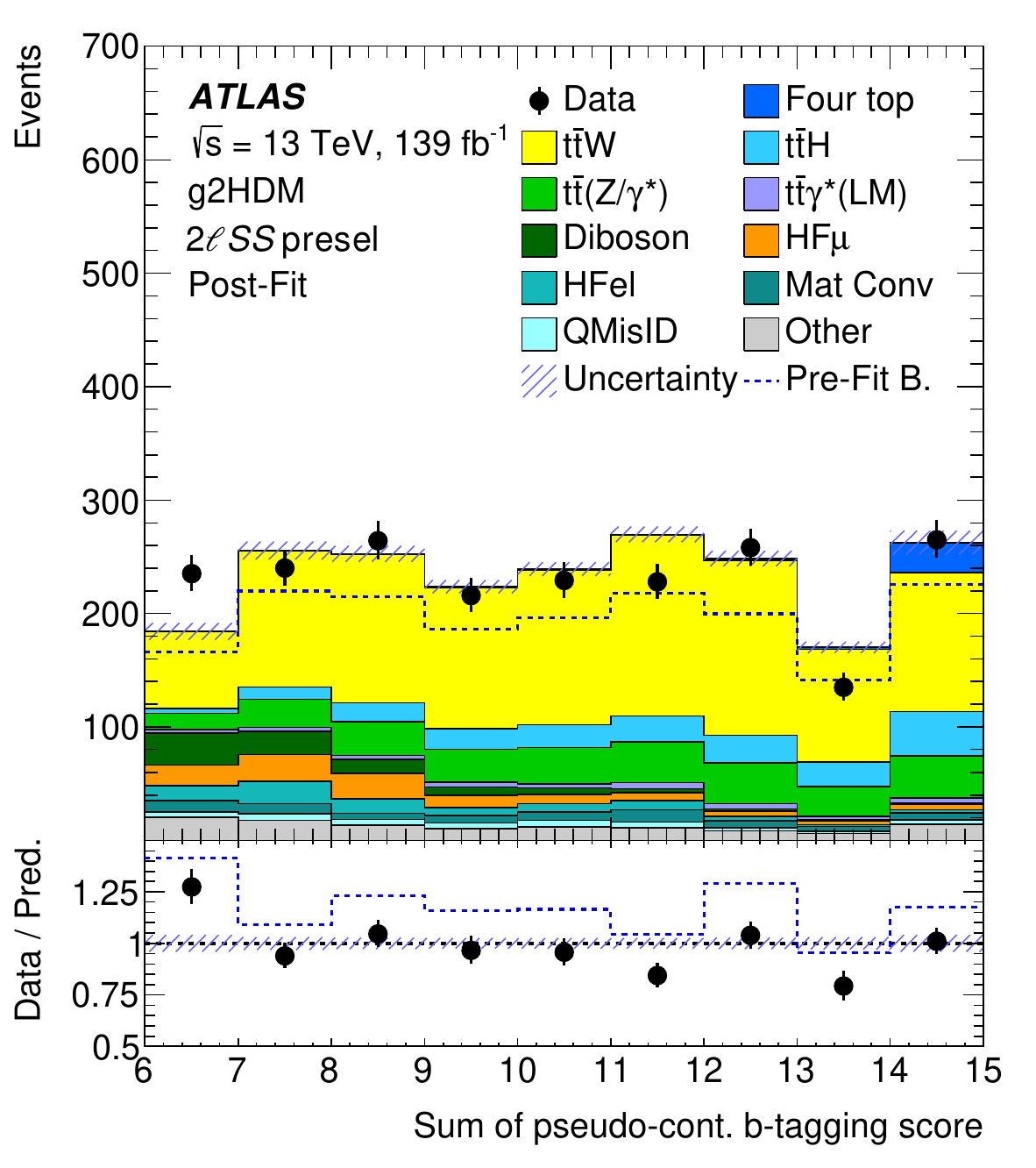}}
\subfloat[]{\includegraphics[width=0.33\textwidth]{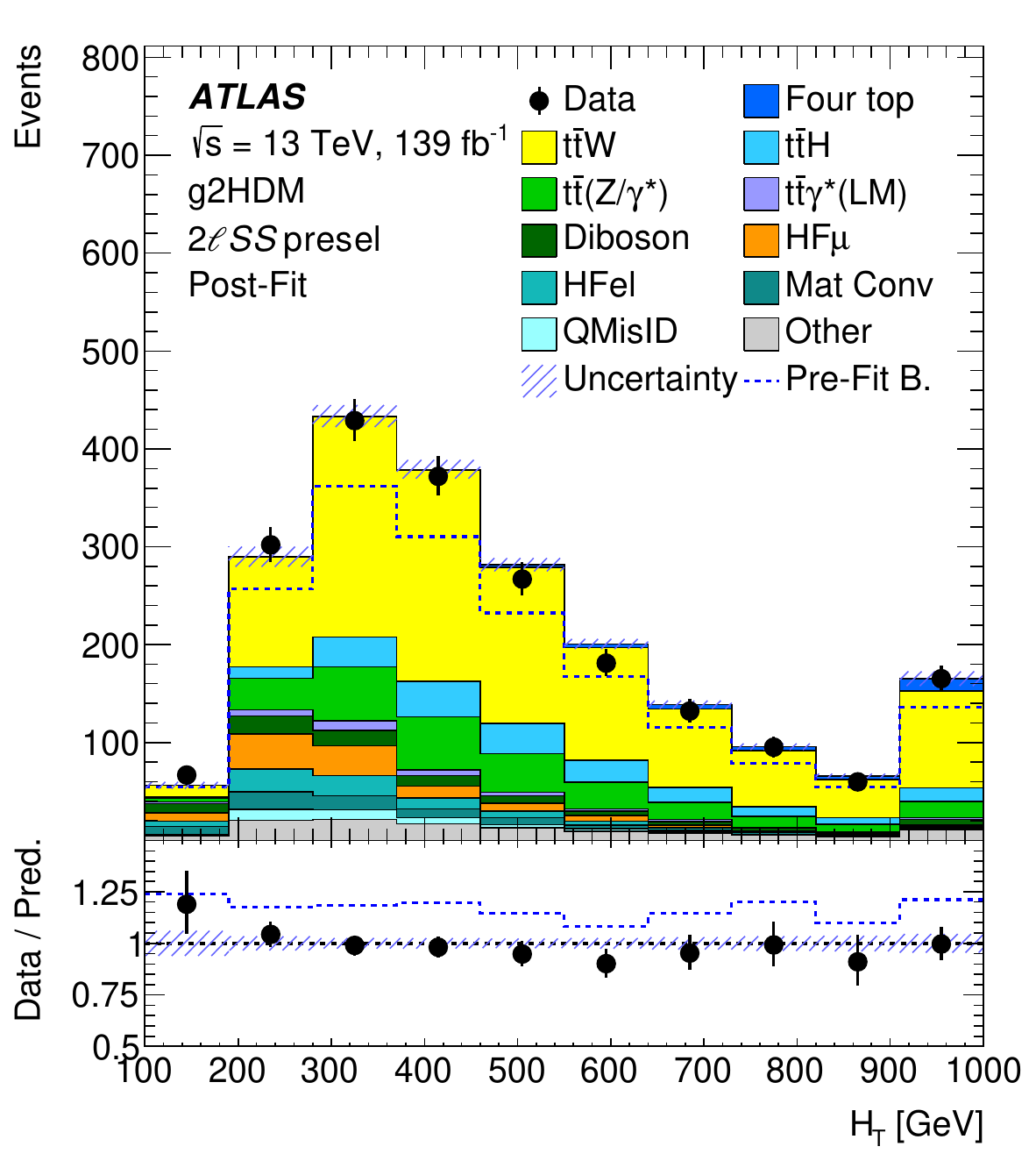}} \\
\subfloat[]{\includegraphics[width=0.33\textwidth]{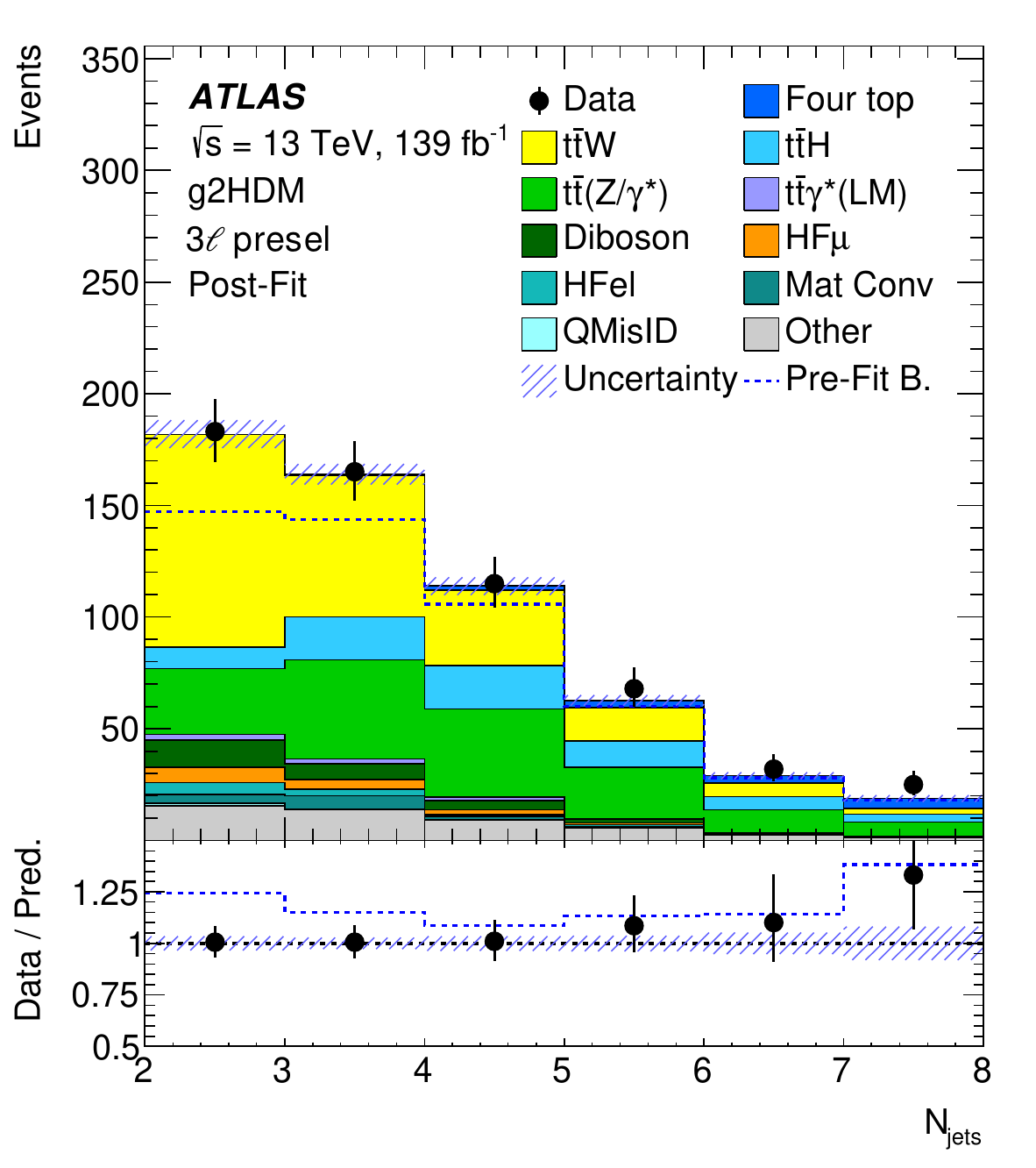}}
\subfloat[]{\includegraphics[width=0.33\textwidth]{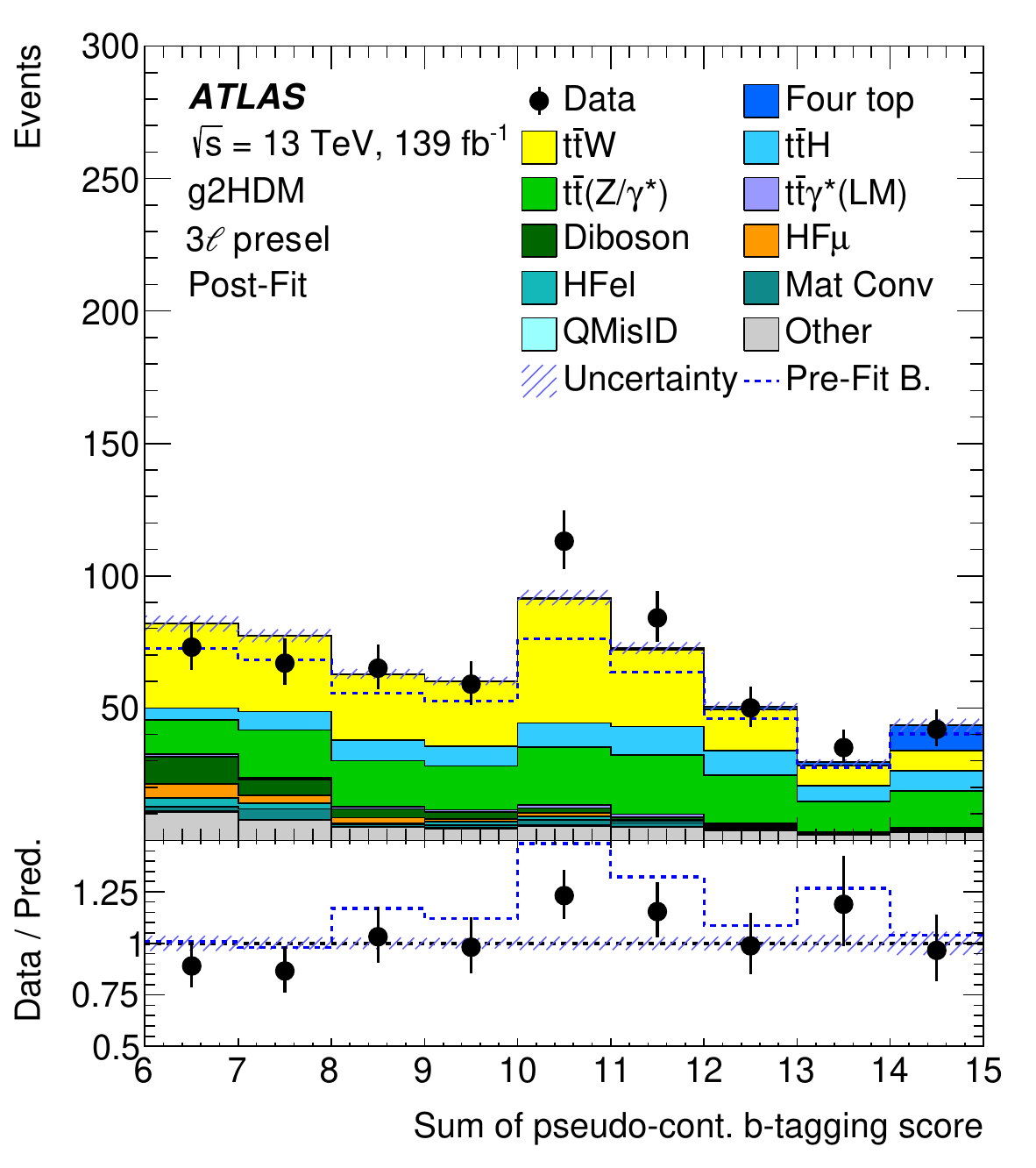}}
\subfloat[]{\includegraphics[width=0.33\textwidth]{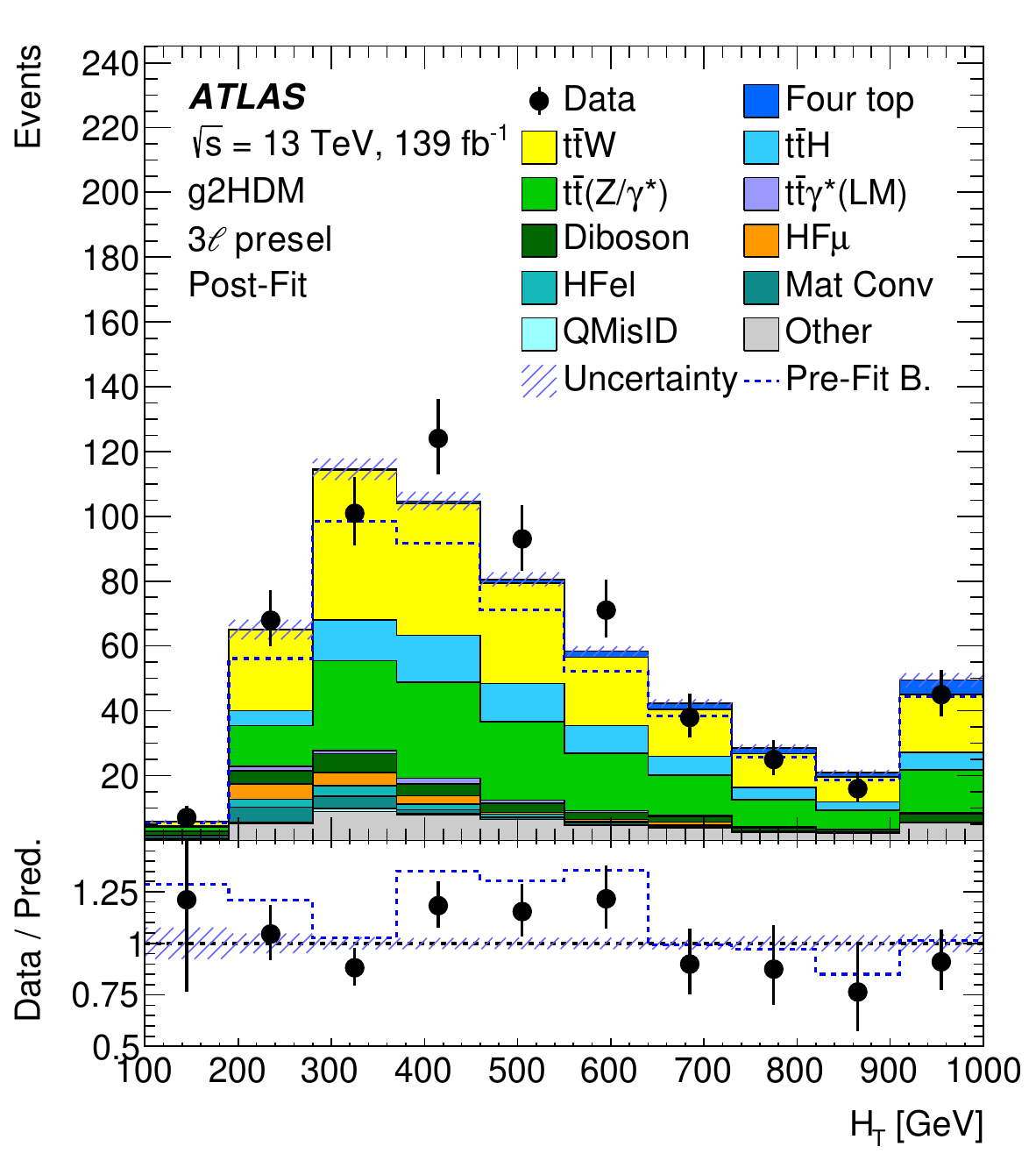}}
\caption{
Comparison between data and the background prediction for the distribution of the jet multiplicity (a, d), sum of pseudo-continuous $b$-tagging scores of jets (b, e) and the sum of \pt of the jets and leptons (c, f), in the \ll (a-c) and \lll (d-f) signal region selections before any categorisation based on \dnncat.
The expected signal for $m_{H}=900~\gev$ and couplings $\rhott$=\bestrhott, $\rhotc$=\bestrhotc, and $\rhotu$=\bestrhotu, along with the background contributions, is shown after the likelihood fit to data (``Post-Fit'') for the background-only hypothesis.
The ratio of the data to the prediction (``Pred.'') is shown in the lower panel, separately for post-fit signal-plus-background (black points) and pre-fit background (dashed blue line).
The size of the combined statistical and systematic uncertainty in the signal-plus-background prediction is indicated by the blue hatched band.
The last bin in each Figure contains the overflow.}
\label{fig:bkgvalidation}
\end{figure}

% End of text imported from the .//Tex/background.tex input file

\section{Systematic uncertainties}
\label{sec:systematics}

% The next lines are included from the .//Tex/systematics.tex input file
The signal and background yields in each signal and control region may be affected by several sources of
systematic uncertainty, described in the following subsections.
Given the low background yields and good signal-to-background separation provided by the final discriminating variable
used in the signal-rich event categories, the search sensitivity is determined by the limited number of data
events rather than by the systematic uncertainties on the background estimate.
The final uncertainty in the background estimate in the SRs is dominated by the uncertainty in the fitted
background normalisations, in particular \ttW.
A summary of all systematic uncertainties included in the analysis is given in Table~\ref{tab:systematics}.
 
\subsection{Experimental uncertainties}
The combined 2015--2018 integrated luminosity, obtained using the LUCID-2 detector~\cite{LUCID2} for the primary luminosity measurements, has an uncertainty of 1.7\% \cite{ATLAS-CONF-2019-021}.
 
Uncertainties associated with the lepton selection arise from the trigger, reconstruction, identification and
isolation efficiencies, and lepton momentum scale and resolution~\cite{MUON-2018-03,EGAM-2018-01,PERF-2017-01}.
Uncertainties associated with the jet selection arise from the jet energy scale (JES), the JVT requirement
and the jet energy resolution (JER)~\cite{JETM-2018-05,PERF-2014-03}.
 
The efficiency of the flavour-tagging algorithm is measured for each jet flavour using control samples in data
and in simulation. From these measurements, correction factors are derived to correct the tagging rates
in the simulation~\cite{FTAG-2020-08,FTAG-2018-01,ATLAS-CONF-2018-006}.
These systematic uncertainties are taken as uncorrelated between $b$-jets, $c$-jets, and
light-flavour jets.
An additional uncertainty is assigned to account for the extrapolation of the $b$-tagging
efficiency measurement from the \pt region used to determine the correction factors to regions with higher
transverse momentum~\cite{ATL-PHYS-PUB-2021-003}.
This uncertainty is the leading experimental uncertainty in the analysis, with relative variations up to 25\% on the signal.
 
The treatment of the uncertainties associated with reconstructed objects is common to all analysis channels
and applies to all signal and background samples, except for the background from electrons with misidentified charge,
and thus these are considered as fully correlated among different analysis regions and samples.
 
\subsection{Theoretical uncertainties}
The modelling uncertainties on the main irreducible backgrounds are assessed through comparisons with alternative MC samples, as listed in Table~\ref{tab:mcconfig}. Additional uncertainties are evaluated from renormalisation and factorisation scale variations by a factor of 0.5 and 2, relative to the nominal scales, for the \ttW, \ttZ, and diboson samples.  An additional 20\% uncertainty is assigned to the \ttW electroweak contribution~\cite{PhysRevD.103.094014}.
An additional 50\% uncertainty is assigned to \ttW, \ttZ, and \ttbar events with additional heavy-flavour jets, following Ref.~\cite{TOPQ-2018-05}. This normalisation uncertainty is not applied to diboson events with heavy-flavour since its normalisation is fit to data.
The statistical uncertainty on the fitted parameters for the $VV$ jet-multiplicity correction is propagated as an uncertainty on the diboson background.
The leading theoretical uncertainties arise from \ttW modelling and additional heavy-flavour uncertainties.
 
Finally, additional normalisation uncertainties are included for all processes whose normalisation is not obtained from the fit. The \tttt, \ttH, and $tZ$ processes are assigned an uncertainties of 20\%~\cite{Frederix:2017wme}, 11\%~\cite{deFlorian:2016spz}, and 5\%~\cite{TOPQ-2018-01}, respectively. As a conservative estimate, a 50\% cross section uncertainty is assigned to the \ttt, $tWZ$, \ttWW, and triboson backgrounds, which are small backgrounds with low impact on the search.
 
Uncertainties on the modelling of the signal samples are evaluated through independent variations of the factorisation and renormalisation scales by a factor of two. Additional uncertainties due to PDF effects are estimated through an ensemble of eigenvariations of the NNPDF set, and by taking the differences with respect to alternative PDF sets~\cite{Butterworth:2015oua}.
 
\subsection{Reducible background uncertainties}
The normalisation of HF non-prompt leptons is obtained from regions including at least one $M_{\mathrm{ex}}$ lepton and extrapolated to the signal regions where the same-sign leptons fulfil the $T$ or $M$ identification requirements. An uncertainty of 20\% on the extrapolation from $M_{\mathrm{ex}}$ to $T$ leptons is applied from the comparison of the relative efficiency between nominal and alternative \ttbar\ MC samples.
An additional 50\% uncertainty is assigned to events originating from \ttbar+$\geq 1 b$ and \ttbar+$\geq 1 c$, decorrelated between flavours.
Validation regions with looser lepton requirements and further enriched in non-prompt leptons are defined. A good agreement between data and background prediction is observed in all kinematic variables except for the number of $b$-jets. Based on this disagreement, an \nbjets-dependent uncertainty is added to the HF non-prompt background ranging from 6\%--40\% for 1--3 additional $b$-jets in the non-prompt muon regions, and 10\%--80\% in the non-prompt electron regions.
 
The modelling of internal and material conversions is tested in dedicated validation regions with two tight same-sign leptons, requiring one of them to be a conversion candidate. Additional uncertainties of 10\% and 50\% are assigned to the material and internal conversion backgrounds, respectively, evaluated from the data to background agreement in the validation regions.
 
A systematic uncertainty of 10\%--60\% is assigned to the background from electrons with misidentified charge. The uncertainty increases with electron \pt\ and decreases with $|\eta|$. The uncertainty is assessed combining the uncertainties from the measurement of the charge misassignment rate, the difference in rates from varying the $m_Z$ window selection, and the different rates measured in data and $Z\rightarrow e^+e^-$ MC.
 
\begin{table}
\centering
\caption{Sources of systematic uncertainty considered in the analysis.
``N'' means that the uncertainty is taken as normalisation-only for all processes and channels affected.
Some of the systematic uncertainties are split into several components, as indicated by the number in the rightmost column.
}
\resizebox{\textwidth}{!}{
\begin{tabular}[t]{lc}
\hline\hline
Systematic uncertainty                       & Components  	      \\
\hline
\hline
\textbf{Signal modelling}                      &                     \\
\ \ QCD scale                                   & 1		      \\
\ \ PDFs+$\alpha_\mathrm{S}$                    & 3           \\
\textbf{\ttW modelling}                        &		      \\
\ \ QCD scale                                   & 3		      \\
\ \ Generator                                	& 2		      \\
\ \ Electroweak cross section                   & 1		      \\
\ \ Additional heavy-flavour                    & 1		      \\
\textbf{$\ttdy$ modelling}         &		      \\
\ \ QCD scale                                   & 2		      \\
\ \ Generator                    	            & 2		      \\
\ \ Additional heavy-flavour                    & 1		      \\
\textbf{\ttH modelling}                        &		      \\
\ \ Cross section (N)                          	& 1		      \\
\ \ Parton shower and hadronisation model       & 1		      \\
\ \ Generator                                   & 1		      \\
\ \ QCD scale                                   & 1		      \\
\ \ Additional heavy-flavour                    & 1		      \\
\textbf{$WZ$ modelling}    					&		      \\
\ \ QCD scale                                   & 1		      \\
\ \ Cross section (N)                           & 1		      \\
\ \ Extra-jets correction        			    & 1		      \\
\textbf{$\tttt$ modelling}    					&		      \\
\ \ Generator                                   & 1		      \\
\ \ Cross section (N)                           & 1		      \\
\textbf{Other background modelling}            &		      \\
\ \ Cross section (N)                           & 6		      \\
\hline
Total (Signal and background modelling)         & 32~~		     \\
\hline\hline
\end{tabular}
\begin{tabular}[t]{lc}
\hline\hline
Systematic uncertainty                       & Components  	      \\
\hline
\hline
Luminosity                                   & 1		      \\
Pile-up reweighting                          & 1		      \\
\textbf{Physics objects}                    &		      \\
\ \ Electron                                 & 6		      \\
\ \ Muon                                     & 15~~	      \\
\ \ Electron Non-prompt BDT                  & 14~~		      \\
\ \ Muon Non-prompt BDT                      & 20~~	      \\
\ \ Jet energy scale                         & 30~~                  \\
\ \ Jet energy resolution                    & 12~~                  \\
\ \ Jet vertex fraction                      & 1		      \\
\ \ Jet flavour tagging                      & 62~~	      \\
\ \ \met                                     & 3		      \\
\hline
Total (Experimental)                         & 165~~~~		     \\
\hline
\hline
\textbf{Data-driven reducible background estimates}   &		     \\
\ \ Material conversions modelling             & 1		     \\
\ \ Internal conversions modelling             & 1		      \\
\ \ Charge misassignment                       & 1		      \\
\ \ HF non-prompt                         	   & 8		      \\
\ \ \ttbar additional heavy-flavour            & 2		      \\
\hline
Total (Data-driven reducible background)       & 13~~		   \\
\hline\hline
Total (Overall)                                & 210~~~~	      \\
\hline
\hline
\end{tabular}
}
\label{tab:systematics}
\end{table}

% End of text imported from the .//Tex/systematics.tex input file

\section{Results}
\label{sec:results}

% The next lines are included from the .//Tex/results.tex input file
A maximum-likelihood fit is performed on all bins in the 27 signal and control regions considered in this search to simultaneously determine the background and the signal yields that are most consistent with the data. The \dnnsb\ is used as the discriminating variable in the signal regions, whereas the $\nbjets$, fake-lepton-candidate $\pt$ and event yields are fit in the control regions. The sum of all the g2HDM signal processes studied here (sstt, ttq, ttt, tttq, tttt) is considered as a single signal template and its acceptance in each category is predicted by the simulation. The same procedure is followed for the SUSY signals.

The likelihood function ${\cal L}(\mu,\vec{\lambda}, \vec{\theta})$ is constructed as a product of Poisson probability terms over all bins considered in the search, and depends on the signal-strength parameter, $\mu$, a multiplicative factor applied to the predicted yield for the g2HDM signal (depending on the coupling configuration {$\rhott$, $\rhotc$, $\rhotu$} and on the assumed mass $m_H$), $\vec{\lambda}$, the normalisation factors for several backgrounds (see Section~\ref{sec:background}), and $\vec{\theta}$, a set of nuisance parameters (NP) encoding systematic uncertainties in the signal and background expectations~\cite{Cranmer:2012sba}. Systematic uncertainties can impact the estimated signal and background rates, the migration of events between categories, and the shape of the fitted distributions; they are summarised in Table~\ref{tab:systematics}.
Both $\mu$ and $\vec{\lambda}$ are treated as free parameters in the likelihood fit. The NPs $\vec{\theta}$ allow variations of the expectations for signal and background according to the systematic uncertainties, subject to Gaussian or Poisson constraints in the likelihood fit.
Their fitted values represent the deviations from the nominal expectations that globally provide the best fit to the data.
Statistical uncertainties in each bin due to the limited size of the simulated samples are taken into account by dedicated parameters using the Beeston--Barlow ``lite'' technique~\cite{Conway:2011}.
 
The test statistic $q_\mu$ is defined as the profile likelihood ratio:
$q_\mu = -2\ln({\cal L}(\mu,{\hat{\vec{\lambda}}}_\mu, {\hat{\vec{\theta}}}_\mu)/{\cal L}(\hat{\mu},\hat{\vec{\lambda}}_{\hat{\mu}}, \hat{\vec{\theta}}_{\hat{\mu}}))$,
where $\hat{\mu}$, $\hat{\vec{\lambda}}_{\hat{\mu}}$, and $\hat{\vec{\theta}}_{\hat{\mu}}$ are the values of the parameters that maximise the likelihood function,
and ${\hat{\vec{\lambda}}}_\mu$ and ${\hat{\vec{\theta}}}_\mu$ are the values of the parameters that maximise the likelihood function for a given value of $\mu$.
The test statistic $q_\mu$ is evaluated with the {\textsc RooFit} package~\cite{Verkerke:2003ir}.
A related statistic is used to determine the probability that the observed data are incompatible with the background-only hypothesis (i.e.~the discovery test) by setting $\mu=0$ in the profile likelihood ratio ($q_0$).
The $p$-value (referred to as $p_0$) representing the probability of the data being compatible with the background-only hypothesis is estimated by integrating the distribution of $q_0$ from background-only pseudo-experiments, approximated using the asymptotic formulae given in Ref.~\cite{Cowan:2010js}, above the observed value of $q_0$. Some model dependence exists in the estimation of the $p_0$, as a given signal scenario must be assumed in the calculation of the denominator of $q_0$, even if the overall signal normalisation is allowed to float and is fit to data. The observed $p_0$ is checked for each explored signal scenario.
Upper limits on the signal production cross section for each of the signal scenarios considered are derived by using $q_\mu$ in the CL$_{\textrm{s}}$ method~\cite{Junk:1999kv,Read:2002hq}.
For a given signal scenario, values of the production cross section (parameterised by $\mu$) yielding CL$_{\textrm{s}} < 0.05$, where CL$_{\textrm{s}}$ is computed using the asymptotic approximation~\cite{Cowan:2010js}, are excluded at $\geq 95\%$ confidence level (CL).
 
The smallest $p_0$ value is observed when assuming a signal with $m_{H}=900~\gev$ and couplings $\rhott$=\bestrhott, $\rhotc$=\bestrhotc, and $\rhotu$=\bestrhotu, corresponding to a local significance of \bestsignificance\ standard deviations.
The fitted signal strength is $\mu=0.07\pm0.03$, pointing to an incompatibility of the model prediction with the size of the excess or else the need for additional undetected decay modes taking up 93\% of the branching ratio.
The signal cross section resulting from the fit to data for this g2HDM signal hypothesis is 154 fb, with fractional contributions of 55\% ttq, 31\% sstt, and 14\% ttt.
The signal with the fitted signal strength closest to unity ($\mu=0.9\pm0.4$) corresponds to $m_{H}=900~\gev$ and couplings $\rhott$=0.2, $\rhotc$=0.4, and $\rhotu$=0.4 and a local significance of 2.4 $\sigma$.
Figure~\ref{fig:triangle_significance_900} shows the local significance as a function of the three couplings normalised to the sum of the couplings. This normalisation eliminates one degree of freedom related to the total normalisation of the signal, which is not relevant for the computation of the significance.
However, a residual dependency on the actual value of the coupling remains as the normalization of the sstt process scales as the fourth power of the couplings, while the rest of the processes scale as a function of the couplings squared.
 
\begin{figure}[tp]
\centering
\includegraphics[width=0.6\textwidth]{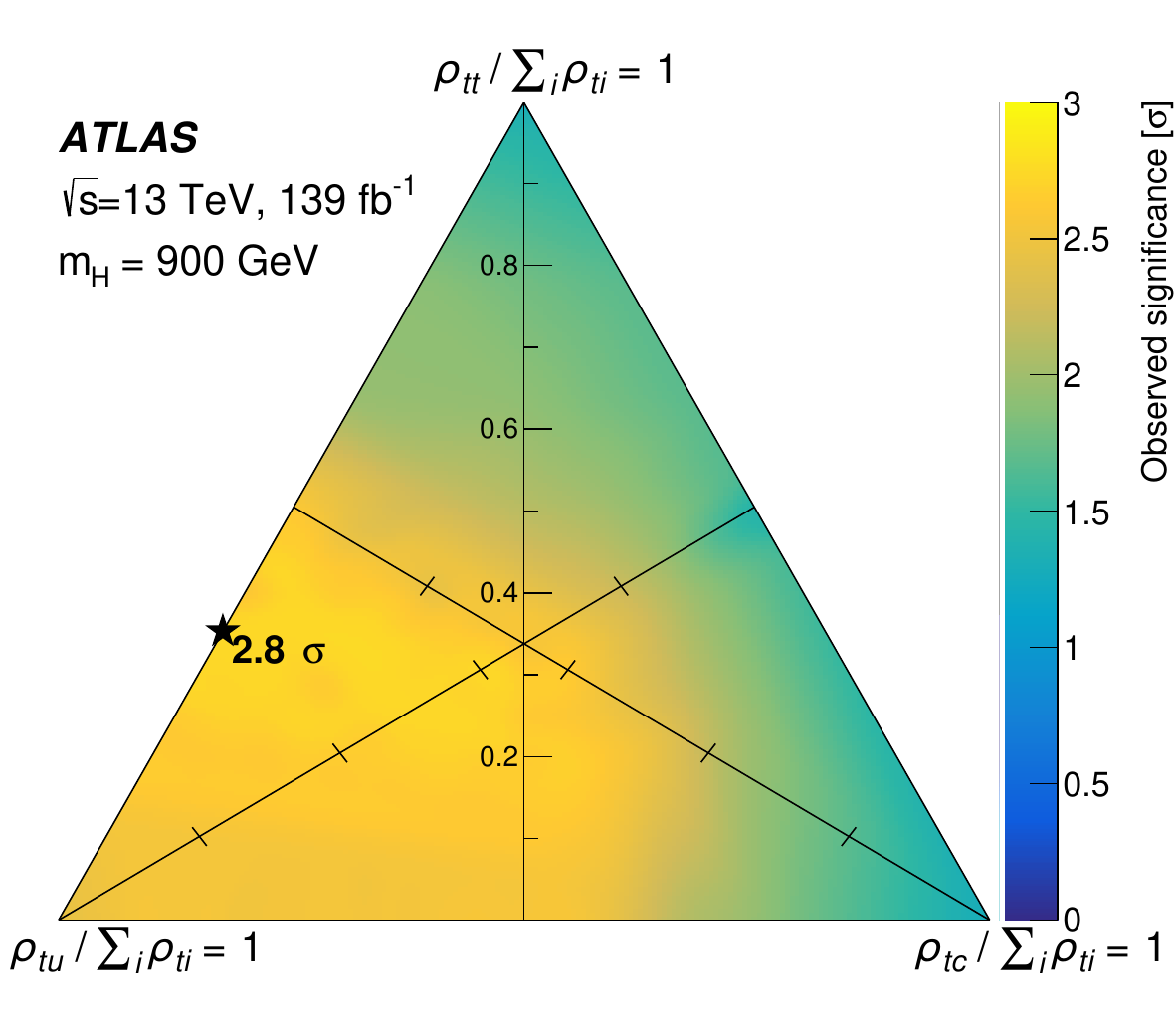} 
\caption{
Observed significance for a heavy scalar with a mass of 900 \gev\ as a function of the three couplings normalised to the sum of the couplings. This normalisation eliminates one degree of freedom related to the total normalisation of the signal, which is not relevant for the computation of the significance. A residual dependency on the actual value of the coupling remains as the normalization of the sstt process scales as the fourth power of the couplings, while the rest of the processes scale as a function of the couplings squared. The star indicates the coupling configuration leading to the highest observed significance of 2.8 standard deviations.
}
\label{fig:triangle_significance_900}
\end{figure}
 
\begin{figure*}[htp]
\centering
\subfloat[]{\includegraphics[width=0.73\textwidth]{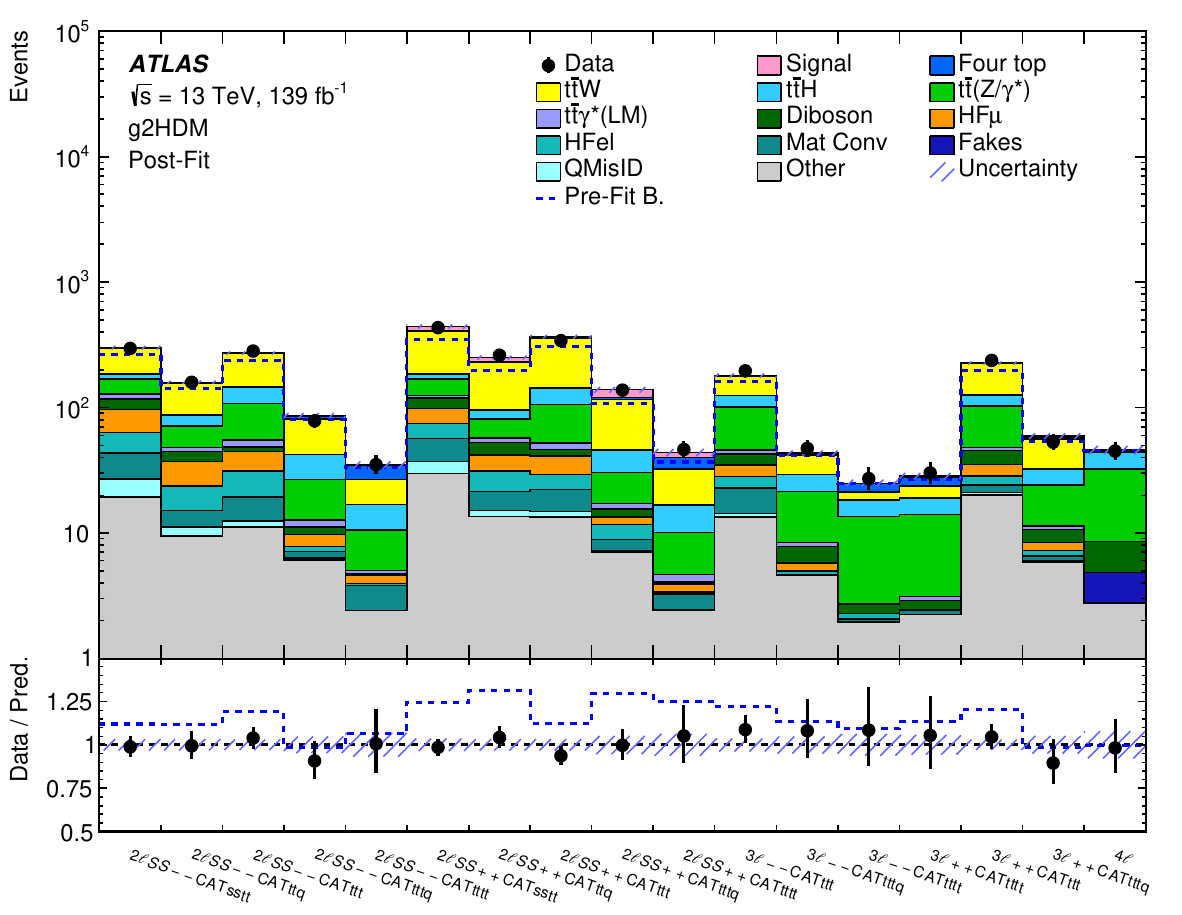}\label{fig:summarySR}} \\
\subfloat[]{\includegraphics[width=0.73\textwidth]{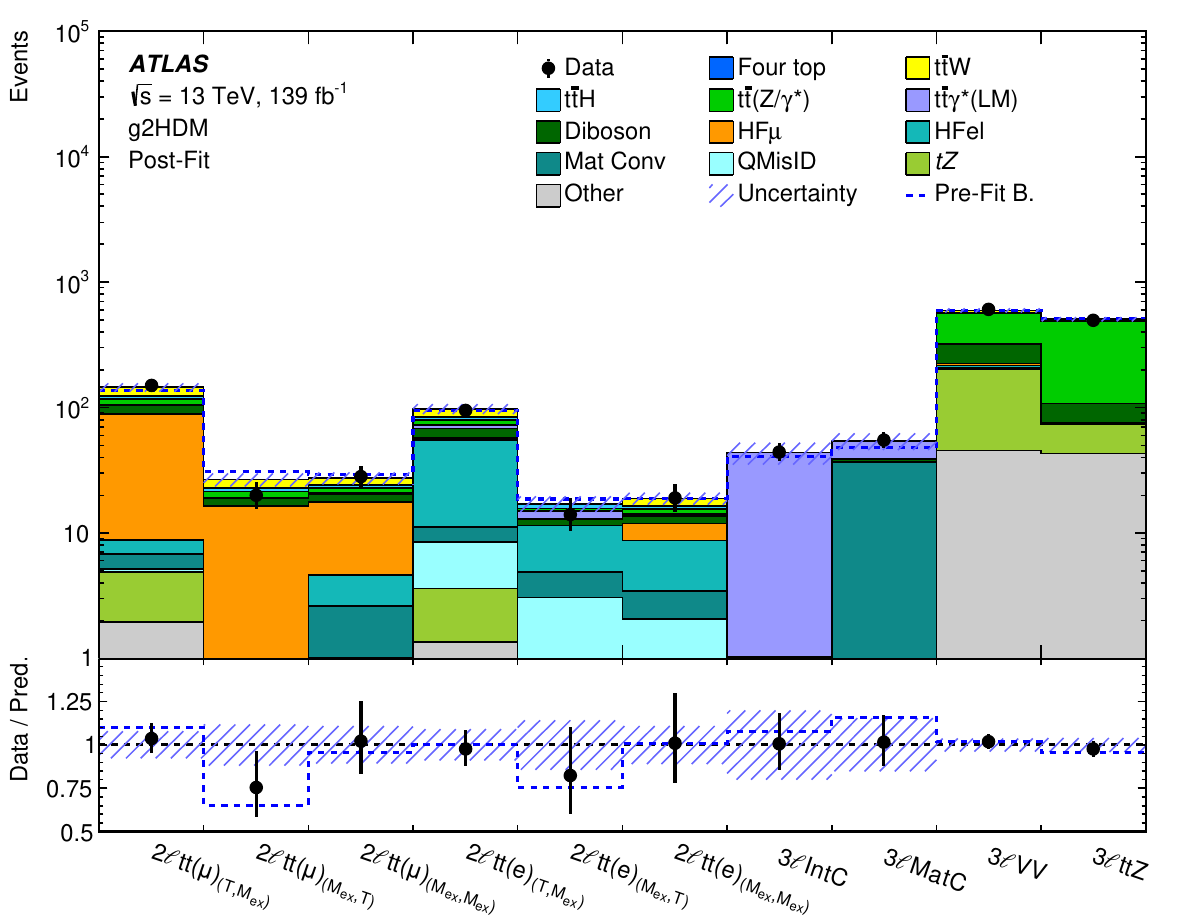}\label{fig:summaryCR}}
\caption{Comparison between data and the background prediction for the event yields in (a) the 17 signal region categories and (b) the 10 control region categories.
The expected signal for $m_{H}=900~\gev$ and couplings $\rhott$=\bestrhott, $\rhotc$=\bestrhotc, and $\rhotu$=\bestrhotu, along with the background contributions, is shown after the likelihood fit to data (``Post-Fit'') for the signal-plus-background hypothesis.
The total background prediction before the likelihood fit to data (``Pre-Fit'') is shown as a dashed blue histogram in the upper panel.
The ratio of the data to the total prediction is shown in the lower panel, separately for post-fit signal-plus-background (black points) and pre-fit background (dashed blue line).
The size of the combined statistical and systematic uncertainty in the background prediction is indicated by the blue hatched band.
}
\end{figure*}
 
A comparison of the distributions of observed and expected yields is shown Figure~\ref{fig:summarySR} for the 17 SRs, and Figure~\ref{fig:summaryCR} for the 10 CRs, after the combined likelihood fit for the signal-plus-background hypothesis.
The corresponding post-fit yields for the SRs can be found in Tables~\ref{tab:postfityields2lpp},~\ref{tab:postfityields2lmm}, and~\ref{tab:postfityields3l4l} for the \llSS positively charged, \llSS negatively charged, and \lll and \llll SRs, respectively.
The signal shown in the figures and tables is the g2HDM signal with couplings $\rhott$=\bestrhott, $\rhotc$=\bestrhotc, and $\rhotu$=\bestrhotu, and mass of $900~\GeV$, which corresponds to the largest observed significance above the background only hypothesis.
Good agreement between the data and fitted signal-plus-background yields is found across all event categories.
 
The systematic uncertainties with the largest impact on the signal strength originate from the modelling of $\ttW$ with and without additional heavy flavour jets, $\ttZ$, $\ttH$, and $\tttt$ processes.
The signal strength is partially anti-correlated with $\lambda_{\ttW}$, with a linear correlation value of $-35\%$.
The search is dominated by statistical uncertainties.
 
Comparisons between data and the background prediction for the \dnnsb distributions used in the different SRs are shown in Figures~\ref{fig:dnnspectrum1} and~\ref{fig:dnnspectrum2}. The binning used for the \dnnsb distributions in the different SRs represents a compromise between preserving enough discrimination in the fit between the background and the signal for the different values of the heavy $H$ mass considered and keeping the MC statistical uncertainty of the background prediction per bin well below 30\%.
The signal regions with the largest pre-fit tension between data and the background yields (shown in the blue dashed line) at high values of the \dnnsb are the \twopptttq, the \twopptttt, the \twopptt, and the \twoppttq\ regions.
Within this model, the $\hat{\lambda}_{\ttW}$ remains higher than 1, as observed by other analyses~\cite{TOPQ-2016-11,CMS-TOP-21-011,EXOT-2019-26,CMS-TOP-18-003,CMS-HIG-19-008}.
Since the largest discrepancies between data and the background expectation before the fit are observed in signal categories with positive total lepton charge, this tension cannot be explained by the lepton-charge-symmetric SM \tttt production. The goodness-of-fit based on the saturated model~\cite{SaturatedModel} for the best fit g2HDM signal plus background is 62\%, which shows a better fit to data than the background-only hypothesis with a goodness-of-fit of 45\%.
 
Exclusion limits on the heavy Higgs boson mass are set for different choices of the couplings, as shown in Figure~\ref{fig:mass_limit}.
Masses of an additional scalar boson $m_{H}$ between $200$-$620$ ($200$-$840$) $\GeV$ with couplings $\rho_{tt}=0.4$, $\rho_{tc}=0.2$, and $\rho_{tu}=0.2$ are observed (expected) to be excluded at 95\% confidence level.
Limits on the heavy Higgs boson mass are also set for a scenario without coupling to two top quarks, $\rhott=0$, $\rhotc=0.2$, $\rhotu=0.2$, resulting on an observed (expected) limit of 200--320 (200--560) \gev\ on the heavy Higgs boson mass. No limits can be set on scenarios without off-diagonal couplings, leading only to four-top final states with a coupling set $\rhott=1$, $\rhotc=0$, $\rhotu=0$. The sensitivity of the analysis on the four-top final state is similar to previous ATLAS analyses~\cite{ATLAS-CONF-2022-008}.
The excluded mass is also presented as a function of two couplings for different assumptions, as shown in Figure~\ref{fig:coupling_limit}.
 
The search is also used to set limits on RPV SUSY models using the existing DNNs that were trained for the g2HDM model. Figures~\ref{fig:limit_higgsino} and~\ref{fig:limit_wino} show the exclusion limits obtained on the higgsino and wino models, respectively. Higgsinos (winos) with masses up to 585 (670) \gev\ are excluded. Figure~\ref{fig:SUSY_limits2D} shows limits on the smuon-bino model. Smuon masses up to 460 \gev\ are excluded, with weaker exclusion limits for small mass splittings between the smuon and the lightest SUSY particle (LSP), or for LSP masses close to the top-quark threshold.

\begin{table}[htpb]
\begin{center}
\caption{Post-fit yields of the \llSS positively charged signal regions. The best-fit signal for $m_{H}=900~\gev$ and couplings $\rhott$=\bestrhott, $\rhotc$=\bestrhotc, and $\rhotu$=\bestrhotu, is shown for a signal strength $\mu = 0.07 \pm 0.03$. }
\label{tab:postfityields2lpp}
\scriptsize
\begin{tabular}{lr@{}lr@{}lr@{}lr@{}lr@{}l}
\hline
& \multicolumn{2}{c}{\twopptt} & \multicolumn{2}{c}{\twoppttq} & \multicolumn{2}{c}{\twoppttt} & \multicolumn{2}{c}{\twopptttq} & \multicolumn{2}{c}{\twopptttt} \\
\hline
Signal   & 35 $\pm$& 13 & 19 $\pm$& 7 & 5.8 $\pm$& 2.2 & 18 $\pm$& 7 & 3.6 $\pm$& 1.4 \\
$\tttt$   & 0.19 $\pm$& 0.05 & 0.45 $\pm$& 0.09 & 2.3 $\pm$& 0.5 & 4.7 $\pm$& 0.9 & 7.9 $\pm$& 1.4 \\
$t\bar{t}W$   & 220 $\pm$& 20 & 135 $\pm$& 10 & 214 $\pm$& 16 & 70 $\pm$& 7 & 15.7 $\pm$& 2.3 \\
$t\bar{t}H$   & 17.2 $\pm$& 2.6 & 15.4 $\pm$& 2.6 & 37 $\pm$& 5 & 15.2 $\pm$& 3.2 & 6.3 $\pm$& 1.5 \\
$t\bar{t}Z/\gamma*$   & 43 $\pm$& 5 & 23.6 $\pm$& 2.8 & 53 $\pm$& 5 & 13.4 $\pm$& 1.4 & 5.5 $\pm$& 0.8 \\
$t\bar{t}\gamma*$ (LM)  & 5.6 $\pm$& 3.1 & 4.5 $\pm$& 2.7 & 5.5 $\pm$& 3.0 & 1.5 $\pm$& 1.0 & 0.6 $\pm$& 0.5 \\
$VV$   & 22 $\pm$& 7 & 11 $\pm$& 4 & 5.2 $\pm$& 2.0 & 2.2 $\pm$& 0.9 & 0.16 $\pm$& 0.08 \\
$tZ$   & 23.7 $\pm$& 1.4 & 9.3 $\pm$& 0.6 & 5.3 $\pm$& 0.4 & 0.72 $\pm$& 0.06 & 0.007 $\pm$& 0.006 \\
Non-prompt $\ell$    & 42 $\pm$& 15 & 20 $\pm$& 7 & 19 $\pm$& 6 & 4.7 $\pm$& 2.7 & 0.6 $\pm$& 0.6 \\
Mat Conv   & 19 $\pm$& 5 & 6.1 $\pm$& 1.6 & 7.3 $\pm$& 2.2 & 1.7 $\pm$& 0.7 & 0.86 $\pm$& 0.24 \\
QMisID   & 7.4 $\pm$& 2.7 & 1.7 $\pm$& 0.6 & 1.4 $\pm$& 0.5 & 0.20 $\pm$& 0.08 & 0.021 $\pm$& 0.009 \\
Other   & 5.9 $\pm$& 2.0 & 4.1 $\pm$& 1.4 & 8.1 $\pm$& 2.2 & 6.2 $\pm$& 1.9 & 2.4 $\pm$& 0.6 \\
\hline
Total  & 441 $\pm$& 16 & 250 $\pm$& 8 & 365 $\pm$& 12 & 138 $\pm$& 7 & 43.6 $\pm$& 2.8 \\
\hline
Data   & \multicolumn{1}{r}{434~~} & & \multicolumn{1}{r}{261~~} & & \multicolumn{1}{r}{342~~} & & \multicolumn{1}{r}{138~~} & & \multicolumn{1}{r}{46~~} & \\
\hline
\end{tabular}
\end{center}
\end{table}
 
\begin{table}[htpb]
\begin{center}
\caption{Post-fit yields of the \ll negatively charged signal regions. The best-fit signal for $m_{H}=900~\gev$ and couplings $\rhott$=\bestrhott, $\rhotc$=\bestrhotc, and $\rhotu$=\bestrhotu, is shown for a signal strength $\mu = 0.07 \pm 0.03$. }
\label{tab:postfityields2lmm}
\scriptsize
\begin{tabular}{lr@{}lr@{}lr@{}lr@{}lr@{}l}
\hline
& \multicolumn{2}{c}{\twonntt} & \multicolumn{2}{c}{\twonnttq} & \multicolumn{2}{c}{\twonnttt} & \multicolumn{2}{c}{\twonntttq} & \multicolumn{2}{c}{\twonntttt} \\
\hline
Signal   & 3.2 $\pm$& 2.1 & 1.4 $\pm$& 0.7 & 0.48 $\pm$& 0.26 & 0.9 $\pm$& 0.4 & 0.20 $\pm$& 0.09 \\
$\tttt$   & 0.20 $\pm$& 0.05 & 0.49 $\pm$& 0.11 & 2.4 $\pm$& 0.6 & 4.7 $\pm$& 0.9 & 7.9 $\pm$& 1.4 \\
$t\bar{t}W$   & 110 $\pm$& 9 & 70 $\pm$& 5 & 124 $\pm$& 9 & 38 $\pm$& 4 & 9.8 $\pm$& 1.5 \\
$t\bar{t}H$   & 17.1 $\pm$& 2.5 & 15.3 $\pm$& 2.2 & 37 $\pm$& 6 & 15.5 $\pm$& 3.2 & 6.2 $\pm$& 1.4 \\
$t\bar{t}Z/\gamma*$   & 42 $\pm$& 5 & 23.4 $\pm$& 2.6 & 53 $\pm$& 5 & 13.7 $\pm$& 1.5 & 5.5 $\pm$& 0.8 \\
$t\bar{t}\gamma*$ (LM)   & 10 $\pm$& 5 & 3.7 $\pm$& 2.1 & 7 $\pm$& 4 & 1.6 $\pm$& 1.0 & 0.31 $\pm$& 0.26 \\
$VV$   & 21 $\pm$& 6 & 7.5 $\pm$& 2.6 & 3.7 $\pm$& 1.5 & 1.4 $\pm$& 0.5 & 0.11 $\pm$& 0.05 \\
$tZ$   & 13.0 $\pm$& 0.8 & 5.53 $\pm$& 0.33 & 3.25 $\pm$& 0.28 & 0.274 $\pm$& 0.033 & 0.045 $\pm$& 0.017 \\
Non-prompt $\ell$   & 54 $\pm$& 16 & 22 $\pm$& 8 & 25 $\pm$& 9 & 2.7 $\pm$& 0.9 & 0.9 $\pm$& 0.6 \\
Mat Conv   & 16 $\pm$& 5 & 4.0 $\pm$& 1.1 & 6.9 $\pm$& 1.9 & 0.8 $\pm$& 0.4 & 1.3 $\pm$& 0.6 \\
QMisID   & 7.4 $\pm$& 2.7 & 1.7 $\pm$& 0.6 & 1.4 $\pm$& 0.5 & 0.19 $\pm$& 0.08 & 0.021 $\pm$& 0.009 \\
Other   & 6.5 $\pm$& 2.1 & 3.9 $\pm$& 1.2 & 7.8 $\pm$& 2.3 & 5.8 $\pm$& 1.8 & 2.3 $\pm$& 0.6 \\
\hline
Total  & 300 $\pm$& 10 & 159 $\pm$& 5 & 271 $\pm$& 8 & 86 $\pm$& 4 & 34.6 $\pm$& 2.4 \\
\hline
Data   & \multicolumn{1}{r}{296~~} & & \multicolumn{1}{r}{158~~} & & \multicolumn{1}{r}{282~~} & & \multicolumn{1}{r}{78~~} & & \multicolumn{1}{r}{35~~} & \\
\hline
\end{tabular}
\end{center}
\end{table}

\begin{table}[htpb]
\begin{center}
\caption{Post-fit yields of the \lll and \llll signal regions. The best-fit signal for $m_{H}=900~\gev$ and couplings $\rhott$=\bestrhott, $\rhotc$=\bestrhotc, and $\rhotu$=\bestrhotu, is shown for a signal strength $\mu = 0.07 \pm 0.03$. }
\label{tab:postfityields3l4l}
\scriptsize
\begin{tabular}{lr@{}lr@{}lr@{}lr@{}lr@{}lr@{}lr@{}l}
\hline
& \multicolumn{2}{c}{\threepptttt} & \multicolumn{2}{c}{\threeppttt} & \multicolumn{2}{c}{\threepptttq} & \multicolumn{2}{c}{\threenntttt} & \multicolumn{2}{c}{\threennttt} & \multicolumn{2}{c}{\threenntttq} & \multicolumn{2}{c}{\fourlep} \\
\hline
Signal   & 0.9 $\pm$& 0.4 & 1.5 $\pm$& 0.6 & 2.2 $\pm$& 0.9 & 0.049 $\pm$& 0.022 & 0.08 $\pm$& 0.04 & 0.10 $\pm$& 0.04 & 0.034 $\pm$& 0.013 \\
$\tttt$   & 3.8 $\pm$& 0.7 & 0.80 $\pm$& 0.17 & 1.74 $\pm$& 0.32 & 3.8 $\pm$& 0.7 & 0.87 $\pm$& 0.21 & 1.71 $\pm$& 0.32 & 1.4 $\pm$& 0.9 \\
$t\bar{t}W$   & 4.9 $\pm$& 0.6 & 97 $\pm$& 8 & 23.1 $\pm$& 3.0 & 2.8 $\pm$& 0.4 & 54 $\pm$& 4 & 12.4 $\pm$& 1.5 & 0.50 $\pm$& 0.11 \\
$t\bar{t}H$   & 4.9 $\pm$& 1.0 & 23.0 $\pm$& 3.1 & 8.0 $\pm$& 1.6 & 4.8 $\pm$& 1.0 & 23.1 $\pm$& 3.2 & 7.9 $\pm$& 1.5 & 11.4 $\pm$& 2.8 \\
$t\bar{t}Z/\gamma*$   & 10.9 $\pm$& 1.1 & 55 $\pm$& 6 & 12.9 $\pm$& 1.9 & 10.8 $\pm$& 1.2 & 55 $\pm$& 6 & 13.0 $\pm$& 1.7 & 24.2 $\pm$& 2.9 \\
$t\bar{t}\gamma*$ (LM)   & 0.23 $\pm$& 0.26 & 2.8 $\pm$& 1.6 & 0.7 $\pm$& 0.4 & 0.0 $\pm$& 0.0 & 3.5 $\pm$& 2.0 & 0.5 $\pm$& 0.4 & 0 (0) \\
$VV$   & 0.46 $\pm$& 0.20 & 10 $\pm$& 4 & 2.3 $\pm$& 0.9 & 0.43 $\pm$& 0.18 & 7.6 $\pm$& 3.0 & 2.1 $\pm$& 0.6 & 3.6 $\pm$& 1.3 \\
$tZ$   & 0.268 $\pm$& 0.028 & 12.5 $\pm$& 0.8 & 2.53 $\pm$& 0.17 & 0.106 $\pm$& 0.013 & 7.2 $\pm$& 0.4 & 1.19 $\pm$& 0.08 & 0.0$\pm$& 0.0 \\
Non-prompt $\ell$    & 0.0 $\pm$& 0.0 & 11 $\pm$& 4 & 1.9 $\pm$& 0.8 & 0.23 $\pm$& 0.16 & 12 $\pm$& 6 & 1.2 $\pm$& 0.6 & 2.09 $\pm$& 0.35 \\
Mat Conv   & 0.21 $\pm$& 0.07 & 3.3 $\pm$& 1.1 & 0.49 $\pm$& 0.26 & 0.10 $\pm$& 0.06 & 9 $\pm$& 6 & 0.037 $\pm$& 0.010 & 0 (0) \\
Other   & 2.0 $\pm$& 0.6 & 7.6 $\pm$& 2.3 & 3.3 $\pm$& 1.0 & 1.8 $\pm$& 0.5 & 6.1 $\pm$& 2.1 & 3.4 $\pm$& 1.0 & 2.7 $\pm$& 0.9 \\
\hline
Total  & 28.5 $\pm$& 1.6 & 226 $\pm$& 7 & 59.2 $\pm$& 3.1 & 24.9 $\pm$& 1.6 & 179 $\pm$& 7 & 43.5 $\pm$& 2.2 & 46 $\pm$& 4 \\
\hline
Data   & \multicolumn{1}{r}{30~~}  & & \multicolumn{1}{r}{236~~} & & \multicolumn{1}{r}{53~~} & & \multicolumn{1}{r}{27~~} & & \multicolumn{1}{r}{195~~} & & \multicolumn{1}{r}{47~~} & & \multicolumn{1}{r}{45~~} &  \\
\hline
\end{tabular}
\end{center}
\end{table}
 
\clearpage

\begin{figure*}[tp]
\centering
\subfloat[]{\includegraphics[width=0.249\textwidth,trim={0 0 0.3cm 0},clip]{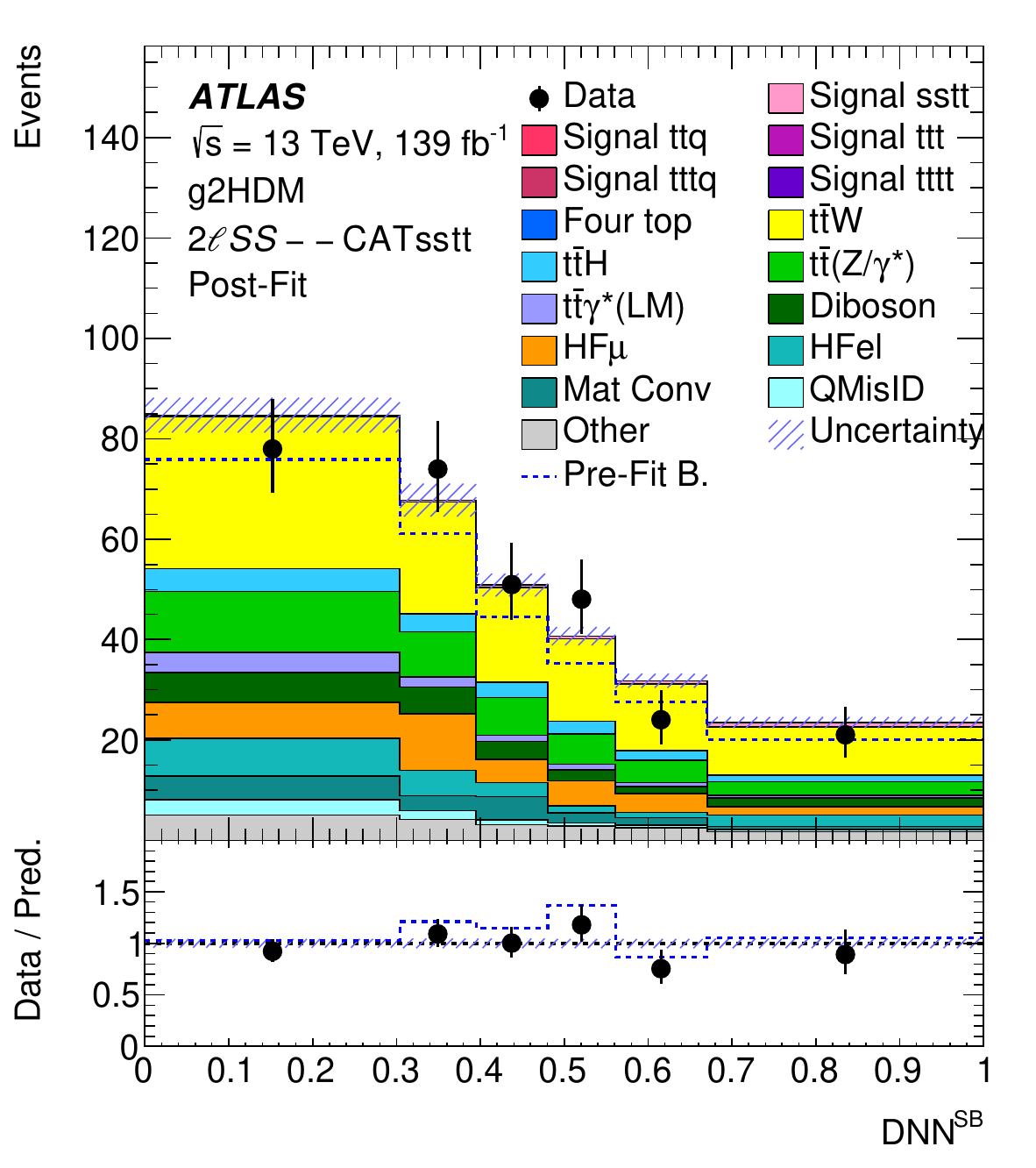}}
\subfloat[]{\includegraphics[width=0.249\textwidth,trim={0 0 0.3cm 0},clip]{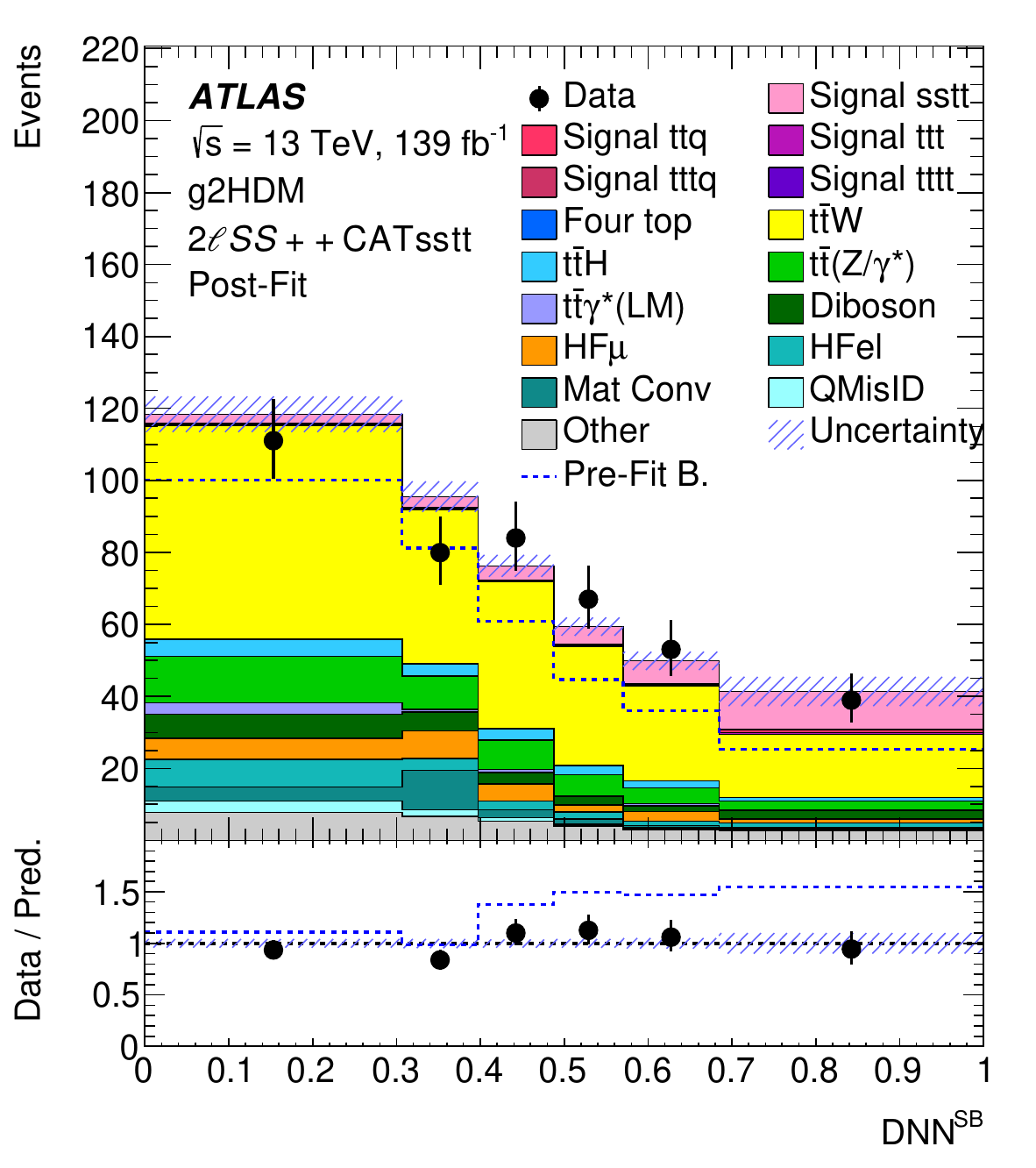}}
\subfloat[]{\includegraphics[width=0.249\textwidth,trim={0 0 0.3cm 0},clip]{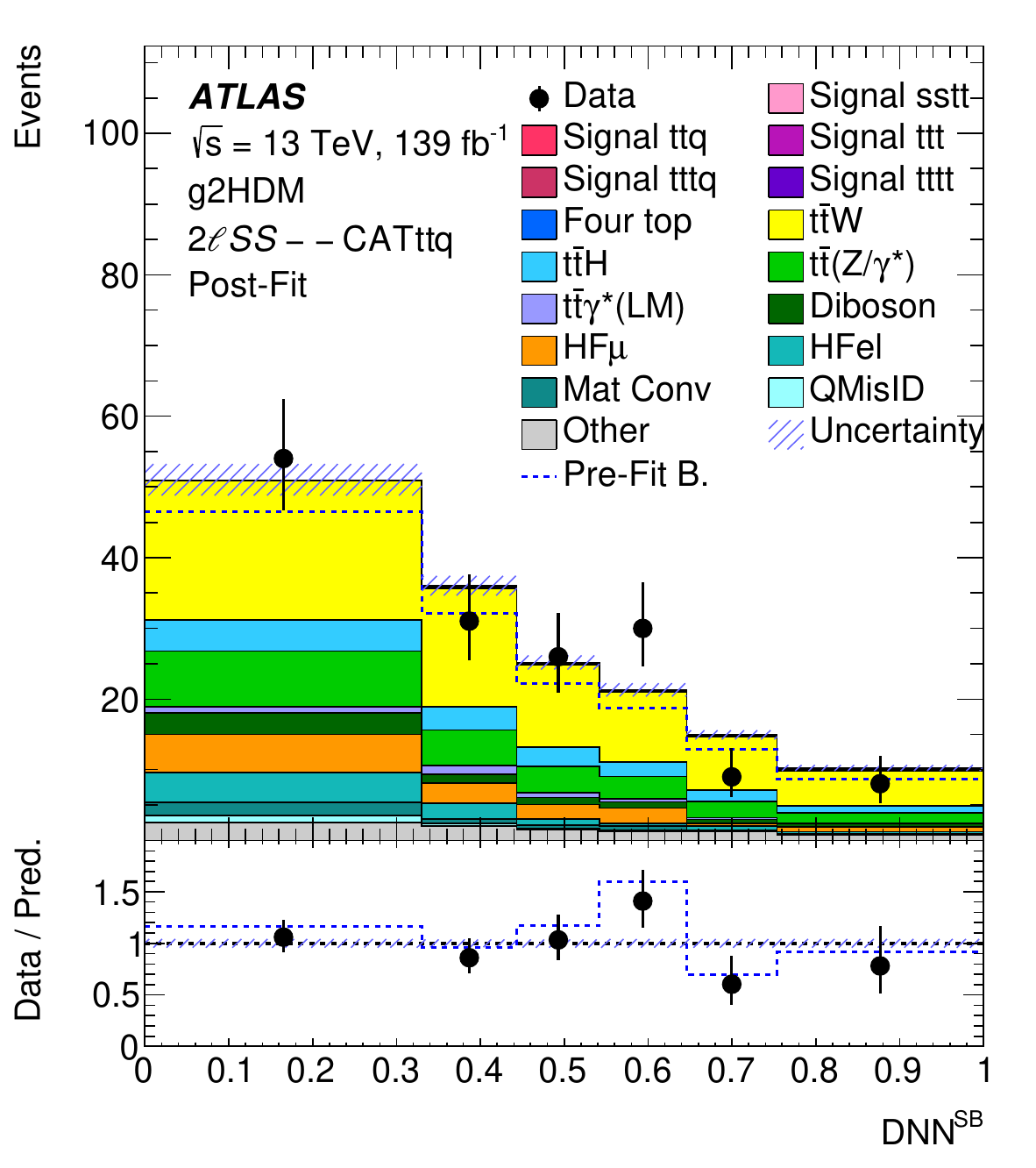}}
\subfloat[]{\includegraphics[width=0.249\textwidth,trim={0 0 0.3cm 0},clip]{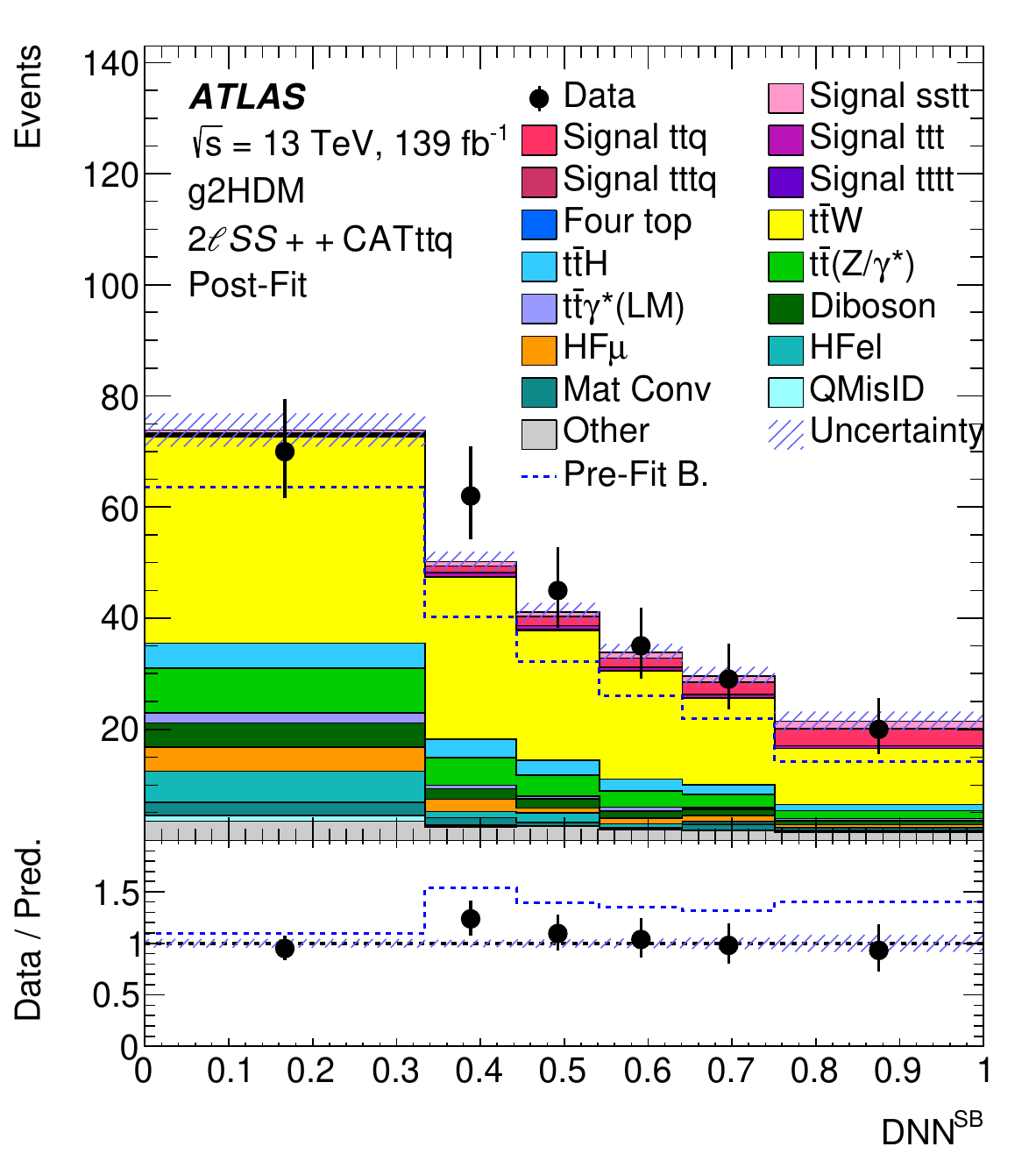}}\\
\subfloat[]{\includegraphics[width=0.249\textwidth,trim={0 0 0.3cm 0},clip]{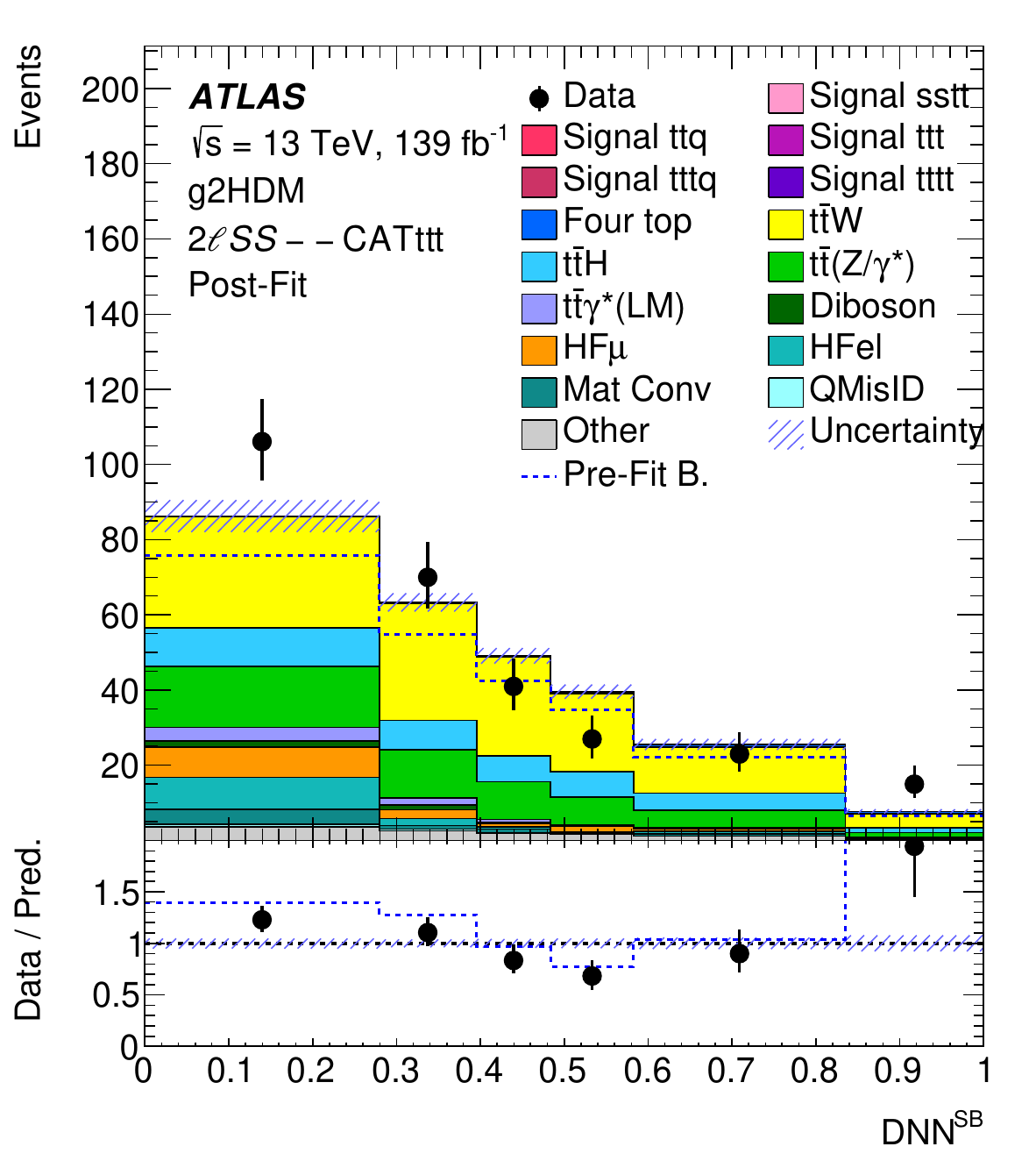}}
\subfloat[]{\includegraphics[width=0.249\textwidth,trim={0 0 0.3cm 0},clip]{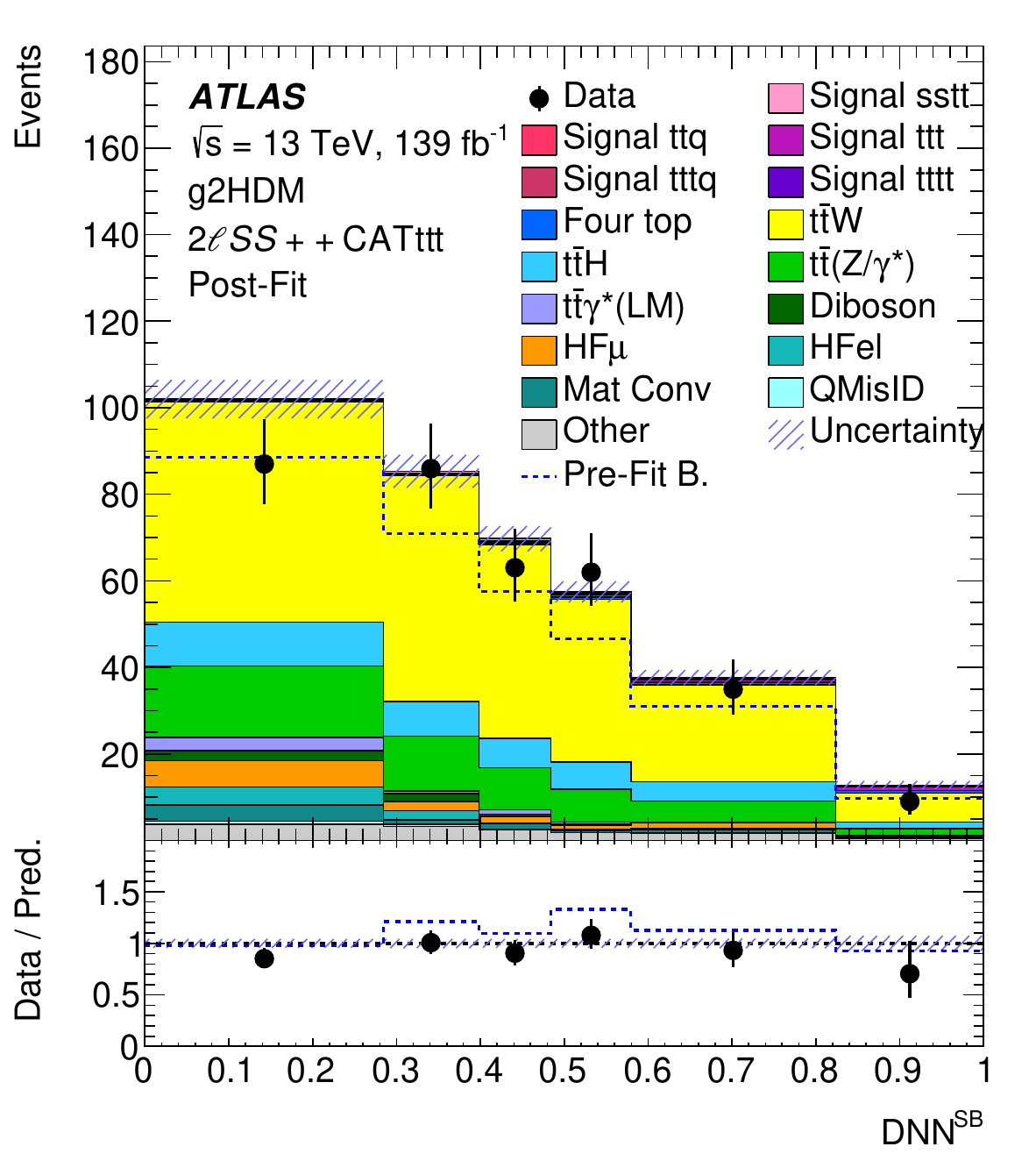}}
\subfloat[]{\includegraphics[width=0.249\textwidth,trim={0 0 0.3cm 0},clip]{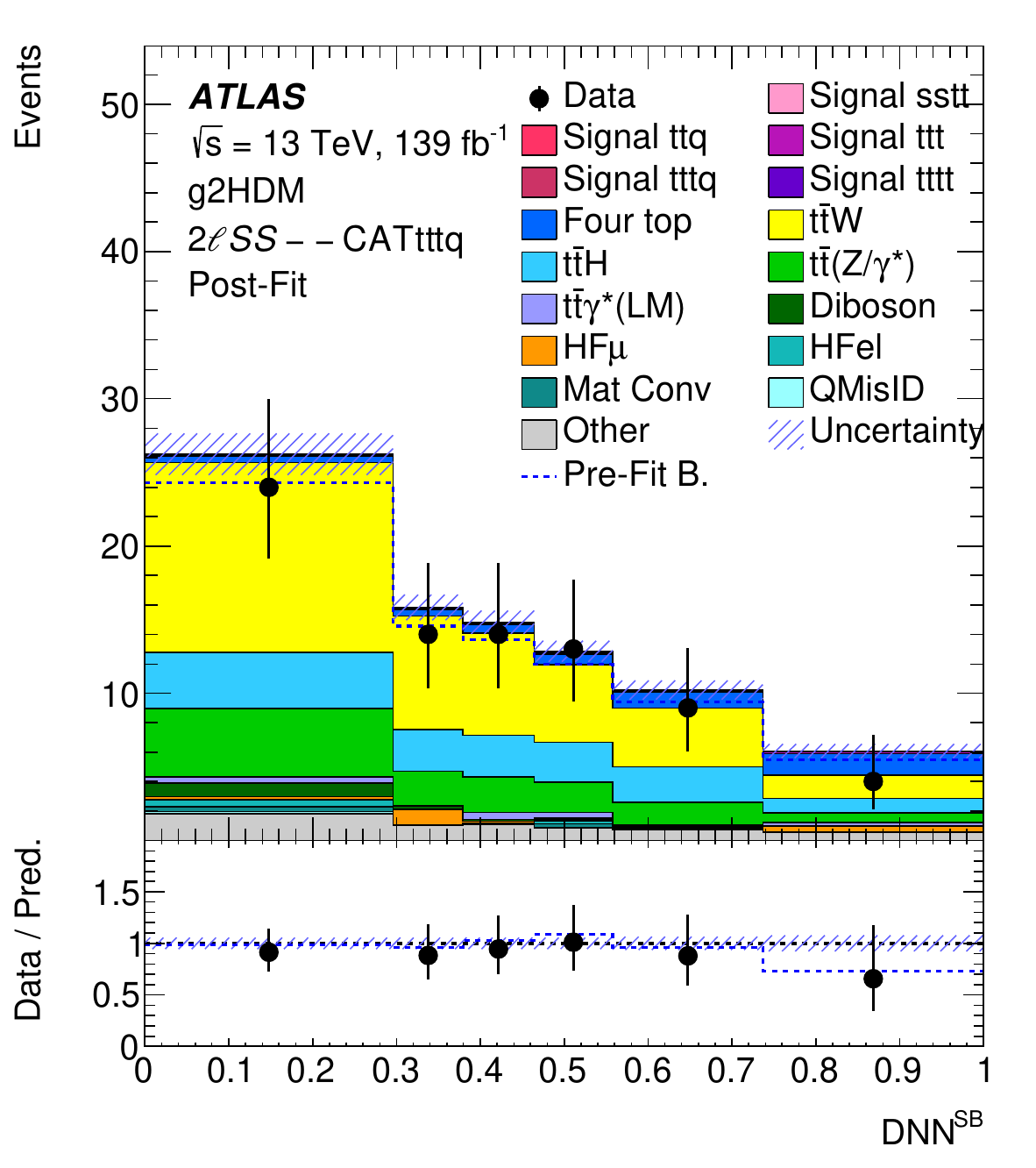}}
\subfloat[]{\includegraphics[width=0.249\textwidth,trim={0 0 0.3cm 0},clip]{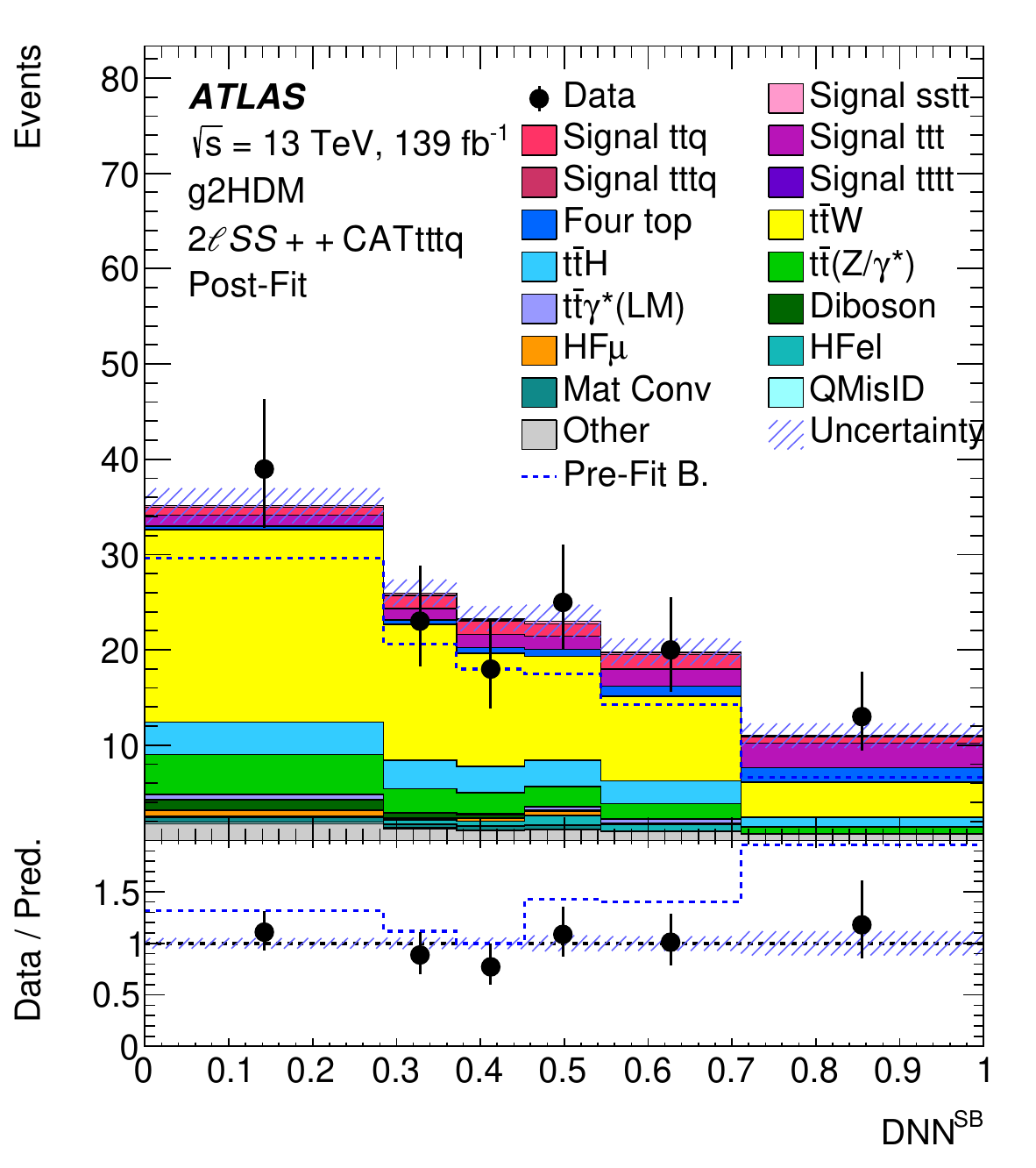}}\\
\subfloat[]{\includegraphics[width=0.249\textwidth,trim={0 0 0.3cm 0},clip]{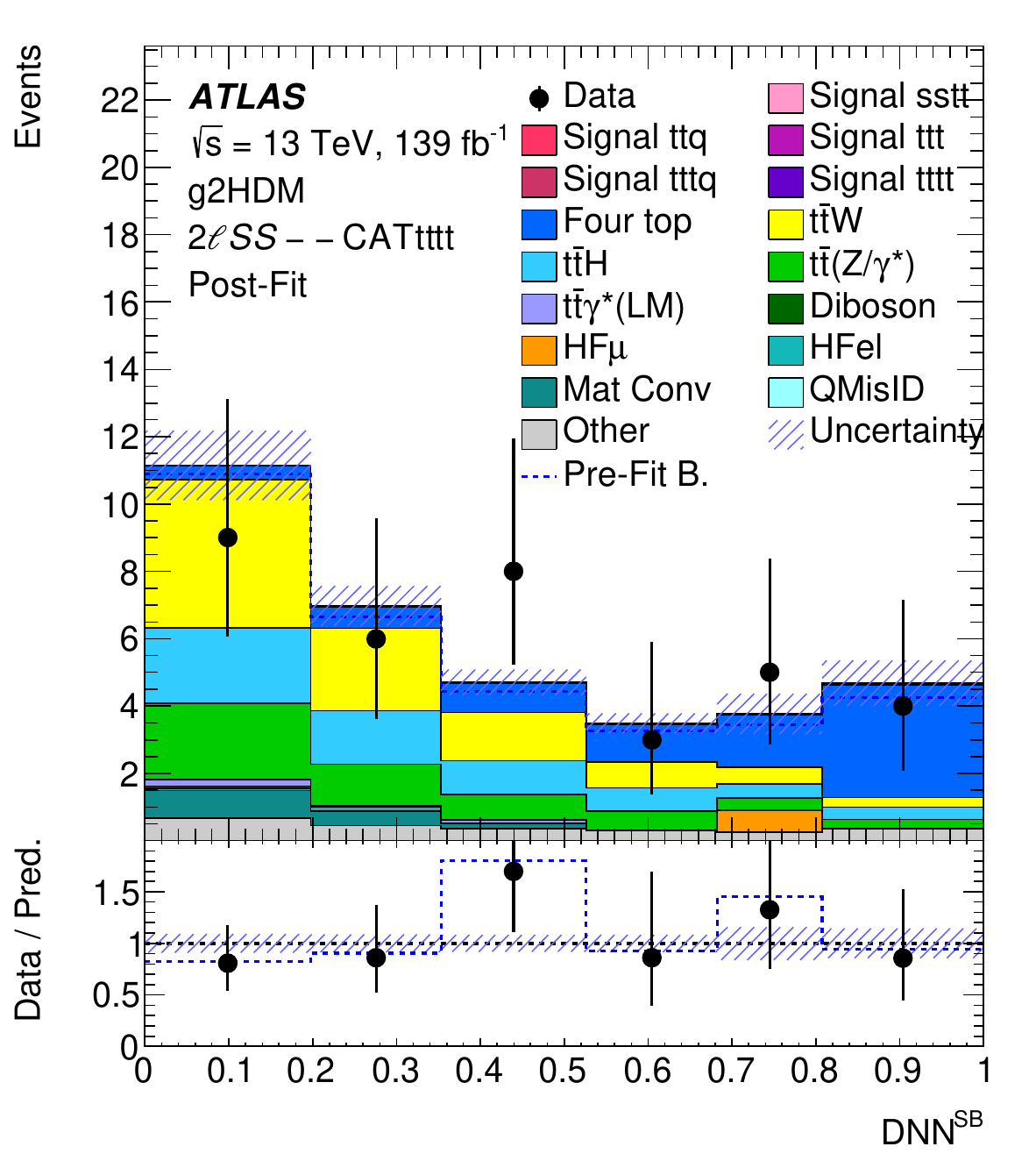}}
\subfloat[]{\includegraphics[width=0.249\textwidth,trim={0 0 0.3cm 0},clip]{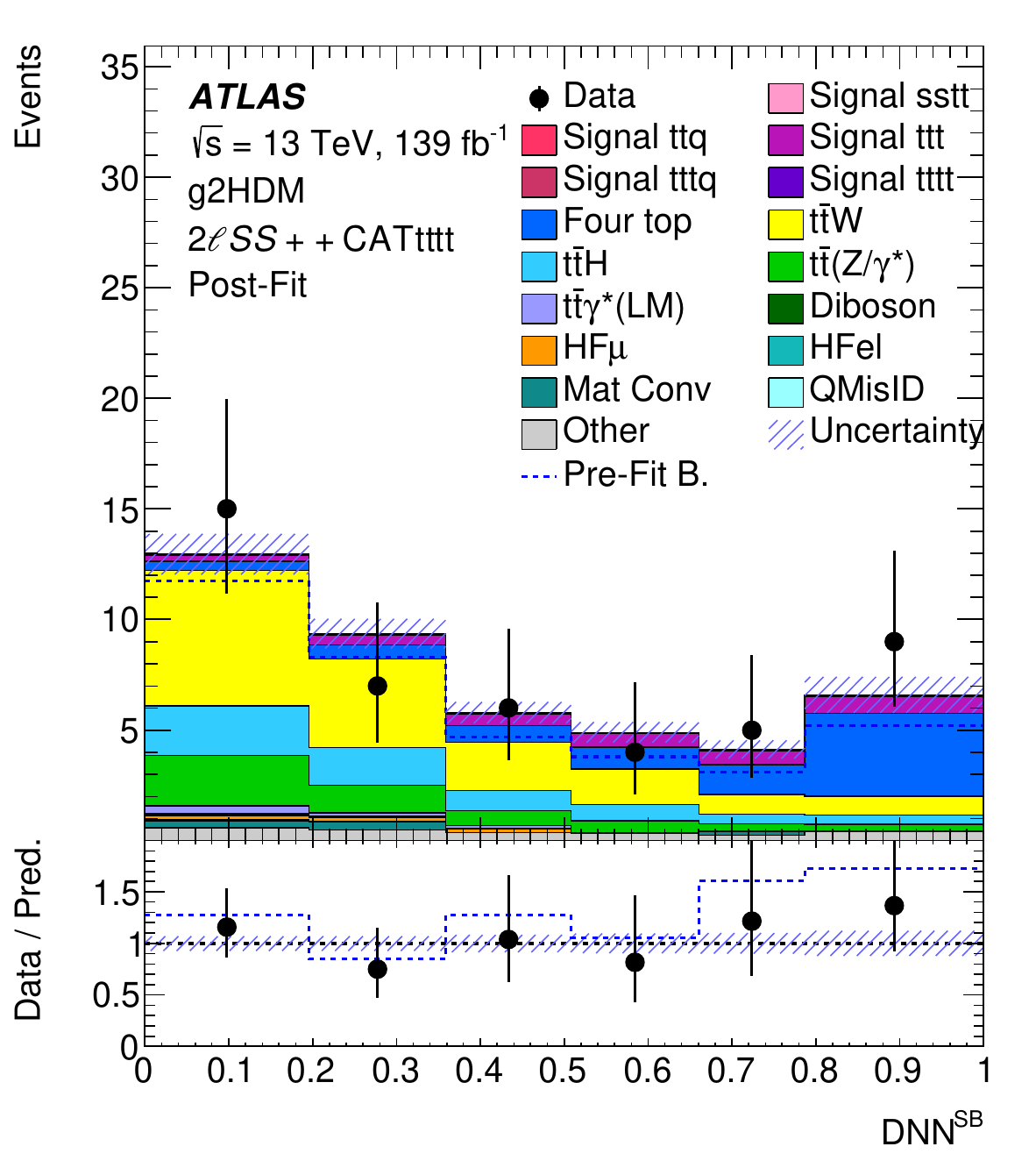}}
\caption{
Comparison between data and prediction for the \dnnsb distribution used in different signal region categories of the \llSS channel:
(a) \twonntt, (b) \twopptt,  (c) \twonnttq, (d) \twoppttq, (e) \twonnttt, (f) \twoppttt, (g) \twonntttq, (h) \twopptttq, (i) \twonntttt, and (j) \twopptttt.
The expected signal for $m_{H}=900~\gev$ and couplings $\rhott$=\bestrhott, $\rhotc$=\bestrhotc, and $\rhotu$=\bestrhotu, along with the background contributions, is shown after the likelihood fit to data (``Post-Fit'') for the signal-plus-background hypothesis.
The ratio of the data to the prediction (``Pred.'') is shown in the lower panel, separately for post-fit signal-plus-background (black points) and pre-fit background (dashed blue line).
The size of the combined statistical and systematic uncertainty in the signal-plus-background prediction is indicated by the blue hatched band.
}
\label{fig:dnnspectrum1}
\end{figure*}
 
\begin{figure*}[tp]
\centering
\subfloat[]{\includegraphics[width=0.249\textwidth,trim={0 0 0.3cm 0},clip]{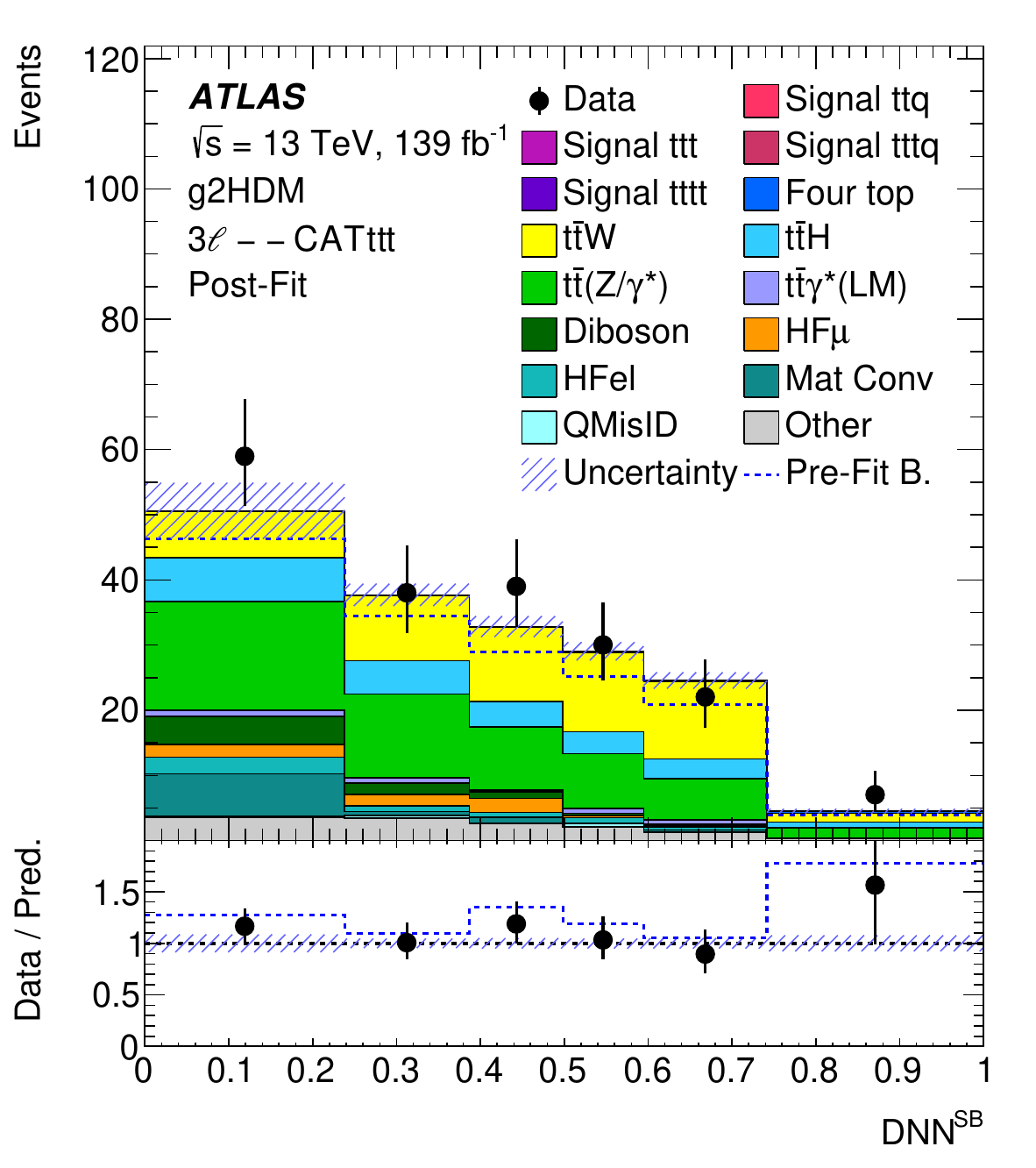}}
\subfloat[]{\includegraphics[width=0.249\textwidth,trim={0 0 0.3cm 0},clip]{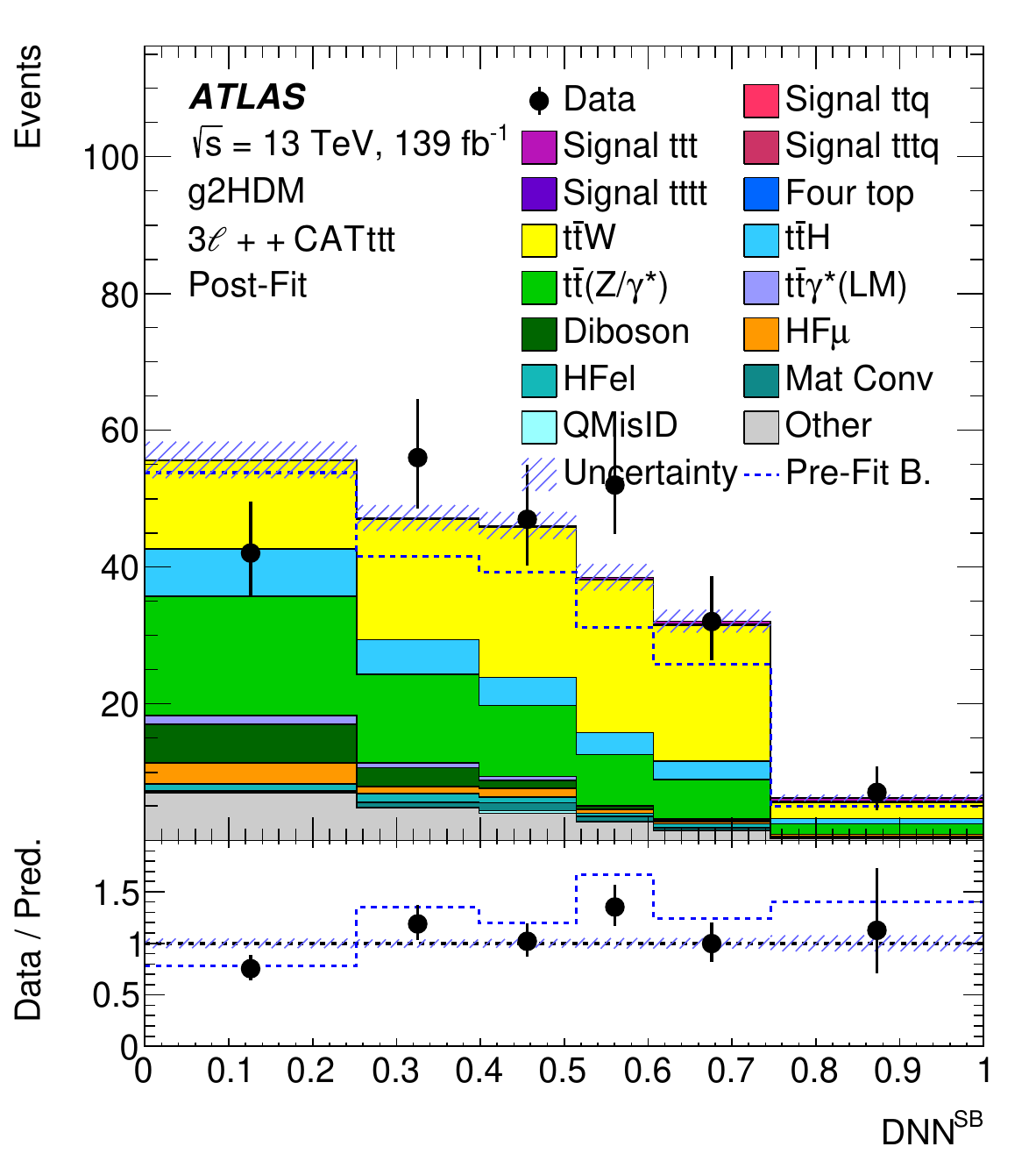}}
\subfloat[]{\includegraphics[width=0.249\textwidth,trim={0 0 0.3cm 0},clip]{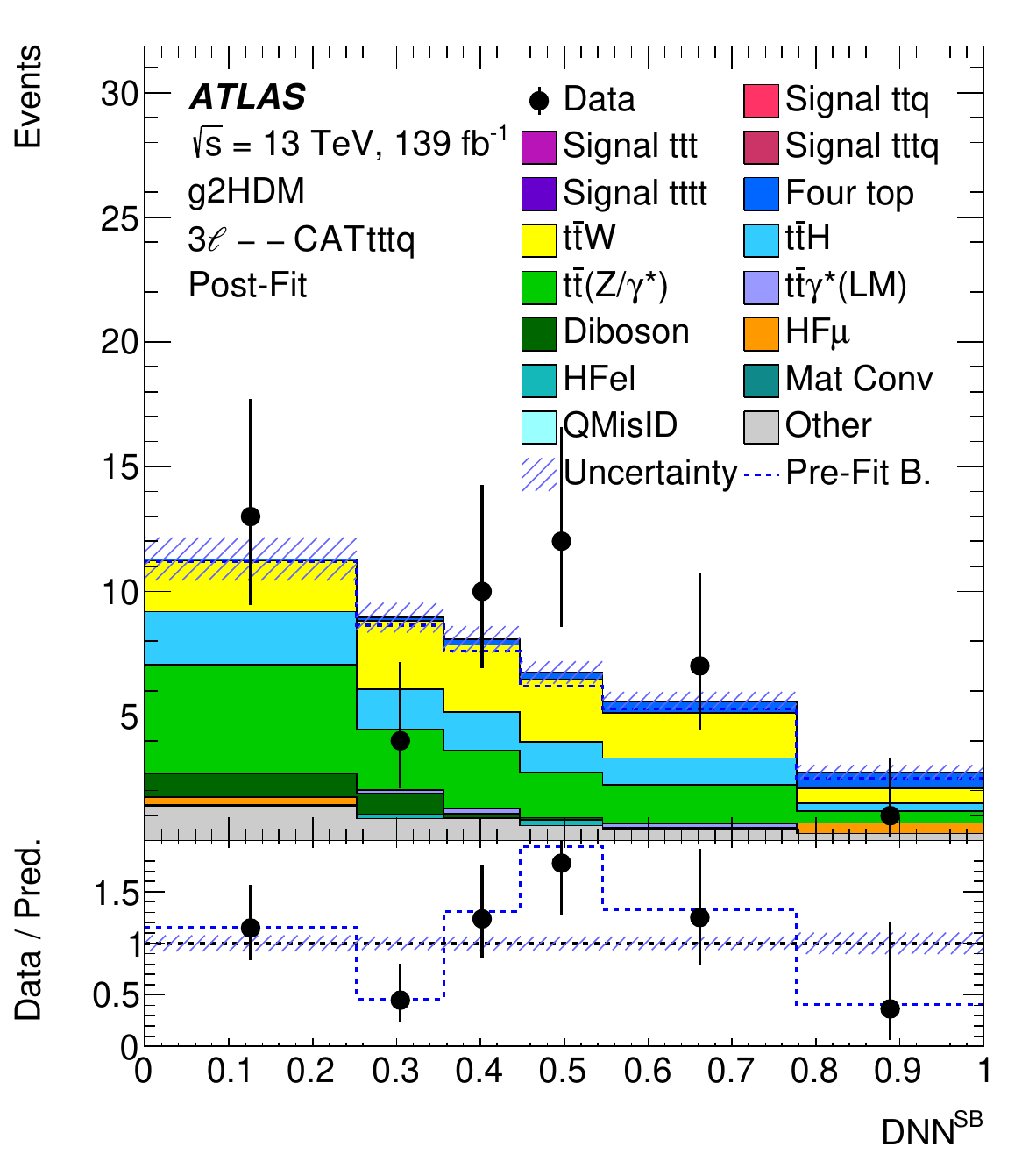}}
\subfloat[]{\includegraphics[width=0.249\textwidth,trim={0 0 0.3cm 0},clip]{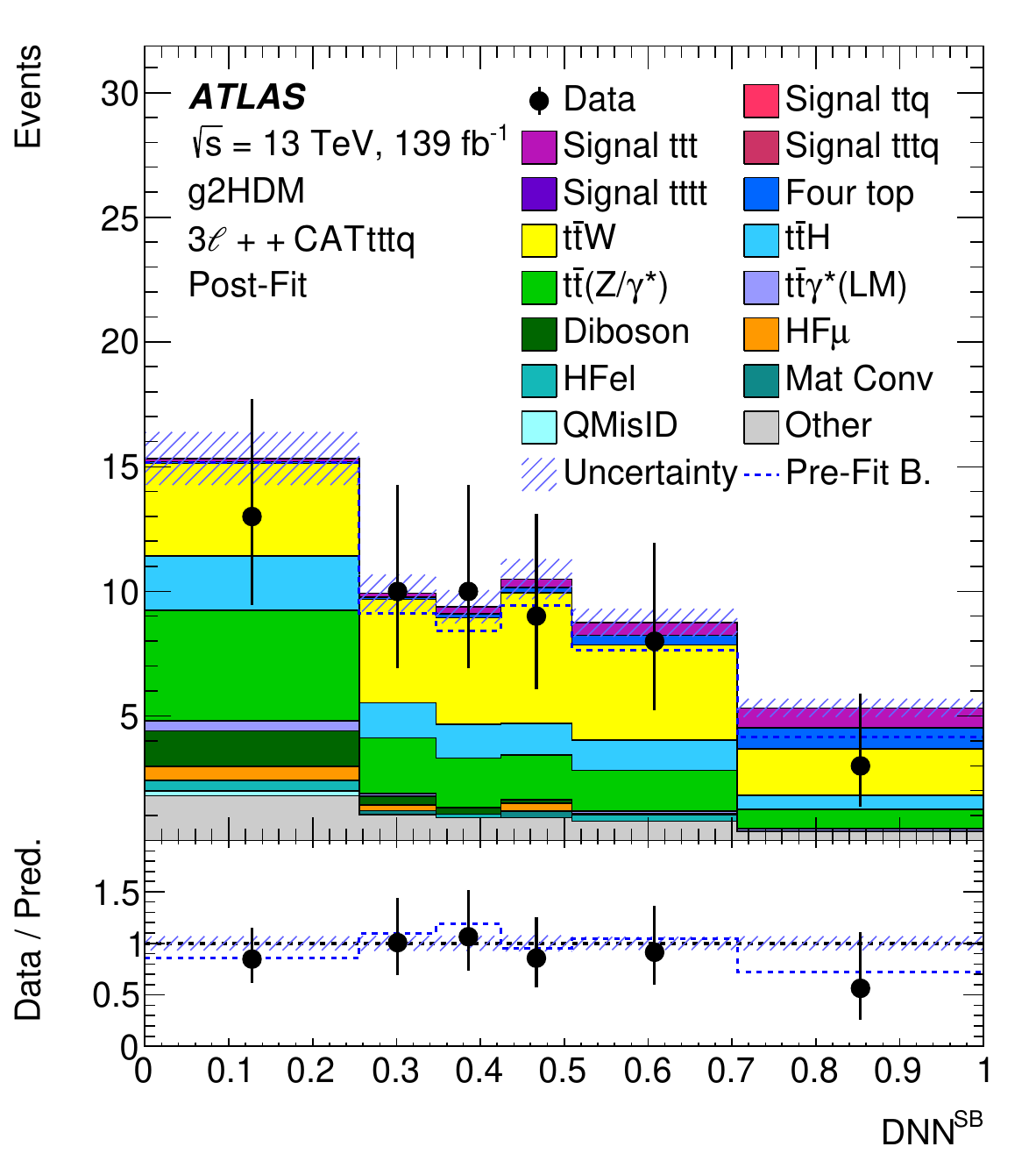}}\\
\subfloat[]{\includegraphics[width=0.249\textwidth,trim={0 0 0.1cm 0},clip]{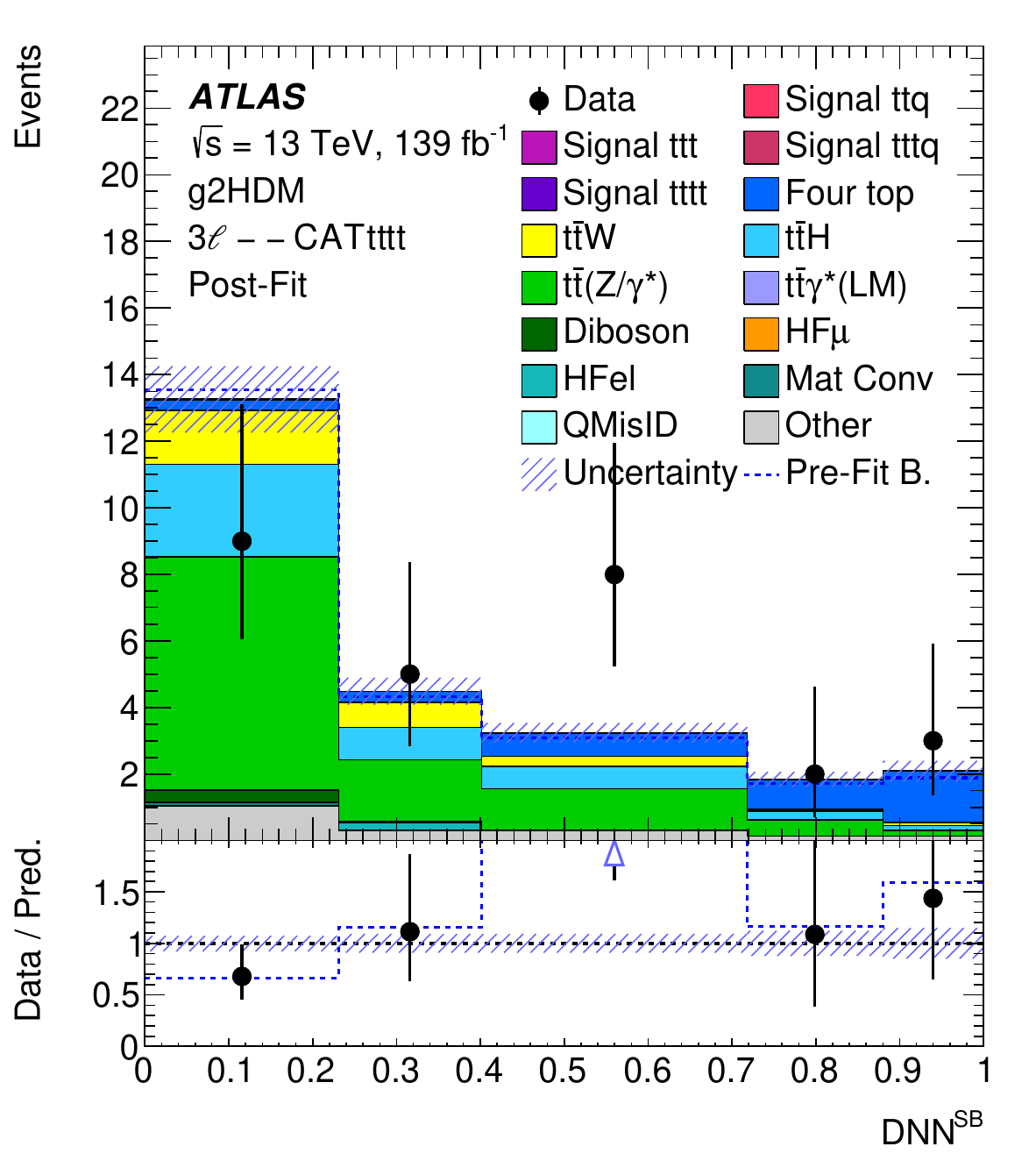}}
\subfloat[]{\includegraphics[width=0.249\textwidth,trim={0 0 0.1cm 0},clip]{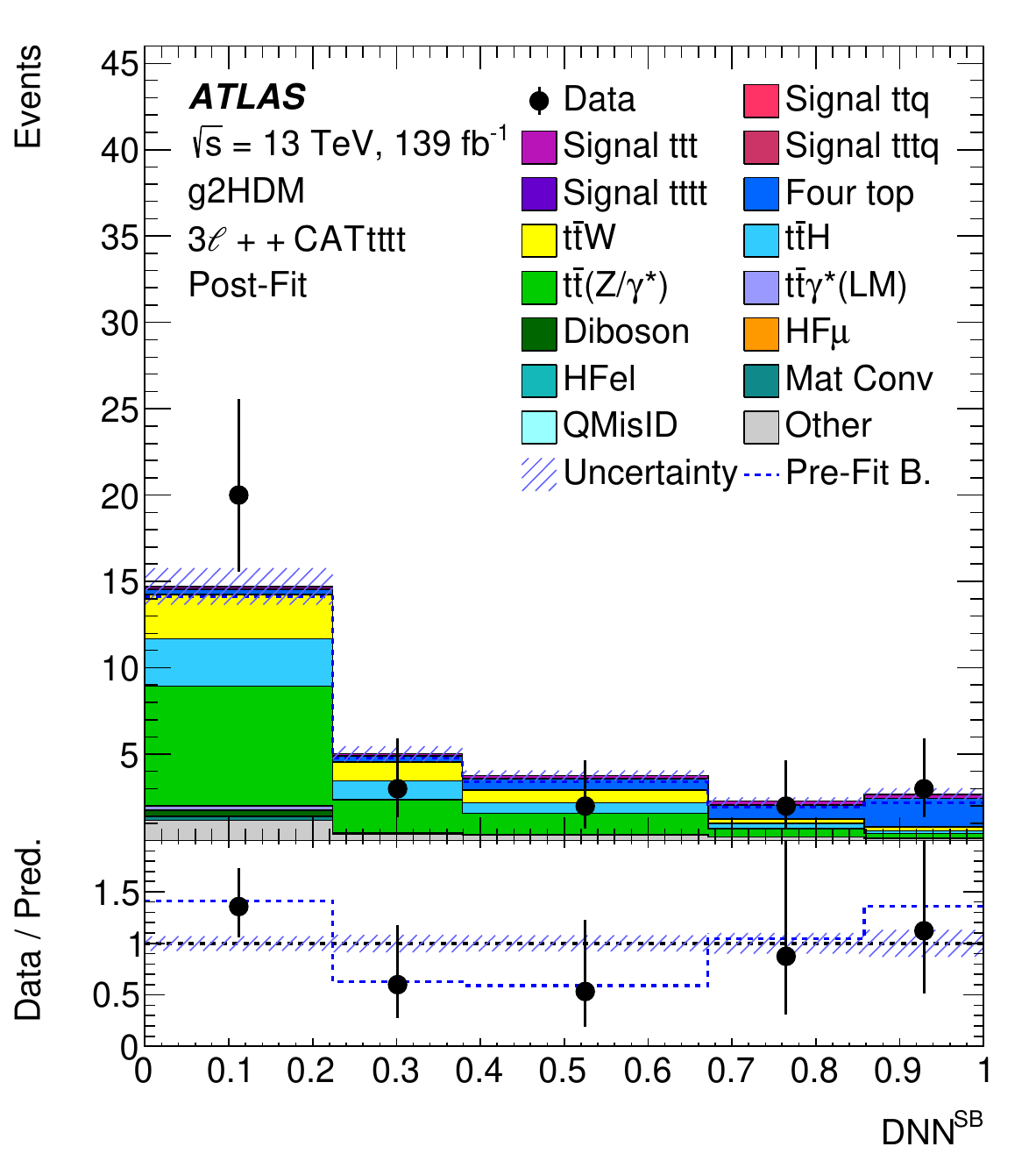}}
\subfloat[]{\includegraphics[width=0.249\textwidth,trim={0 0 0.1cm 0},clip]{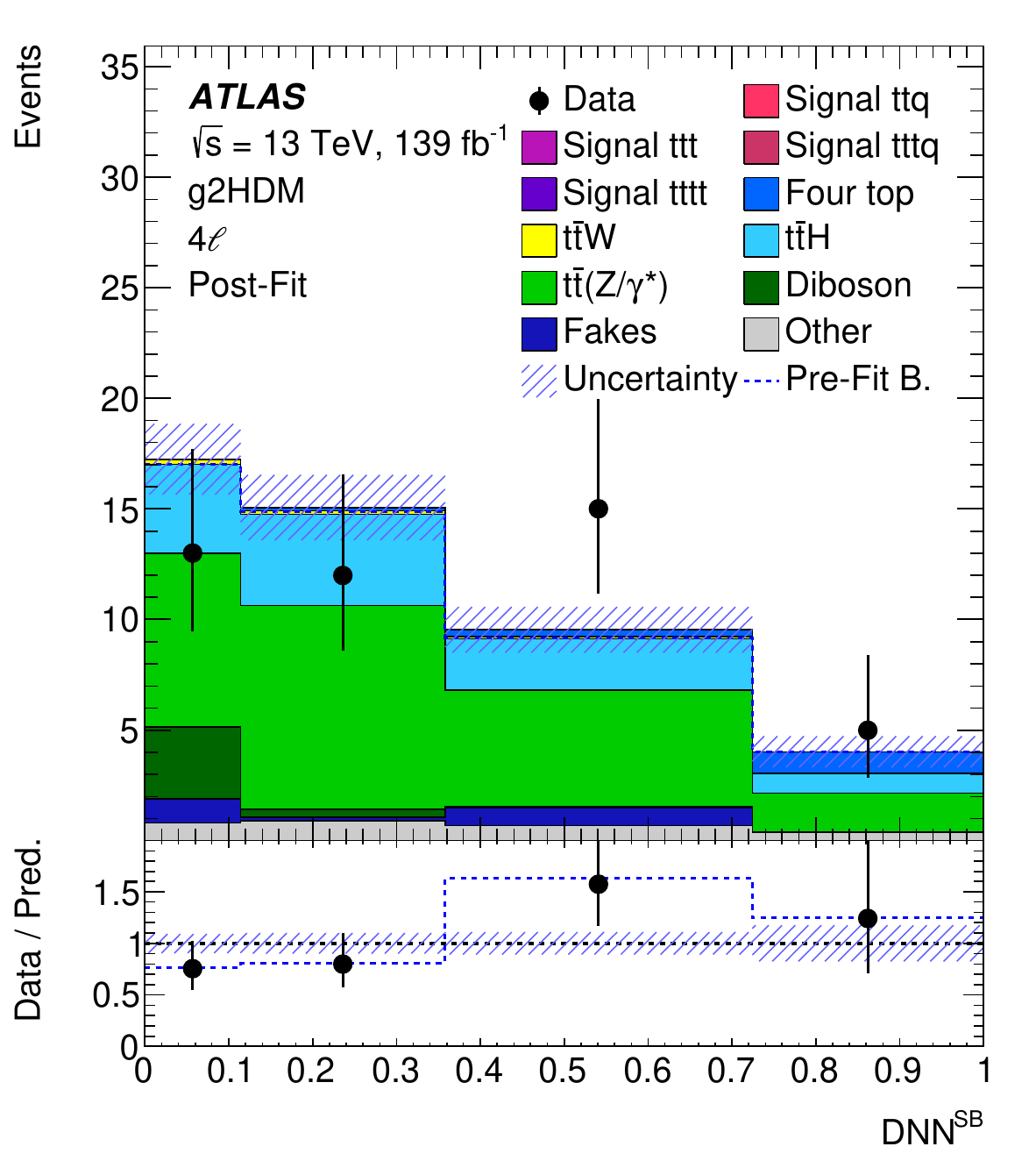}}
\caption{
Comparison between data and prediction for the \dnnsb distribution used in different signal region categories of the \lll and \llll channels:
(a) \threennttt, (b) \threeppttt, (c) \threenntttq, (d) \threepptttq, (e) \threenntttt, (f) \threepptttt, and (g) \llll.
The expected signal for $m_{H}=900~\gev$ and couplings $\rhott$=\bestrhott, $\rhotc$=\bestrhotc, and $\rhotu$=\bestrhotu, along with the background contributions, is shown after the likelihood fit to data (``Post-Fit'') for the signal-plus-background hypothesis.
The ratio of the data to the prediction (``Pred.'') is shown in the lower panel, separately for post-fit signal-plus-background (black points) and pre-fit background (dashed blue line).
The size of the combined statistical and systematic uncertainty in the signal-plus-background prediction is indicated by the blue hatched band.
}
\label{fig:dnnspectrum2}
\end{figure*}
 
\begin{figure}[tp]
\centering
\subfloat[]{\includegraphics[width=0.49\textwidth]{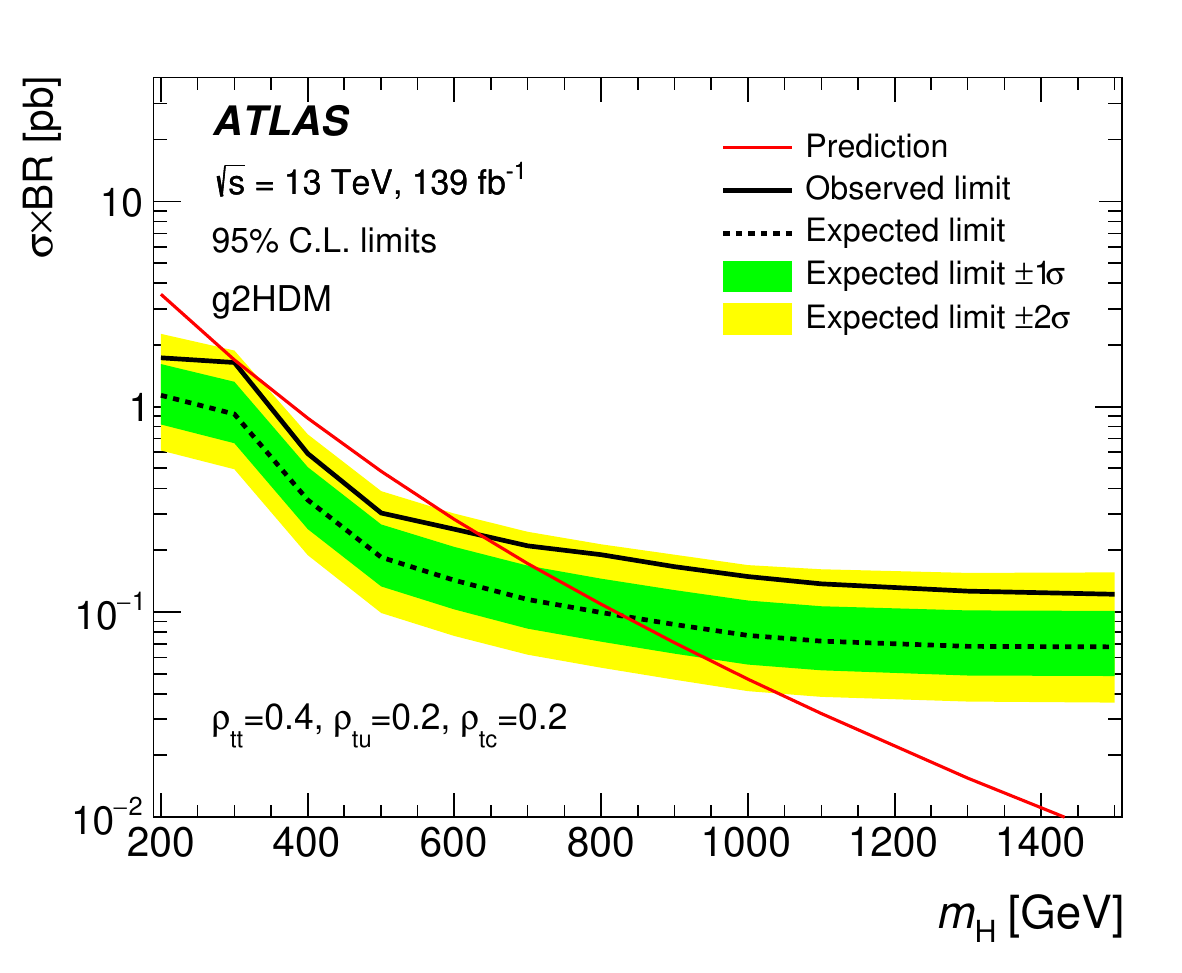}\label{fig:limit_obs_benc}}
\subfloat[]{\includegraphics[width=0.49\textwidth]{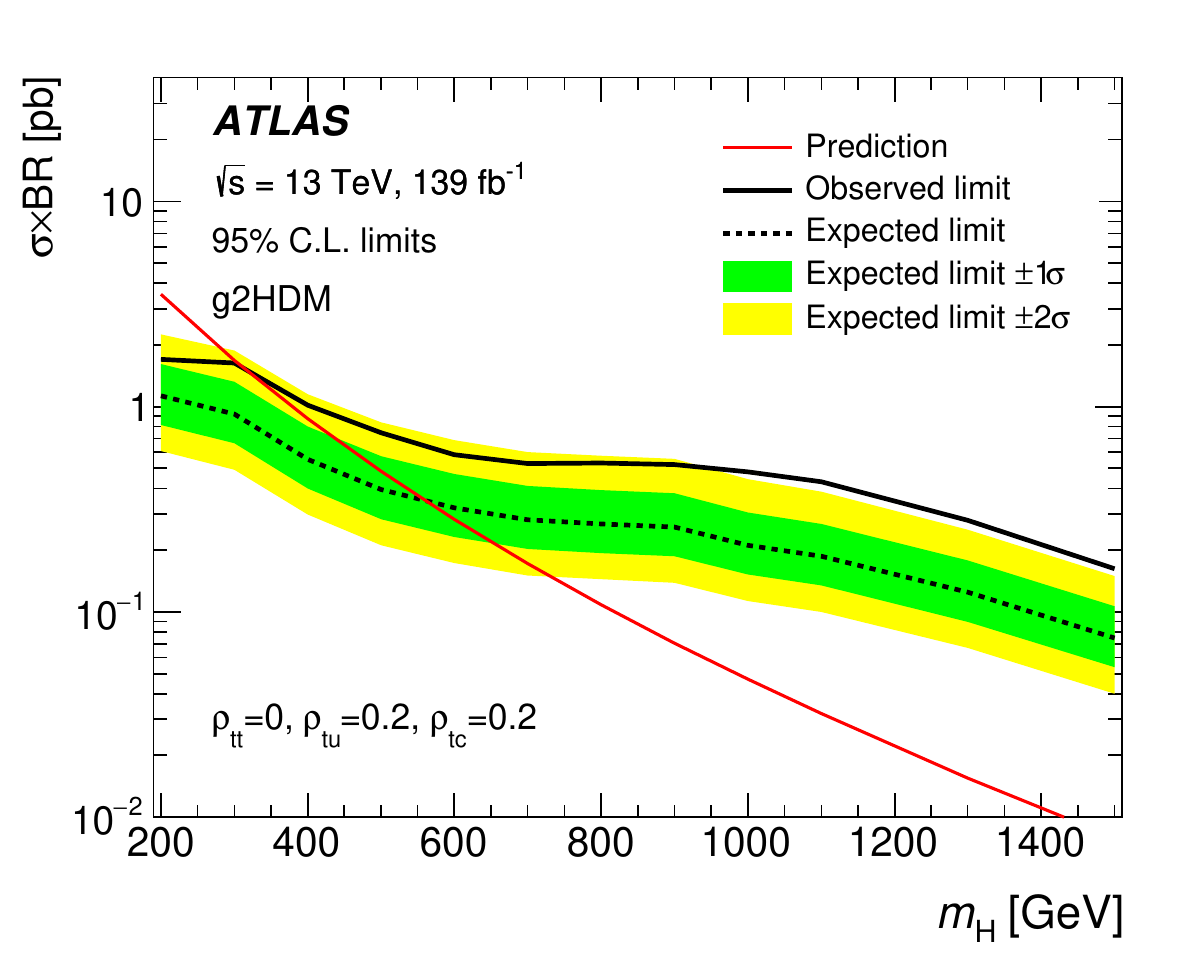}\label{fig:limit_obs_benc0tt}} \\
\subfloat[]{\includegraphics[width=0.49\textwidth]{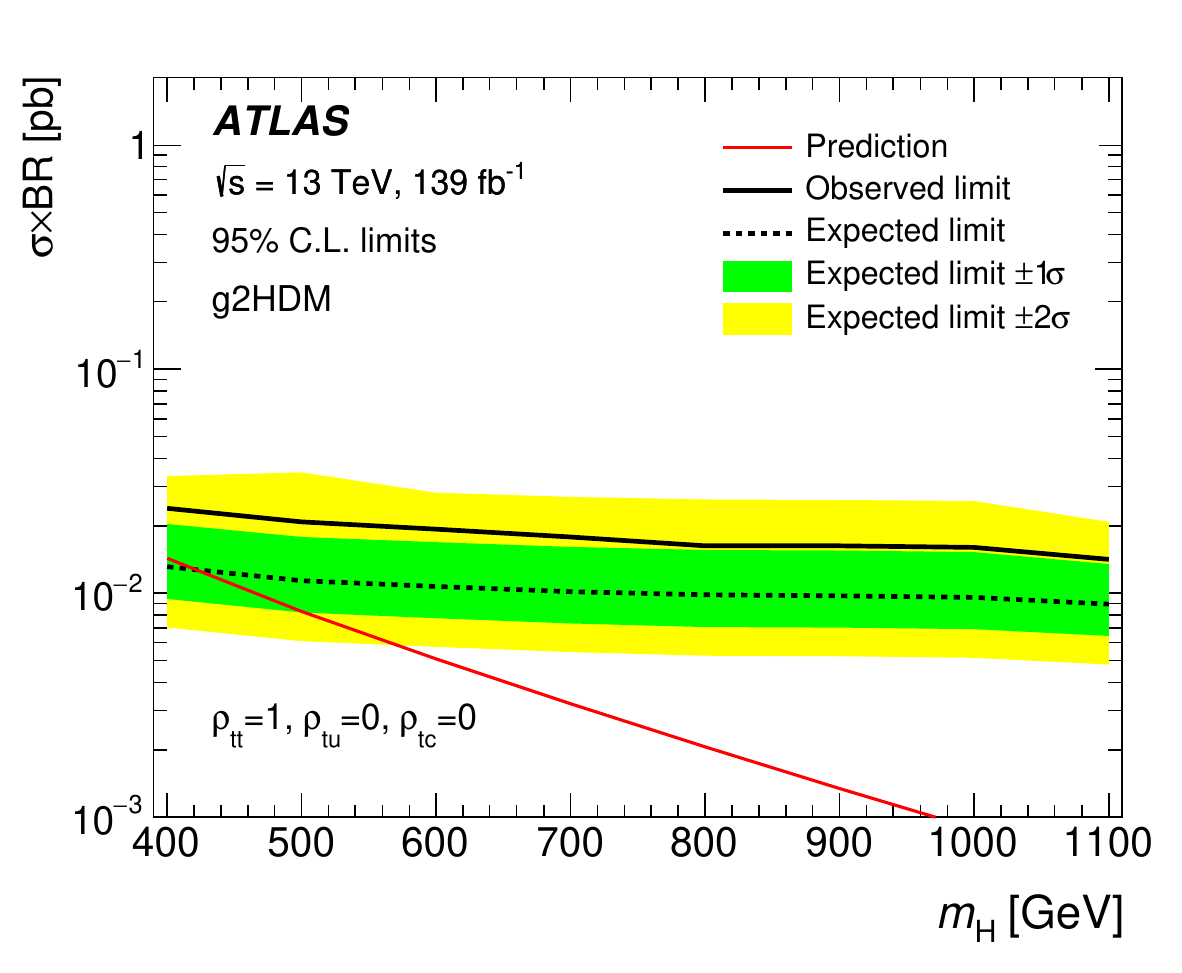}\label{fig:limit_gtt1}}
\subfloat[]{\includegraphics[width=0.49\textwidth]{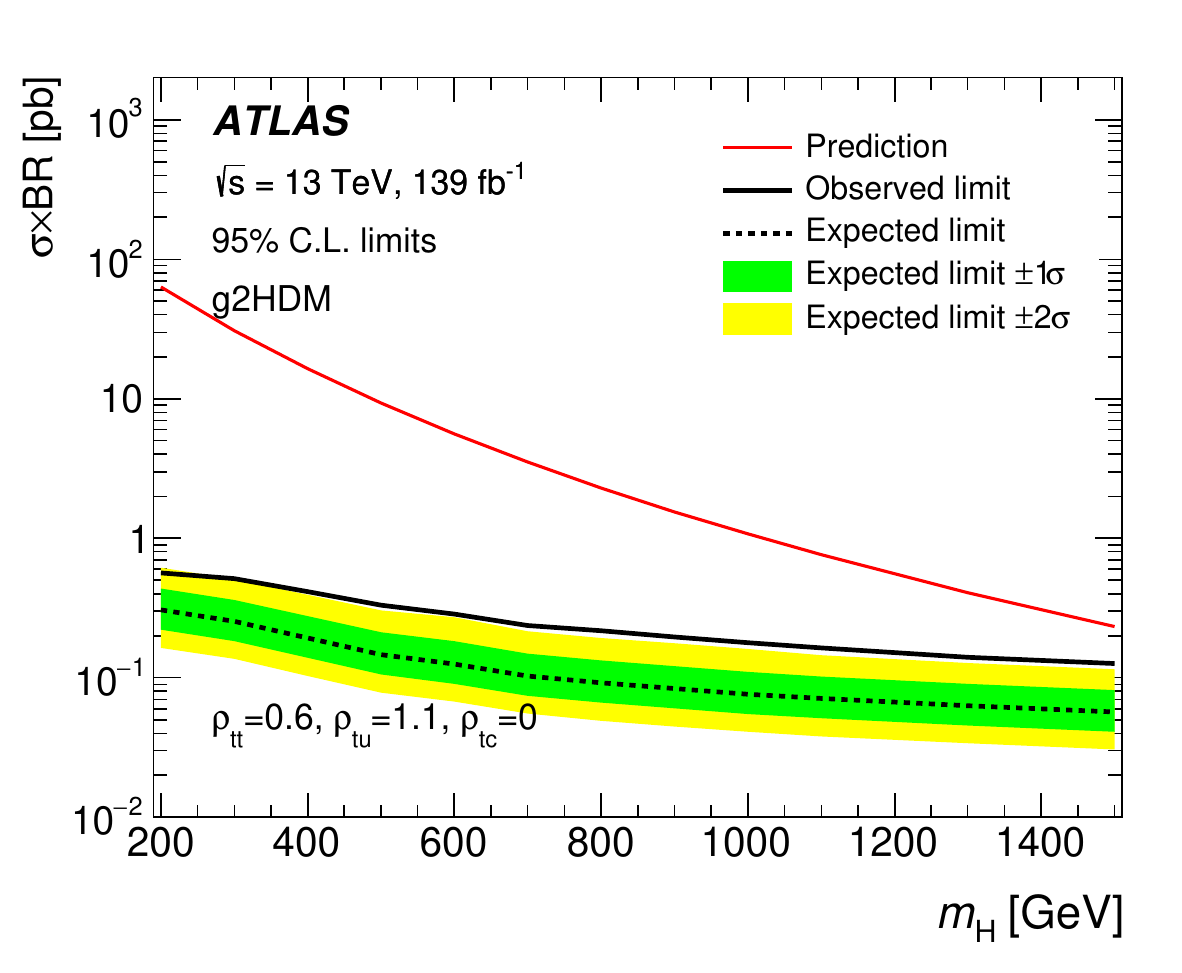}\label{fig:limit_mostsig}}
\caption{
Observed and expected exclusion limits at 95\% confidence level on the heavy Higgs boson mass for the g2HDM signal model for different couplings choices: (a) $\rhott=0.4$, $\rhotc=0.2$, $\rhotu=0.2$, (b) $\rhott=0$, $\rhotc=0.2$, $\rhotu=0.2$, (c) $\rhott=1$, $\rhotc=0$, $\rhotu=0$, and (d) $\rhott=0.6$, $\rhotc=0$, $\rhotu=1.1$, the latter corresponding to the couplings yielding the most significant excess.
The yellow and green contours of the band around the expected limit are the $\pm 1 \sigma$ and $\pm 2 \sigma$ variations including all uncertainties, respectively.
The theoretical prediction for the signal production cross section is also shown as a red line.
The production cross section is the sum of the five production modes considered in the search.
}
\label{fig:mass_limit}
\end{figure}

\begin{figure}[tbp]
\centering
\subfloat[]{\includegraphics[width=0.49\textwidth]{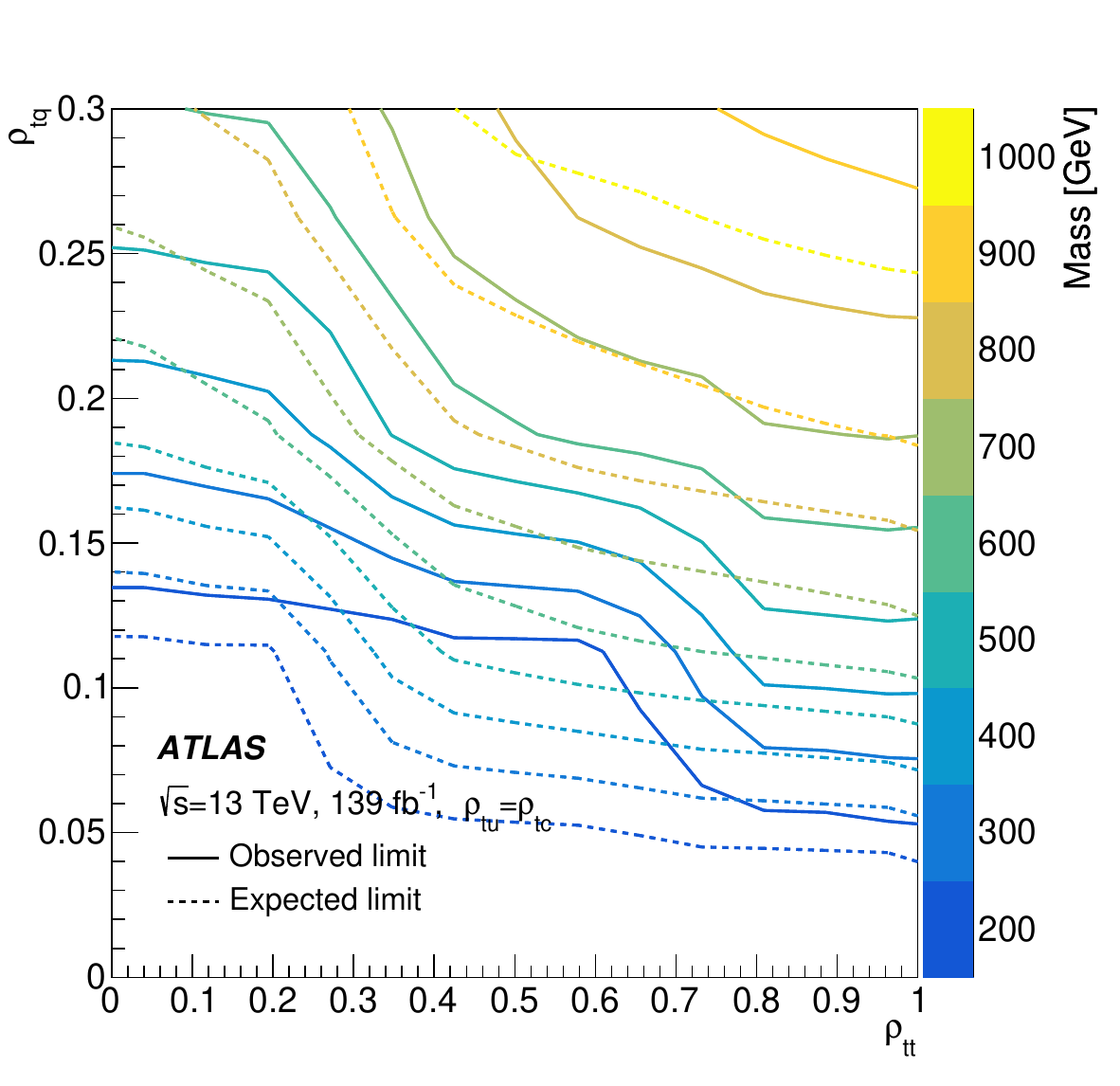}\label{fig:limit_coupling1}}
\subfloat[]{\includegraphics[width=0.49\textwidth]{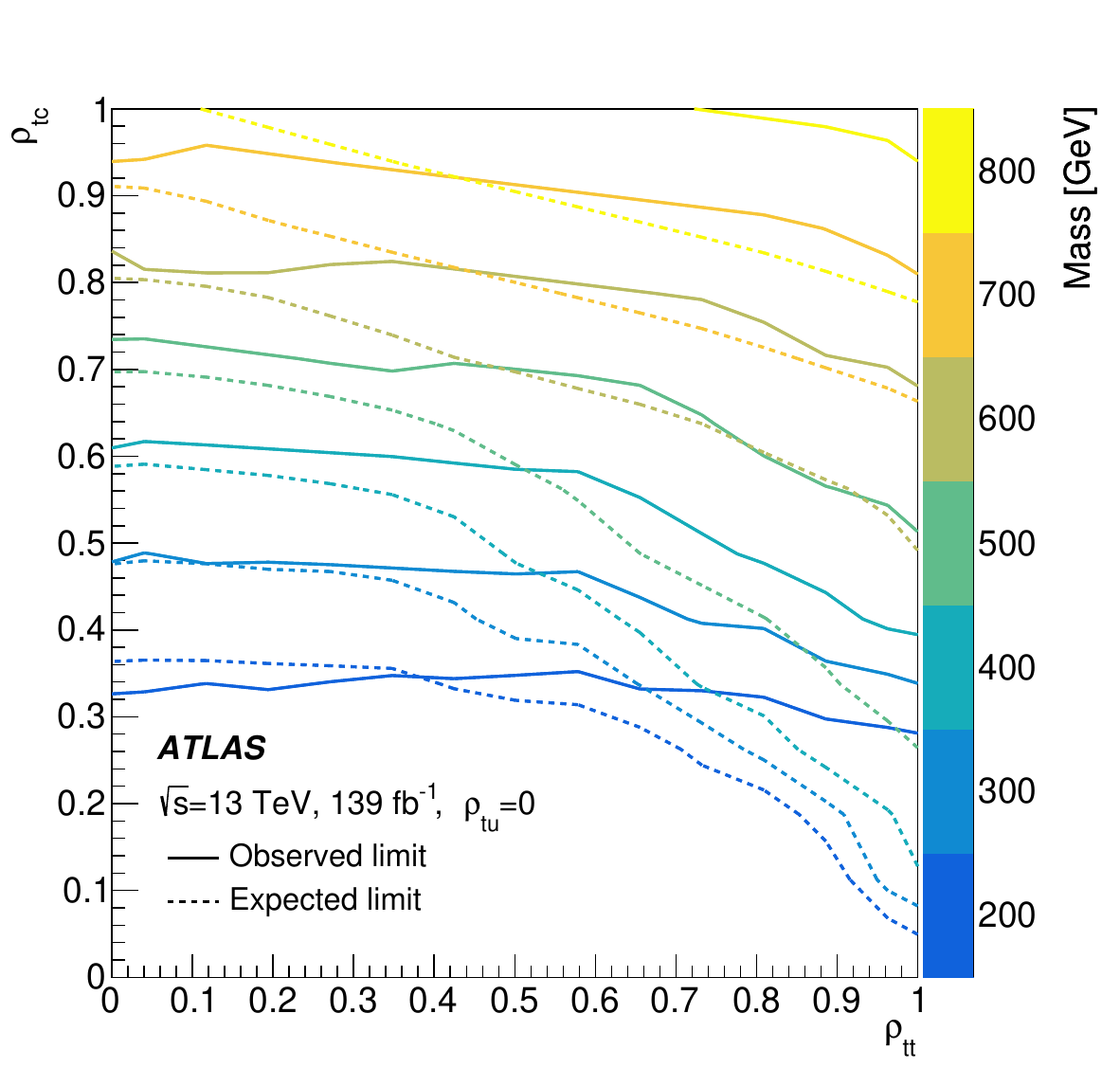}\label{fig:limit_coupling2}}
\\
\subfloat[]{\includegraphics[width=0.49\textwidth]{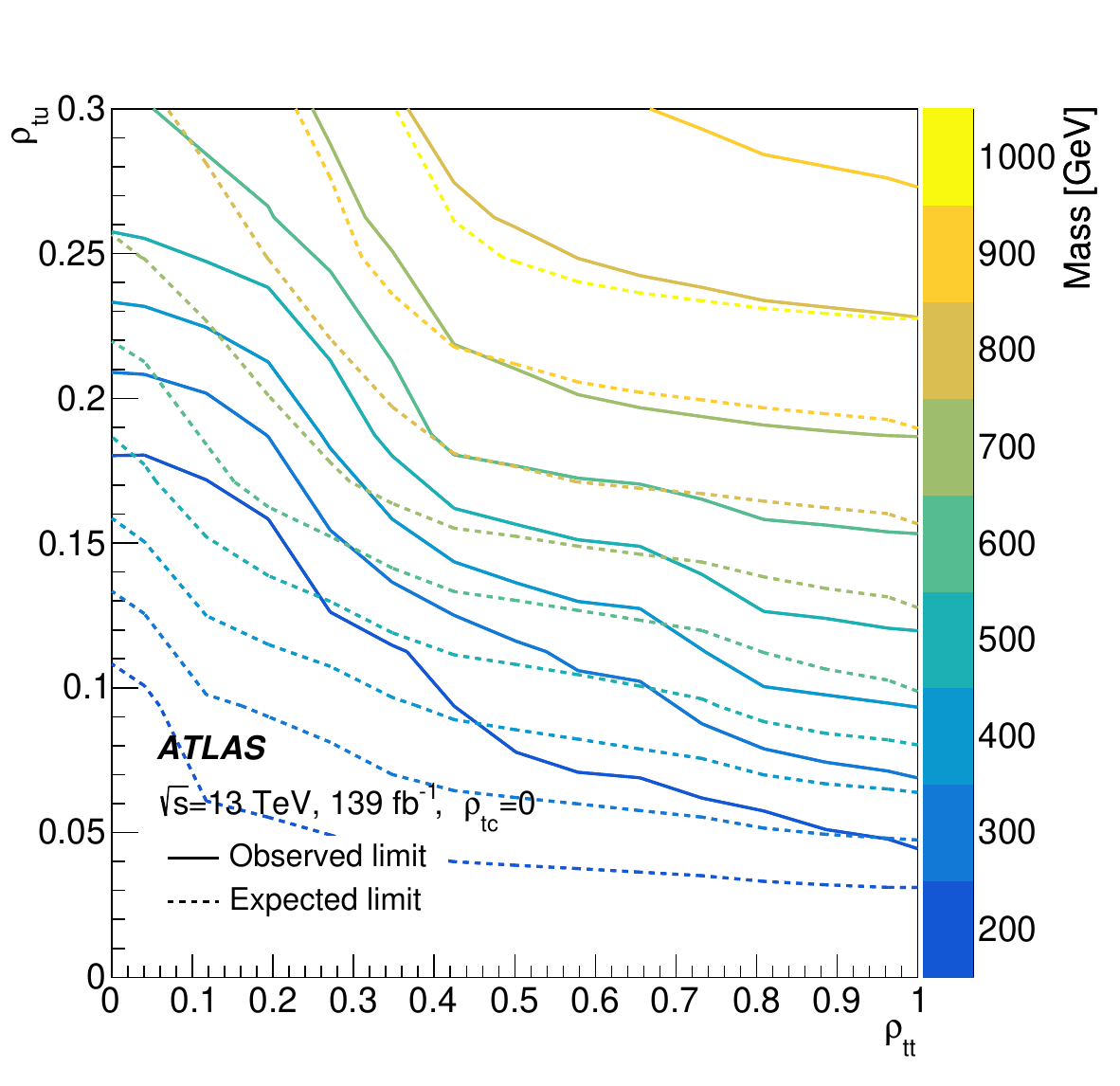}\label{fig:limit_coupling3}}
\subfloat[]{\includegraphics[width=0.49\textwidth]{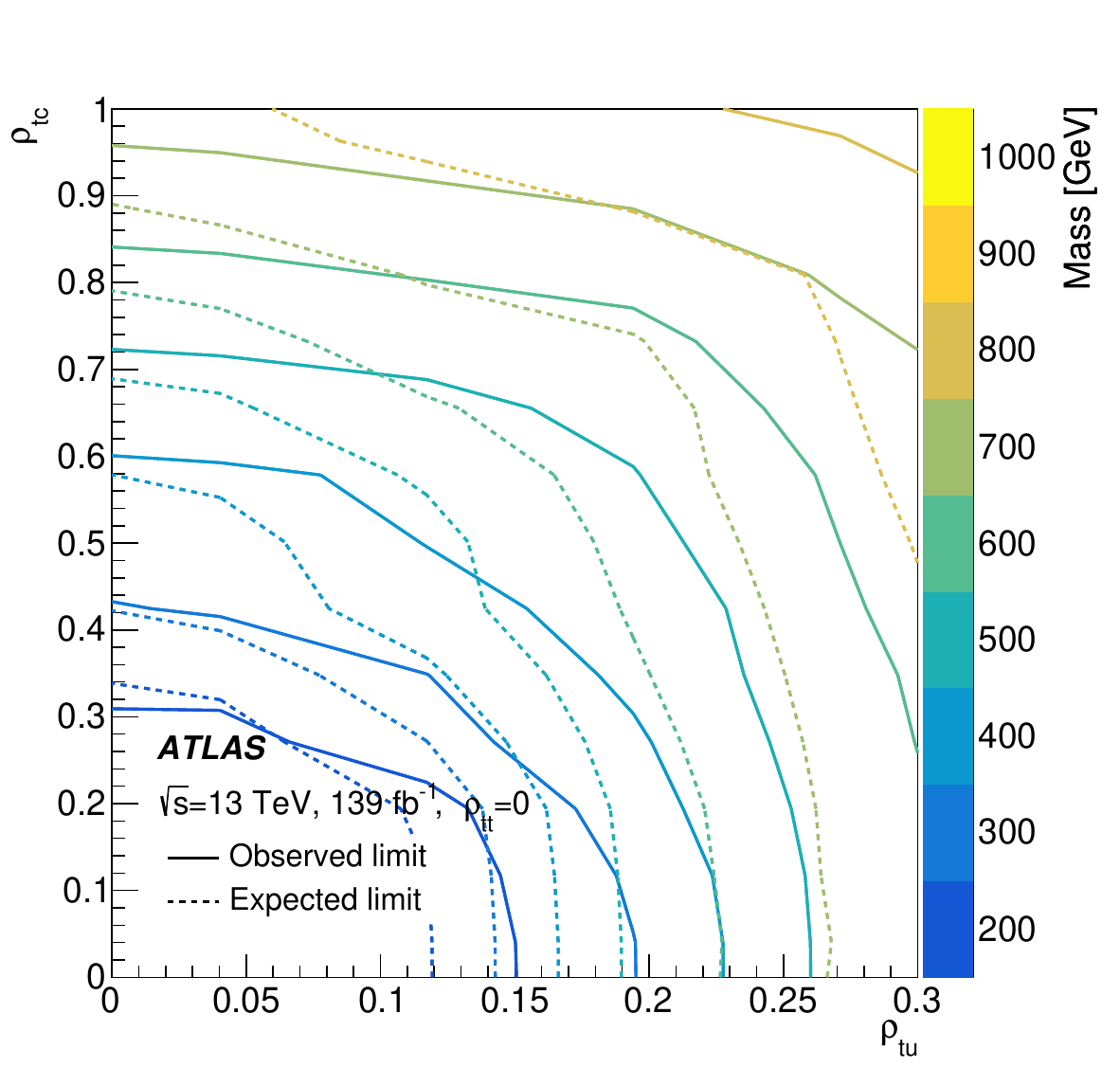}\label{fig:limit_coupling4}}
\caption{
Observed (solid line) and expected (dashed line) exclusion limits on the scalar mass as a function of the coupling for different assumptions: (a) $\rhotc=\rhotu$, (b) $\rhotu=0$, (c) $\rhotc=0$, and (d) $\rhott=0$.
}
\label{fig:coupling_limit}
\end{figure}
 
\clearpage
 
\begin{figure}[tbp]
\centering
\subfloat[]{\includegraphics[width=0.49\textwidth]{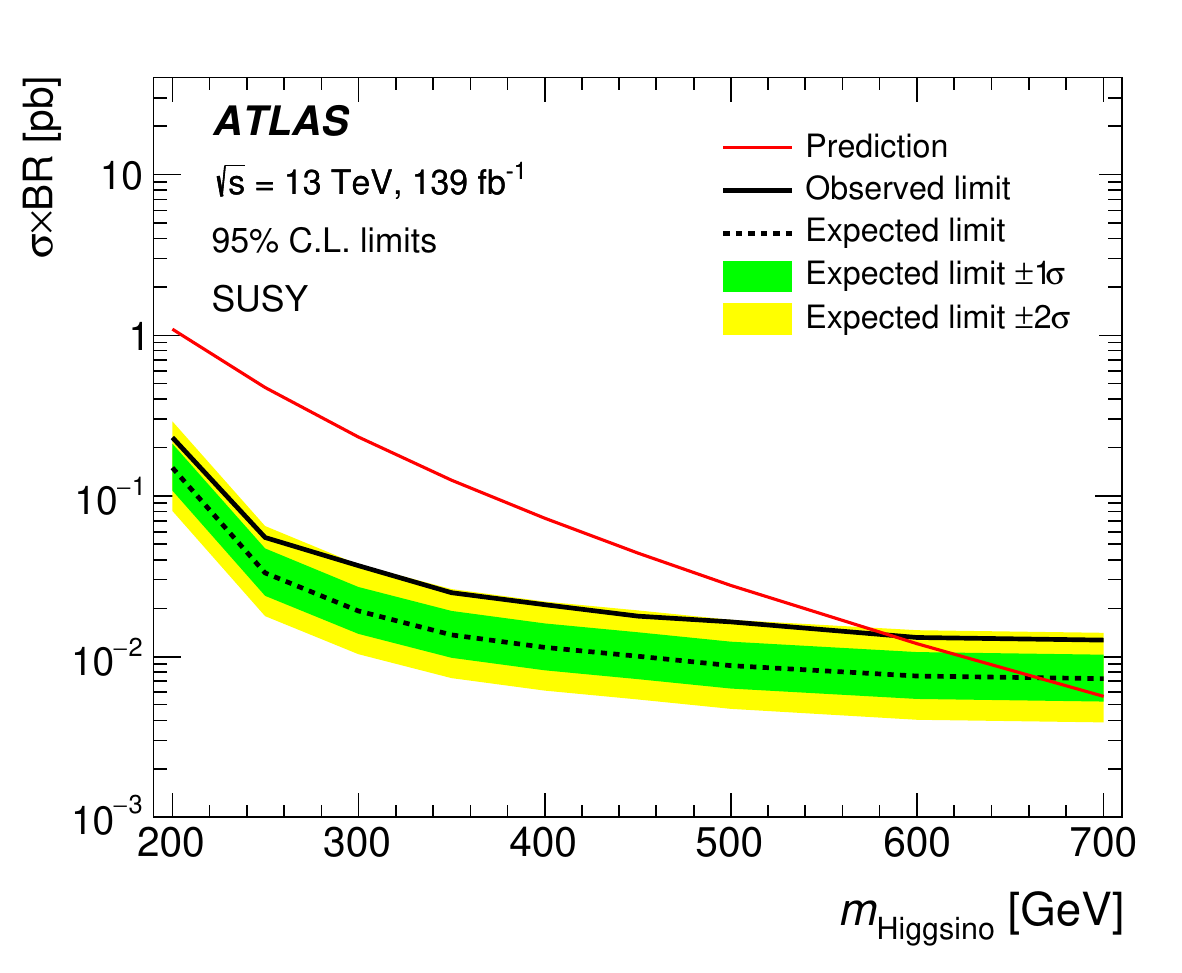}\label{fig:limit_higgsino}}
\subfloat[]{\includegraphics[width=0.49\textwidth]{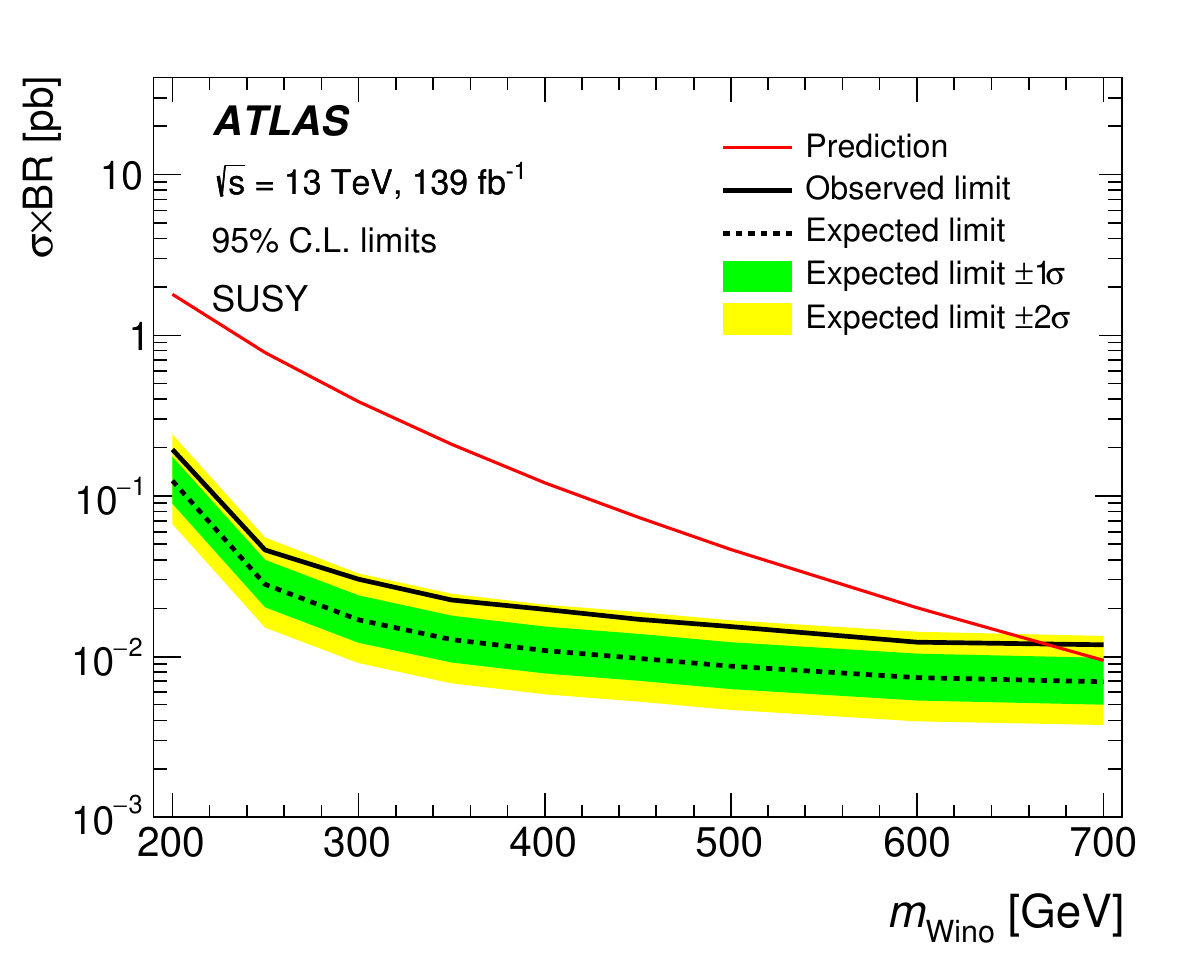}\label{fig:limit_wino}}
\caption{
Expected and observed exclusion limits on the cross section $\times$ branching ratio of electroweakino production in RPV SUSY models as a function of the sparticle masses: (a) higgsino model, and (b) wino model.
The yellow and green contours of the band around the expected limit are the $\pm 1 \sigma$ and $\pm 2 \sigma$ variations including all uncertainties, respectively.
The theoretical prediction for the signal production cross section is also shown as a red line.
}
\label{fig:SUSY_limits1D}
\end{figure}
 
\begin{figure}[tbp]
\centering
\includegraphics[width=0.6\textwidth]{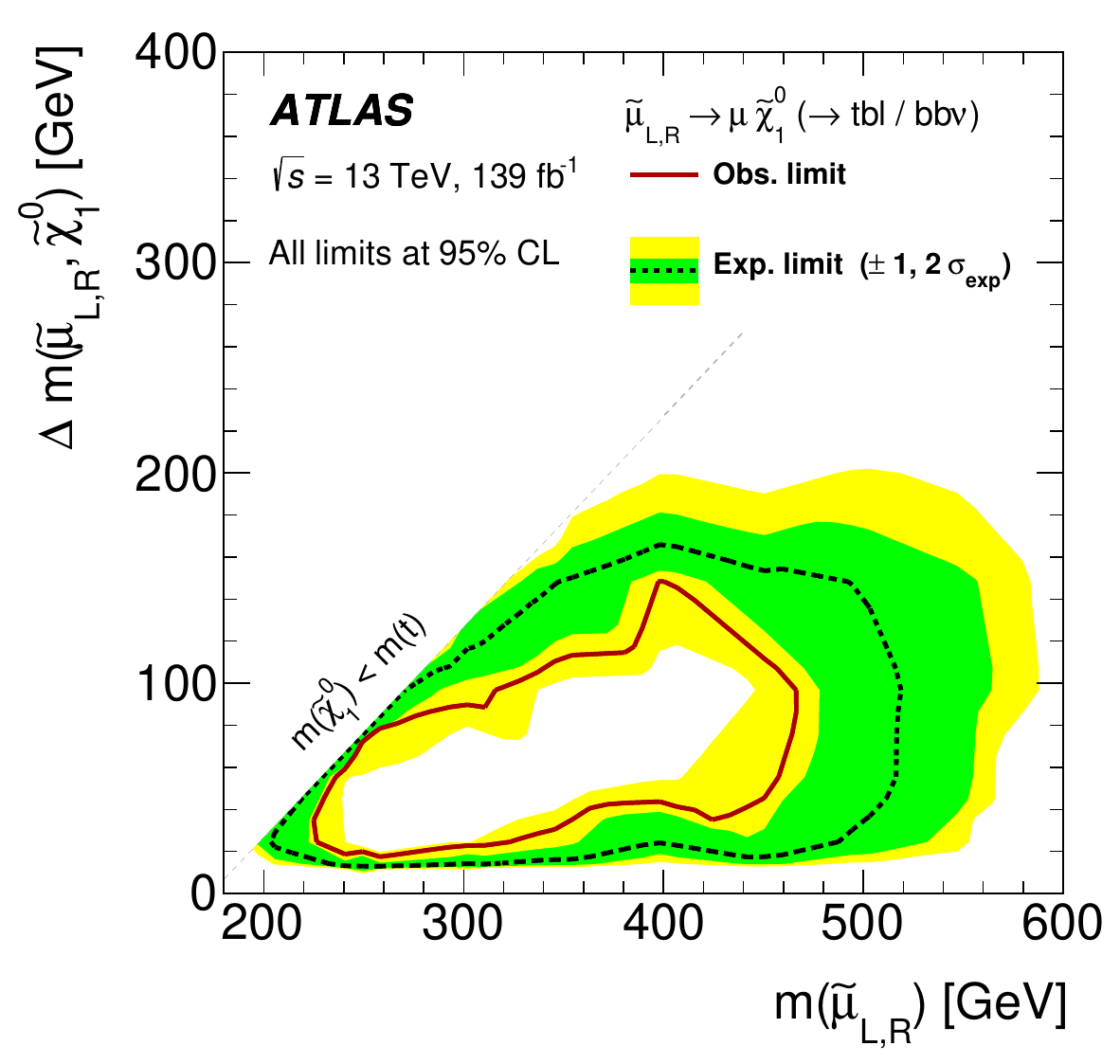}\label{fig:limit_smuon}
\caption{
Expected and observed exclusion limits on the smuon plus bino RPV SUSY model.
The yellow and green contours of the band around the expected limit are the $\pm 1 \sigma$ and $\pm 2 \sigma$ variations including all uncertainties, respectively.
The diagonal grey lines indicates the allowed kinematic limit for the decays. The neutralino is assumed to decay to $tbl$ or $bb\nu$ with equal probability.
}
\label{fig:SUSY_limits2D}
\end{figure}

% End of text imported from the .//Tex/results.tex input file

\FloatBarrier
 
\section{Conclusion}
\label{sec:conclusion}

% The next lines are included from the .//Tex/conclusion.tex input file
A search for a general two Higgs doublet model is presented, where the heavy Higgs bosons feature flavour changing couplings.
Such couplings allow for same-sign top and three-top production among others, with a sizeable charge asymmetry.
The targeted final state is characterised by multiple leptons and multiple $b$-jets.
To improve the sensitivity of the search, events are categorised according to the lepton multiplicity, total lepton charge, and a multi-output deep neural network classifier.
The dominant backgrounds originate from \ttW, \ttZ, and \ttbar, and are estimated from Monte-Carlo simulation and normalised to data.
The analysis is performed with proton--proton collision data at $\sqrt{s} = 13$ \tev\ collected from 2015 to 2018 with the ATLAS
detector at the LHC, corresponding to an integrated luminosity of 139 fb$^{-1}$.
This search is the first collider result on general two Higgs doublet model with flavour violation. It also represents the first search to target explicitly beyond-the-standard-model production of three top quarks.
 
The largest deviation observed with respect to the Standard Model expectation corresponds to a local significance of \bestsignificance\ standard deviations for a signal with $m_{H}=900~\gev$ and couplings $\rhott$=\bestrhott, $\rhotc$=\bestrhotc, and $\rhotu$=\bestrhotu.
Exclusion limits are set on the mass and couplings of the heavy Higgs bosons, where an additional scalar boson with couplings $\rhott=0.4$, $\rhotc=0.2$, and $\rhotu=0.2$ is excluded at 95\% confidence level observed (expected) for masses $m_{H}$ between 200--620 (200--840)~\GeV. Additional mass limits are set for different coupling choices. For a fixed mass of $m_{H}=400~\gev$, exclusion limits are set on the allowed coupling strengths as low as $\rhott=0.3$, $\rhotc=\rhotu=0.18$. Different assumptions are tested to set 2-dimensional exclusion limits on the three couplings.
Additional models based on $R$-parity-violating supersymmetry with the lepton-number-violating coupling $\lambda^\prime_{i33}$ (with $i \in {2,3}$), are used to further interpret the results of the search. Scenarios with direct electroweak production of higgsinos (winos) are excluded for masses between 200--585 (200--670)~\gev. Smuons with masses between 225 and 460~\gev\ are excluded in a model with direct smuon production and decay to a bino-like neutralino, which in turn decays via the $\lambda^\prime_{i33}$ coupling.
These are the first collider limits for these models.

% End of text imported from the .//Tex/conclusion.tex input file

\section*{Acknowledgements}

% The next lines are included from the .//acknowledgements/Acknowledgements.tex input file

We thank CERN for the very successful operation of the LHC, as well as the
support staff from our institutions without whom ATLAS could not be
operated efficiently.
 
We acknowledge the support of
ANPCyT, Argentina;
YerPhI, Armenia;
ARC, Australia;
BMWFW and FWF, Austria;
ANAS, Azerbaijan;
CNPq and FAPESP, Brazil;
NSERC, NRC and CFI, Canada;
CERN;
ANID, Chile;
CAS, MOST and NSFC, China;
Minciencias, Colombia;
MEYS CR, Czech Republic;
DNRF and DNSRC, Denmark;
IN2P3-CNRS and CEA-DRF/IRFU, France;
SRNSFG, Georgia;
BMBF, HGF and MPG, Germany;
GSRI, Greece;
RGC and Hong Kong SAR, China;
ISF and Benoziyo Center, Israel;
INFN, Italy;
MEXT and JSPS, Japan;
CNRST, Morocco;
NWO, Netherlands;
RCN, Norway;
MEiN, Poland;
FCT, Portugal;
MNE/IFA, Romania;
MESTD, Serbia;
MSSR, Slovakia;
ARRS and MIZ\v{S}, Slovenia;
DSI/NRF, South Africa;
MICINN, Spain;
SRC and Wallenberg Foundation, Sweden;
SERI, SNSF and Cantons of Bern and Geneva, Switzerland;
MOST, Taiwan;
TENMAK, T\"urkiye;
STFC, United Kingdom;
DOE and NSF, United States of America.
In addition, individual groups and members have received support from
BCKDF, CANARIE, Compute Canada and CRC, Canada;
PRIMUS 21/SCI/017 and UNCE SCI/013, Czech Republic;
COST, ERC, ERDF, Horizon 2020, ICSC-NextGenerationEU and Marie Sk{\l}odowska-Curie Actions, European Union;
Investissements d'Avenir Labex, Investissements d'Avenir Idex and ANR, France;
DFG and AvH Foundation, Germany;
Herakleitos, Thales and Aristeia programmes co-financed by EU-ESF and the Greek NSRF, Greece;
BSF-NSF and MINERVA, Israel;
Norwegian Financial Mechanism 2014-2021, Norway;
NCN and NAWA, Poland;
La Caixa Banking Foundation, CERCA Programme Generalitat de Catalunya and PROMETEO and GenT Programmes Generalitat Valenciana, Spain;
G\"{o}ran Gustafssons Stiftelse, Sweden;
The Royal Society and Leverhulme Trust, United Kingdom.
 
The crucial computing support from all WLCG partners is acknowledged gratefully, in particular from CERN, the ATLAS Tier-1 facilities at TRIUMF (Canada), NDGF (Denmark, Norway, Sweden), CC-IN2P3 (France), KIT/GridKA (Germany), INFN-CNAF (Italy), NL-T1 (Netherlands), PIC (Spain), ASGC (Taiwan), RAL (UK) and BNL (USA), the Tier-2 facilities worldwide and large non-WLCG resource providers. Major contributors of computing resources are listed in Ref.~\cite{ATL-SOFT-PUB-2023-001}.

% End of text imported from the .//acknowledgements/Acknowledgements.tex input file

\clearpage
\printbibliography
 
\clearpage
% ATLAS Collaboration author list
% Reference date of HDBS-2020-03 is 2023-02-07
% Author list last updated on date 14-SEP-23
% Data extracted on 14-Sep-2023 for paper reference HDBS-2020-03
% at 4:39pm
 
\begin{flushleft}
\hypersetup{urlcolor=black}
{\Large The ATLAS Collaboration}

\bigskip

\AtlasOrcid[0000-0002-6665-4934]{G.~Aad}$^\textrm{\scriptsize 102}$,
\AtlasOrcid[0000-0002-5888-2734]{B.~Abbott}$^\textrm{\scriptsize 120}$,
\AtlasOrcid[0000-0002-1002-1652]{K.~Abeling}$^\textrm{\scriptsize 55}$,
\AtlasOrcid[0000-0001-5763-2760]{N.J.~Abicht}$^\textrm{\scriptsize 49}$,
\AtlasOrcid[0000-0002-8496-9294]{S.H.~Abidi}$^\textrm{\scriptsize 29}$,
\AtlasOrcid[0000-0002-9987-2292]{A.~Aboulhorma}$^\textrm{\scriptsize 35e}$,
\AtlasOrcid[0000-0001-5329-6640]{H.~Abramowicz}$^\textrm{\scriptsize 151}$,
\AtlasOrcid[0000-0002-1599-2896]{H.~Abreu}$^\textrm{\scriptsize 150}$,
\AtlasOrcid[0000-0003-0403-3697]{Y.~Abulaiti}$^\textrm{\scriptsize 117}$,
\AtlasOrcid[0000-0003-0762-7204]{A.C.~Abusleme~Hoffman}$^\textrm{\scriptsize 137a}$
% ...
\bigskip
\\

% ...
$^{29}$Physics Department, Brookhaven National Laboratory, Upton NY; United States of America.\\
% ...
$^{35}$$^{(a)}$Facult\'e des Sciences Ain Chock, R\'eseau Universitaire de Physique des Hautes Energies - Universit\'e Hassan II, Casablanca;$^{(b)}$Facult\'{e} des Sciences, Universit\'{e} Ibn-Tofail, K\'{e}nitra;$^{(c)}$Facult\'e des Sciences Semlalia, Universit\'e Cadi Ayyad, LPHEA-Marrakech;$^{(d)}$LPMR, Facult\'e des Sciences, Universit\'e Mohamed Premier, Oujda;$^{(e)}$Facult\'e des sciences, Universit\'e Mohammed V, Rabat;$^{(f)}$Institute of Applied Physics, Mohammed VI Polytechnic University, Ben Guerir; Morocco.\\
% ...
$^{49}$Fakult\"{a}t Physik , Technische Universit{\"a}t Dortmund, Dortmund; Germany.\\
% ...
$^{55}$II. Physikalisches Institut, Georg-August-Universit\"{a}t G\"ottingen, G\"ottingen; Germany.\\
% ...
$^{102}$CPPM, Aix-Marseille Universit\'e, CNRS/IN2P3, Marseille; France.\\
% ...
$^{117}$Department of Physics, New York University, New York NY; United States of America.\\
% ...
$^{120}$Homer L. Dodge Department of Physics and Astronomy, University of Oklahoma, Norman OK; United States of America.\\
% ...
$^{137}$$^{(a)}$Departamento de F\'isica, Pontificia Universidad Cat\'olica de Chile, Santiago;$^{(b)}$Millennium Institute for Subatomic physics at high energy frontier (SAPHIR), Santiago;$^{(c)}$Instituto de Investigaci\'on Multidisciplinario en Ciencia y Tecnolog\'ia, y Departamento de F\'isica, Universidad de La Serena;$^{(d)}$Universidad Andres Bello, Department of Physics, Santiago;$^{(e)}$Instituto de Alta Investigaci\'on, Universidad de Tarapac\'a, Arica;$^{(f)}$Departamento de F\'isica, Universidad T\'ecnica Federico Santa Mar\'ia, Valpara\'iso; Chile.\\
% ...
$^{150}$Department of Physics, Technion, Israel Institute of Technology, Haifa; Israel.\\
$^{151}$Raymond and Beverly Sackler School of Physics and Astronomy, Tel Aviv University, Tel Aviv; Israel.\\
% ...
$^{a}$ Also Affiliated with an institute covered by a cooperation agreement with CERN.\\
$^{b}$ Also at An-Najah National University, Nablus; Palestine.\\
$^{c}$ Also at APC, Universit\'e Paris Cit\'e, CNRS/IN2P3, Paris; France.\\
$^{d}$ Also at Borough of Manhattan Community College, City University of New York, New York NY; United States of America.\\
$^{e}$ Also at Center for High Energy Physics, Peking University; China.\\
$^{f}$ Also at Center for Interdisciplinary Research and Innovation (CIRI-AUTH), Thessaloniki ; Greece.\\
% ...

\end{flushleft}

% Created with Glance <Atlas.Glance@cern.ch>

\end{document}